%% file: main.tex
\documentclass[twocolumn]{aastex631}

\usepackage{import}
\usepackage{xcolor}
\usepackage{comment}
\usepackage{ulem}
\usepackage{xspace}

\graphicspath{ {./Images/} }

\newcommand{\mdot}{$\dot{M}$\xspace}

\newcommand{\rsun}{R$_{\odot}$\xspace}

\newcommand{\rstar}{R$_{\star}$\xspace}
\newcommand{\lacc}{L$_{\mathrm{acc}}$\xspace}
\newcommand{\lline}{L$_{\mathrm{Line}}$\xspace}
\newcommand{\rin}{R$_{in}$\xspace}
\newcommand{\Wr}{W$_r$\xspace}
\newcommand{\Tmax}{T$_{max}$\xspace}

\newcommand{\accrateunits}{$\times$10$^{-8}$ M$_{\odot}$ yr$^{-1}$\xspace}

\newcommand{\ha}{H${\alpha}$\xspace}
\newcommand{\hb}{H${\beta}$\xspace}
\newcommand{\hy}{H${\gamma}$\xspace}
\newcommand{\hd}{H${\delta}$\xspace}

\newcommand{\Bry}{Br${\gamma}$\xspace}
\newcommand{\hst}{\textit{HST}}

\newcommand{\coude}{$\mathrm{Coud\acute{e}}$\xspace}

\defcitealias{RE19}{RE19}

\shorttitle{Accretion Variability in T Tauri Stars. III. Optical Spectra}
\shortauthors{Wendeborn et al.}

\begin{document}

\title{A Multi-wavelength, Multi-epoch Monitoring Campaign of Accretion Variability in T Tauri Stars from the ODYSSEUS Survey. III.~Optical Spectra}\footnote[1]{Based on observations collected at the European Southern Observatory under ESO programme 106.20Z8.}

\author[0000-0002-6808-4066]{John Wendeborn}
\affil{Institute for Astrophysical Research, Department of Astronomy, Boston University, 725 Commonwealth Avenue, Boston, MA 02215, USA}

\author[0000-0001-9227-5949]{Catherine C. Espaillat}
\affil{Institute for Astrophysical Research, Department of Astronomy, Boston University, 725 Commonwealth Avenue, Boston, MA 02215, USA}

\author[0000-0003-4507-1710]{Thanawuth Thanathibodee}
\affil{Institute for Astrophysical Research, Department of Astronomy, Boston University, 725 Commonwealth Avenue, Boston, MA 02215, USA}

\author[0000-0003-1639-510X]{Connor E. Robinson} 
\affil{Department of Physics \& Astronomy, Amherst College, C025 Science Center 25 East Drive, Amherst, MA 01002, USA}

\author[0000-0001-9301-6252]{Caeley V. Pittman}
\affil{Institute for Astrophysical Research, Department of Astronomy, Boston University, 725 Commonwealth Avenue, Boston, MA 02215, USA}

\author[0000-0002-3950-5386]{Nuria Calvet}
\affil{Department of Astronomy, University of Michigan, 1085 South University Avenue, Ann Arbor, MI 48109, USA}

\author[0000-0002-5943-1222]{James Muzerolle}
\affil{Space Telescope Science Institute, 3700 San Martin Drive, Baltimore, MD 21218, USA}

\author[0000-0001-7796-1756]{Fredrick M. Walter}
\affiliation{Department of Physics and Astronomy, Stony Brook University, Stony Brook NY 11794-3800, USA}

\author[0000-0001-6496-0252]{Jochen Eisl\"offel}
\affil{Th\"uringer Landessternwarte, Sternwarte5, D-07778 Tautenburg, Germany}

\author[0000-0002-5261-6216]{Eleonora Fiorellino}
\affil{INAF -- Osservatorio Astronomico di Capodimonte, via Moiariello 16, 80131 Napoli, Italy}

\author[0000-0003-3562-262X]{Carlo F. Manara}
\affil{European Southern Observatory, Karl-Schwarzschild-Strasse 2, 85748 Garching bei M\"unchen, Germany}

\author[0000-0001-7157-6275]{\'{A}gnes K\'{o}sp\'{a}l}
\affil{Konkoly Observatory, HUN-REN Research Centre for Astronomy and Earth Sciences, CSFK, MTA Centre of Excellence, Konkoly-Thege Mikl\'os \'ut 15-17}
\affil{Institute of Physics and Astronomy, ELTE E\"otv\"os Lor\'and University, P\'azm\'any P\'eter s\'et\'any 1/A, 1117 Budapest, Hungary}
\affil{Max Planck Institute for Astronomy, K\"onigstuhl 17, 69117 Heidelberg, Germany}

\author[0000-0001-6015-646X]{P\'eter \'Abrah\'am}
\affil{Konkoly Observatory, HUN-REN Research Centre for Astronomy and Earth Sciences, CSFK, MTA Centre of Excellence, Konkoly-Thege Mikl\'os \'ut 15-17}
\affil{ELTE E\"otv\"os Lor\'and University, Institute of Physics, P\'azm\'any P\'eter s\'et\'any 1/A, 1117 Budapest, Hungary}
\affil{University of Vienna, Dept. of Astrophysics, T\"urkenschanzstr. 17, 1180, Vienna, Austria}

\author[0000-0001-8194-4238]{Rik Claes}
\affil{European Southern Observatory, Karl-Schwarzschild-Strasse 2, 85748 Garching bei München, Germany}

\author[0000-0003-1401-6444]{Elisabetta Rigliaco}
\affil{INAF/Osservatorio Astronomico di Padova, Vicolo dell’osservatorio 5, 35122 Padova, Italy}

\author[0000-0002-4115-0318]{Laura Venuti}
\affil{SETI Institute, 339 Bernardo Avenue, Suite 200, Mountain View, CA94043, USA}

\author[0000-0002-3913-3746]{Justyn Campbell-White}
\affil{SUPA, School of Science and Engineering, University of Dundee, Nethergate, Dundee dd1 4hn, UK}

\author[0000-0001-7476-7253]{Pauline McGinnis}

\author[0000-0002-8364-7795]{Manuele Gangi}
\affil{ASI, Italian Space Agency, Via del Politecnico snc, 00133 Rome, Italy}
\affil{INAF–Osservatorio Astronomico di Roma, Via Frascati 33, 00078 Monte Porzio Catone, Italy}

\author[0000-0001-8284-4343]{Karina Mauco}
\affil{European Southern Observatory, Karl-Schwarzschild-Strasse 2, 85748 Garching bei München, Germany}

\author{Filipe Gameiro}
\affil{6Instituto de Astrofísica e Ciências do Espaço, Universidade do Porto, CAUP, Rua das Estrelas, PT4150-762 Porto, Portugal}
\affil{Departamento de Física e Astronomia, Faculdade de Ciências, Universidade do Porto, Rua do Campo Alegre 687, PT4169-007 Porto, Portugal}

\author[0000-0002-0474-0896]{Antonio Frasca}
\affil{INAF -- Osservatorio Astrofisico di Catania, via S. Sofia 78, 95123 Catania, Italy}

\author[0000-0003-0292-4832]{Zhen Guo}
\affiliation{Instituto de F{\'i}sica y Astronom{\'i}a, Universidad de Valpara{\'i}so, ave. Gran Breta{\~n}a, 1111, Casilla 5030, Valpara{\'i}so, Chile}

\input{Abstract.tex}


\input{Introduction.tex}

\input{Observations.tex}


\input{Results.tex}

\input{Discussion.tex}

\input{Conclusion.tex}

\input{Acknowledgements.tex}

\vspace{5mm}
\facilities{HST (COS), CTIO:1.5m, VLT, TLS}

\software{Python, specutils, astropy}

\bibliography{biblio}

\input{Appendix.tex}


\end{document}

%% file: Abstract.tex
\begin{abstract}

Classical T Tauri Stars (CTTSs) are highly variable stars that possess gas- and dust-rich disks from which planets form. Much of their variability is driven by mass accretion from the surrounding disk, a process that is still not entirely understood. A multi-epoch optical spectral monitoring campaign of four CTTSs (TW~Hya, RU~Lup, BP~Tau, and GM~Aur) was conducted along with contemporaneous $HST$ UV spectra and ground-based photometry in an effort to determine accretion characteristics and gauge variability in this sample. Using an accretion flow model, we find that the magnetospheric truncation radius varies between 2.5--5 \rstar across all of our observations. There is also significant variability in all emission lines studied, particularly \ha, \hb, and \hy. Using previously established relationships between line luminosity and accretion, we find that, on average, most lines reproduce accretion rates consistent with accretion shock modeling of $HST$ spectra to within 0.5 dex. Looking at individual contemporaneous observations, however, these relationships are less accurate, suggesting that variability trends differ from the trends of the population and that these empirical relationships should be used with caution in studies of variability.

\end{abstract}

%% file: Introduction.tex
\section{Introduction} \label{sec:intro}

Mass accretion in young, pre-main sequence stars known as Classical T Tauri Stars (CTTSs) sets the stage for future evolution of the system \citep[see][]{Williams2011, Hartmann2016, PPVII10, Manara2023, PPVII14}. Yet, despite its importance, a comprehensive understanding of this process remains elusive. In particular, measuring the mass accretion rate (\mdot) is complicated by many factors, not least of which is the observational tracer used.

One such observational tracer is the UV-optical continuum emission that is in excess of the underlying photospheric+chromospheric emission. This excess, produced by the energetic shocks of infalling, accreting material \citep{Hartmann2016}, is one of the most direct tracers of accretion and can be modeled to get estimates of the overall accretion rates. These models vary from a simple hydrogen slab model, assuming the excess emission originates from a hot slab of hydrogen on the stellar surface \citep[e.g.][]{Manara2013, Alcala2017, Manara2021}, to a multi-column shock model, assuming the accretion originates from flows of varying densities \citep[e.g.][]{Calvet1998, Muzerolle1998, RE19}. In \citet[][hereafter Paper I]{PaperI}, we model multi-epoch $HST$ observations of four CTTSs (TW~Hya, RU~Lup, BP~Tau, and GM~Aur) from the ULLYSES \citep[UV Legacy Library of Young Stars as Essential Standards,][]{ULLYSES} survey using a multi-column shock model and find significant variability (factor of $\sim$2--5 within several days) in each target, with slightly elevated accretion rates as compared to previous studies, also shown by \citet{Pittman2022}. We find that generally the connection between accretion rates and UV line luminosities is not significant.

Photometry provides an opportunity to probe accretion with higher cadence than spectroscopy can typically provide. Excess $u$-band luminosity has long been associated with accretion in CTTSs \citep{Gullbring1998, RE19}. In \citet[][hereafter Paper II]{PaperII}, we find strong global relationships between accretion luminosity and excess $uBgVriz$ luminosity, though these connections break down for some bands/targets. Additionally, light curve characteristics (time lags, periodicities) can be used to infer characteristics about the accretion such as the shape/distribution of the hotspot \citep{Nature, Herbert2023, PaperII}, hotspot latitude \citep{Siwak2014, Siwak2018}, and structural changes in the flow \citep{Blinova2016, Venuti2017, SiciliaAguilar2020, Zsidi2022}.

Emission lines in optical and NIR spectra provide another opportunity to estimate accretion rates by more directly probing the accretion flows themselves. Empirical relationships between line strength and accretion have been established using large samples of CTTS and (often) simultaneous estimates of accretion rates and line luminosities \citep{Muzerolle1998b, Natta2004, Herczeg2008, Ingleby2013, Alcala2014, Alcala2017}. While the connections between \lacc and \lline are strong, their use in estimating individual CTTS accretion rates introduces significant uncertainty and scatter \citep[e.g., ][]{Fiorellino2021, Bouvier2023, Herczeg2023, Nelissen2023}. Alternatively, \citet{Muzerolle2001, Espaillat2008, Alencar2012, Thanathibodee2019, Thanathibodee2023} use magnetospheric accretion flow models to model the emission profiles of the hydrogen Balmer series. 

Here we employ both accretion flow modeling and empirical \lacc-\lline relationships to estimate and better understand the accretion in a monitoring campaign of four CTTSs: TW~Hya, RU~Lup, BP~Tau, and GM~Aur by the ULLYSES and PENELLOPE \citep{Manara2021} surveys, using contemporaenous data from the ODYSSEUS \citep[Outflows and Disks Around Young Stars: Synergies for the Exploration of ULLYSES Spectra,][]{Espaillat2022} collaboration. See Paper I (Table 1, Section 2) for details on these targets. Here we present our optical observations and data in Section \ref{sec: Observations}. In Section \ref{sec: Analysis and Results}, we describe the accretion flow model and the results of modeling the monitoring data. Next, in Section \ref{sec: Discussion}, we discuss these results in more detail and connect them to the results of our shock modeling (Paper I) and photometry (Paper II). We present our final conclusions and summarize in Section \ref{sec: Conclusion}.

%% file: Observations.tex
\section{Observations and Data Reduction} \label{sec: Observations}

Multi-wavelength, multi-epoch observations of the CTTSs TW~Hya, RU~Lup, BP~Tau, and GM~Aur were carried out in 2021 (Epoch 1/E1) and 2022 (Epoch 2/E2). More background information on the individual objects can be found in Paper I. Here, we present the results of our optical spectral monitoring. Contemporaneous $HST$ UV spectra are presented in Paper I while contemporaneous UV-NIR photometry is presented in Paper II. 

The contemporaneous optical spectra presented here were obtained on a variety of instruments, including SMARTS/CHIRON, VLT/ESPRESSO, VLT/XSHOOTER, VLT/UVES, Haute-Provence/SOPHIE, and Tautenburg/TCES. Details of these observations can be found below and in Table \ref{tab: Optical Spectra Observations}. All spectra have been corrected for radial velocity: TW Hya, 12.3 km s$^{-1}$ \citep{Soubrian2018}; RU~Lup, 3.3 km s$^{-1}$ \citep{Frasca2017}; BP~Tau, 16.6 km s$^{-1}$ \citep{Jonsson2020}; GM~Aur 15.2 km s$^{-1}$ \citep{Nguyen2012}. All spectra have also been de-reddened using the extinction values listed in Table 1 of Paper I and the reddening law of \citet{Whittet2004} assuming R$_\mathrm{V}$=3.1 and constant extinction A$_V$. 

In Paper II we show that variable extinction does not contribute significantly to the variability of our targets. To test if the variability was due to local dust, we compared the observed color slopes (for $u-g$, $B-g$, $g-V$, $g-r$, $g-i$, and $g-z$ versus $g$) to that predicted by a local population of dust grains. We consider a range of grain sizes, from 0.1--10 $\mu$m, including both silicate and graphite grains. We found that no population of dust grains can reproduce the photometric color slopes we see in any object. Further, variable extinction is typically associated with dipper-like events from disk warps/inhomogeneities, which our light curves do not exhibit. The light curves used for these analyses cover a wider time span than our optical spectra, showing that variable extinction should not be significant during our optical monitoring.

\input{Optical_Spectra_Observations_Table.tex}

\subsection{CHIRON} \label{sec:chiron}

High-resolution (R$\sim$80,000) optical spectra of all 4 targets were obtained with the CHIRON \citep{CHIRON} spectrograph on the Small and Medium Aperture Research Telescope System (SMARTS) 1.5m telescope. These spectra cover wavelengths of about 4100--8800 {\AA} and have been reduced using a custom reduction pipeline with improved treatment of background emission and bright emission lines\footnote{$https://www.astro.sunysb.edu/fwalter/SMARTS/\\CHIRON/ch\_reduce.pdf$}.

They were obtained contemporaneously with $HST$ spectra (see Paper I) and photometry (see Paper II) in both E1 and E2, though the number of observations and cadence vary by target/epoch. TW~Hya and RU~Lup were monitored with roughly nightly cadence in both E1 and E2. BP~Tau was observed twice in E1, and both BP~Tau and GM~Aur were monitored with roughly nightly cadence in E2. More details on the timing of these observations can be found in Table \ref{tab: Optical Spectra Observations}.

\subsection{ESPRESSO, X-SHOOTER, and UVES} \label{sec: Other Optical Spectra}

Several other optical spectra were obtained using the Echelle Spectrograph for Rocky Exoplanets and Stable Spectrographic Observations \citep[ESPRESSO,][]{ESPRESSO}, X-Shooter \citep{XSHOOTER}, and UV-visual Echelle Spectrograph \citep[UVES,][]{UVES} instruments on the Very Large Telescope (VLT) alongside the $HST$ observations for all 4 targets in one or both epochs. Based on observations collected at the European Southern Observatory under ESO programmes 106.20Z8.001, 106.20Z8.002, 106.20Z8.003, 106.20Z8.004, 106.20Z8.005, 106.20Z8.006, 106.20Z8.007, 106.20Z8.011 as part of the PENELLOPE Large VLT Program \citep{Manara2021}. Details of these observations can be found in Table \ref{tab: Optical Spectra Observations}.

Our ESPRESSO spectra cover 3800--7900 {\AA} at high resolution (R$\sim$140,000). X-Shooter spectra are obtained using 3 arms (UV, visible, NIR) covering 2989--5560, 5337--10200, and 9940--24790 {\AA}, respectively. Spectral resolution varies by arm: 5,400, 18,400, and 11,600 for the UV, visible and NIR arms. UVES spectra are obtained using two arms (UV, Visible) covering 3282--4563 and 4726--6835 {\AA} with spectral resolutions of 71,000 and 87,400, respectively. Details regarding the reduction of these spectra can be found in \citet{Manara2021} and the data can be downloaded from the PENELLOPE Zenodo website\footnote{https://zenodo.org/communities/odysseus/} and the ESO Archive.

\subsection{SOPHIE} \label{sec: SOPHIE}

We also utilize observations from Observatoire de Haute-Provence using the Spectrographe pour l’Observation des Phénomènes des Intérieurs stellaires et des Exoplanètes (SOPHIE) for GM~Aur. GM~Aur was observed with SOPHIE 15 times roughly nightly between MJD=59499$\sim$59517. These spectra cover 3870--6940 {\AA} at a resolution of about 40,000. See \citet{Bouvier2023} (who first published these spectra) for details of the data reduction and processing.

\subsection{TCES}

GM~Aur was monitored with the Tautenburg \coude Echelle Spectrograph (TCES) on the 2-m Alfred Jensch telescope at Th\"uringer Landessternwarte Tautenburg. All 17 observations were obtained with a 2\arcsec\ wide slit, providing a spectral resolution of R$\sim$67,000 from about 4660--7350 \AA. The data reduction was done using the Tautenburg Spectroscopy Pipeline – $\tau$-spline \citep{Sabotta2019}. This includes the usual steps of bias-subtraction, flat-fielding, removal of cosmic rays, scattered light subtraction, extraction, wavelength calibration, and normalization. The pipeline uses standard IRAF\footnote{IRAF is distributed by the National Optical Astronomy Observatories, which are operated by the Association of Universities for Research in Astronomy, Inc., under cooperative agreement with the National Science Foundation.} and PYRAF\footnote{PYRAF is a product of the Space Telescope Science Institute, which is operated by AURA for NASA} routines and the Cosmic Ray code by Malte Tewes\footnote{https://github.com/grzeimann/Panacea/blob/master/cosmics.py} based on the method by \citet{VanDokkum2001}.

\subsection{Flux Calibration}

In order to properly compare line fluxes at different epochs and to utilize empirical relationships such as those from \citet{Alcala2017}, we need flux-calibrated spectra. To this end, we flux calibrate our spectra by using contemporaneous photometry (see Paper II). For each observation, we find all photometry contemporaneous within 6 hours. Then, for each filter (X) we calculate an average photometric flux (F$_{Phot,X}$) where photometry obtained closer in time to the spectrum is weighted more, as $\frac{1}{\Delta t + 0.1}$, where $\Delta t$ is the time difference (in days) between the spectrum and photometry point. Then for each F$_{Phot,X}$, we calculate a scaling factor $s_X = F_{Phot,X} / F_{Spec, X}$ and scale the entire spectrum by the average $s_X$ for all filters with at least one contemporaneous photometry point. In total, we are able to flux-calibrate 236 of 269 spectra. This includes the XSHOOTER spectra which have already been flux-calibrated, though perform additional calibration for consistency. It is important to note that the final line fluxes are sensitive to the assumed extinction, A$_V$, which here is assumed to be constant throughout all observations for each target.

%% file: Optical_Spectra_Observations_Table.tex
\begin{deluxetable*}{c c c c c c c}[htp]
\setlength{\tabcolsep}{10pt}
\tablecaption{Optical Spectral Observations \label{tab: Optical Spectra Observations}}
\centering
\tablehead{
\colhead{Object} & \colhead{Epoch} & \colhead{Instrument} & \colhead{MJD} & \colhead{Date} & \colhead{\# of} & \colhead{Int. Time$^{\dagger}$} \\
\colhead{} & \colhead{} &\colhead{} & \colhead{[Begin/End]} & \colhead{[Begin/End]} & \colhead{Observations} & \colhead{[s]}
}
\startdata
TW Hya & 1 & CHIRON$^a$ & 59241.3/59314.2 & 2021-01-27/2021-04-10& 26 & 600 \\
 &  & ESPRESSO$^b$ & 59280.3/59313.2 & 2021-03-07/2021-04-09 & 4 & 720 \\
 &  & XSHOOTER$^c$ & 59307.0/59310.2 & 2021-04-03/2021-04-06 & 2 & 140, 50, 40$^*$ \\
\hline
TW Hya & 2 & CHIRON$^a$ & 59653.1/59698.1 & 2022-03-15/2022-04-29 & 42 & 600 \\
 &  & UVES$^d$ &  59667.0/59669.0 & 2022-03-29/2022-03-31 & 2 & 340, 340$^*$ \\
\hline
\hline
RU Lup & 1 & CHIRON$^a$ & 59264.4/59453.0 & 2021-02-19/2021-08-27 & 20 & 1800 \\
 &  & ESPRESSO$^b$ & 59449.0/59458.1 & 2021-08-23/2021-09-01 & 2 & 600 \\
 &  & XSHOOTER$^c$ & 59436.1/59448.1 & 2021-08-10/2021-08-22 & 2 & 140, 50, 30$^*$ \\
\hline
RU Lup & 2 & CHIRON$^a$ & 59676.2/59817.0 & 2022-04-07/2022-08-26 & 27 & 1800 \\
 &  & ESPRESSO$^b$ & 59801.0/59814.0 & 2022-08-10/2022-08-23 & 5 & 600 \\
\hline
\hline
BP Tau & 1 & CHIRON$^a$ & 59467.4/59470.4 & 2021-09-10/2021-09-13 & 2 & 1200 \\
 &  & ESPRESSO$^b$ & 59459.4-59464.4 & 2021-09-02/2021-09-07 & 2 & 1000 \\
 &  & XSHOOTER$^c$ & 59448.4-59460.4 & 2021-08-22/2021-09-03 & 3 & 150, 100, 50$^*$ \\
\hline
BP Tau & 2 & CHIRON$^a$ & 59845.4-59953.1 & 2022-09-23/2023-01-09 & 42 & 1200 \\
 &  & ESPRESSO$^b$ & 59928.1-59936.1 & 2022-12-15/2022-12-23 & 5 & 1000 \\
 &  & XSHOOTER$^c$ & 59928.1-59931.2 & 2022-12-15/2022-12-18 & 2 & 150, 100, 50$^*$\\
\hline
\hline
GM Aur & 1 & ESPRESSO$^b$ & 59509.2-59556.2 & 2021-10-22/2021-12-08 & 5 & 1200 \\
 &  & XSHOOTER$^c$ & 59504.3-59556.2 & 2021-10-17/2021-12-08 & 2 & 390, 300, 100$^*$ \\
 &  & SOPHIE$^e$ & 59499.0-59516.0 & 2021-10-12/2021-10-29 & 15 & 3600 \\
 &  & TCES$^f$ & 59503.0-59559.9 & 2021-10-16/2021-12-11 & 17 & 3600 \\
\hline
GM Aur & 2 & CHIRON$^a$ & 59910.2-59931.1 & 2022-11-27/2022-12-18 & 18 & 1200 \\
 &  & ESPRESSO$^b$ & 59910.2-59916.1 & 2022-11-27/2022-12-03 & 2 & 1200 \\ 
\enddata
\tablenotetext{a}{CHIRON: Spectral coverage $=4100-8000$ \AA, R$\sim80,000$}
\tablenotetext{b}{ESPRESSO: Spectral coverage $=3800-7900$ \AA, R$\sim140,000$}
\tablenotetext{c}{XSHOOTER: Spectral coverage $=3000-24800$ \AA, R$\sim5,400-11,600$}
\tablenotetext{d}{UVES: Spectral coverage $=3800-6800$ \AA, R$\sim71,000-87,400$}
\tablenotetext{e}{SOPHIE: Spectral coverage $=3900-6900$ \AA, R$\sim40,000$}
\tablenotetext{f}{TCES: Spectral coverage $=4660-7350$ \AA, R$\sim67,000$}
\tablenotetext{\dagger}{Exact integration times may vary within and between epochs}
\tablenotetext{*}{Integration times differ between spectral arms}
\end{deluxetable*}

%% file: Results.tex
\section{Analysis and Results} \label{sec: Analysis and Results}

In the following sections we analyze and model the optical spectra presented in Section~\ref{sec: Observations}. We first present fluxes for several optical emission lines, focusing on the Balmer lines, \ha, \hb, \hy, and \hd, plus seven other He lines. We also briefly discuss our measurements of optical veiling. Finally, we present the accretion flow model and its fit to the \ha, \hb, and \hy lines.

\subsection{Optical Lines} \label{sec: Results - Optical Lines}

\begin{figure*}
    \includegraphics[width = 0.98\textwidth]{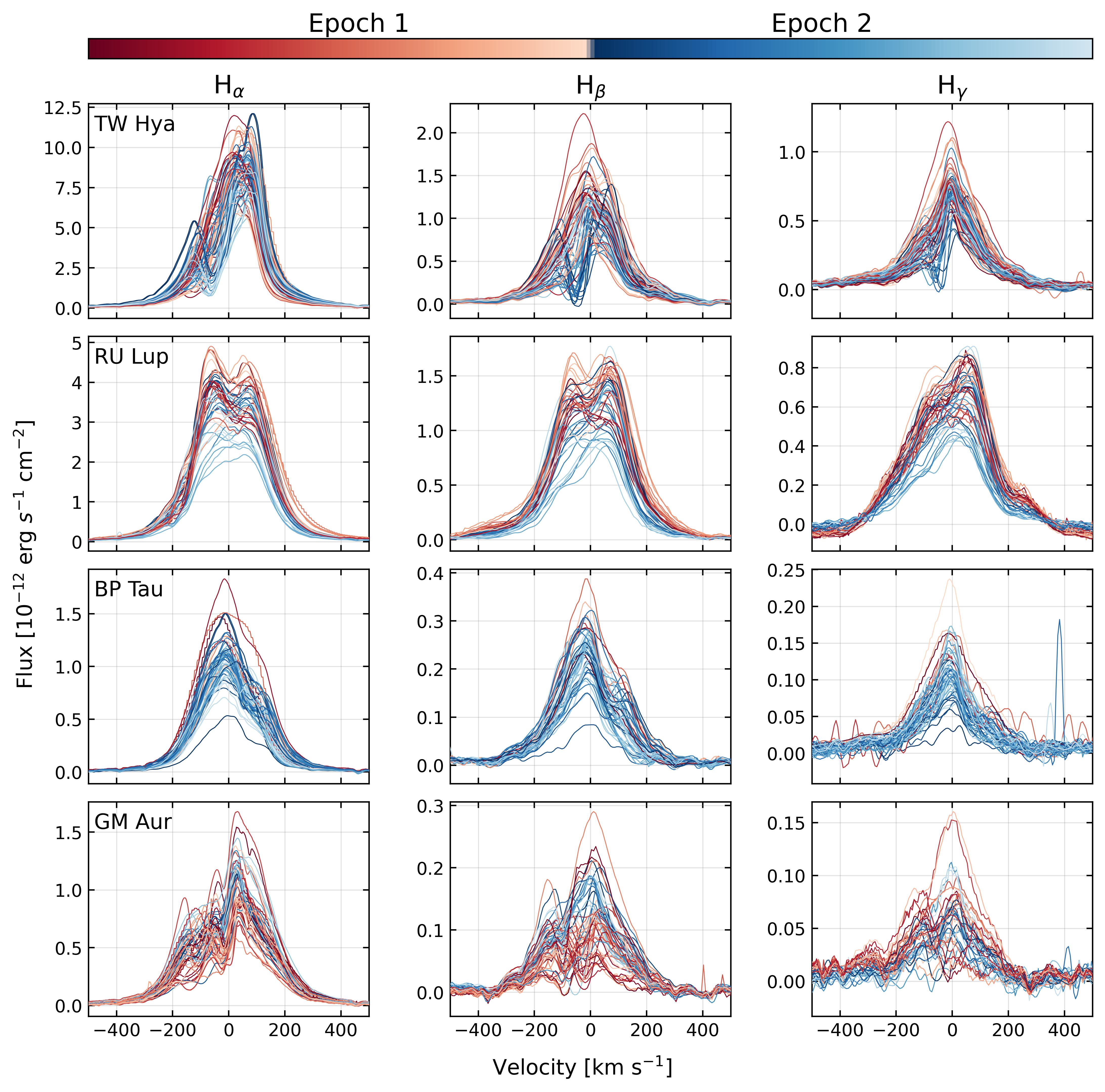}
    \caption{Continuum-subtracted line profiles for \ha, \hb, and \hy\ (left to right) for TW~Hya, RU~Lup, BP~Tau, and GM~Aur (top to bottom). Red lines are spectra from E1, while blue lines are spectra from E2. Darker lines denote spectra obtained earlier in their respective epoch. Spectra have been smoothed using a Savitzky-Golay filter for clarity.}
    \label{fig: Balmer Profiles}
\end{figure*}

\begin{figure*}
    \includegraphics[width = 0.97\textwidth]{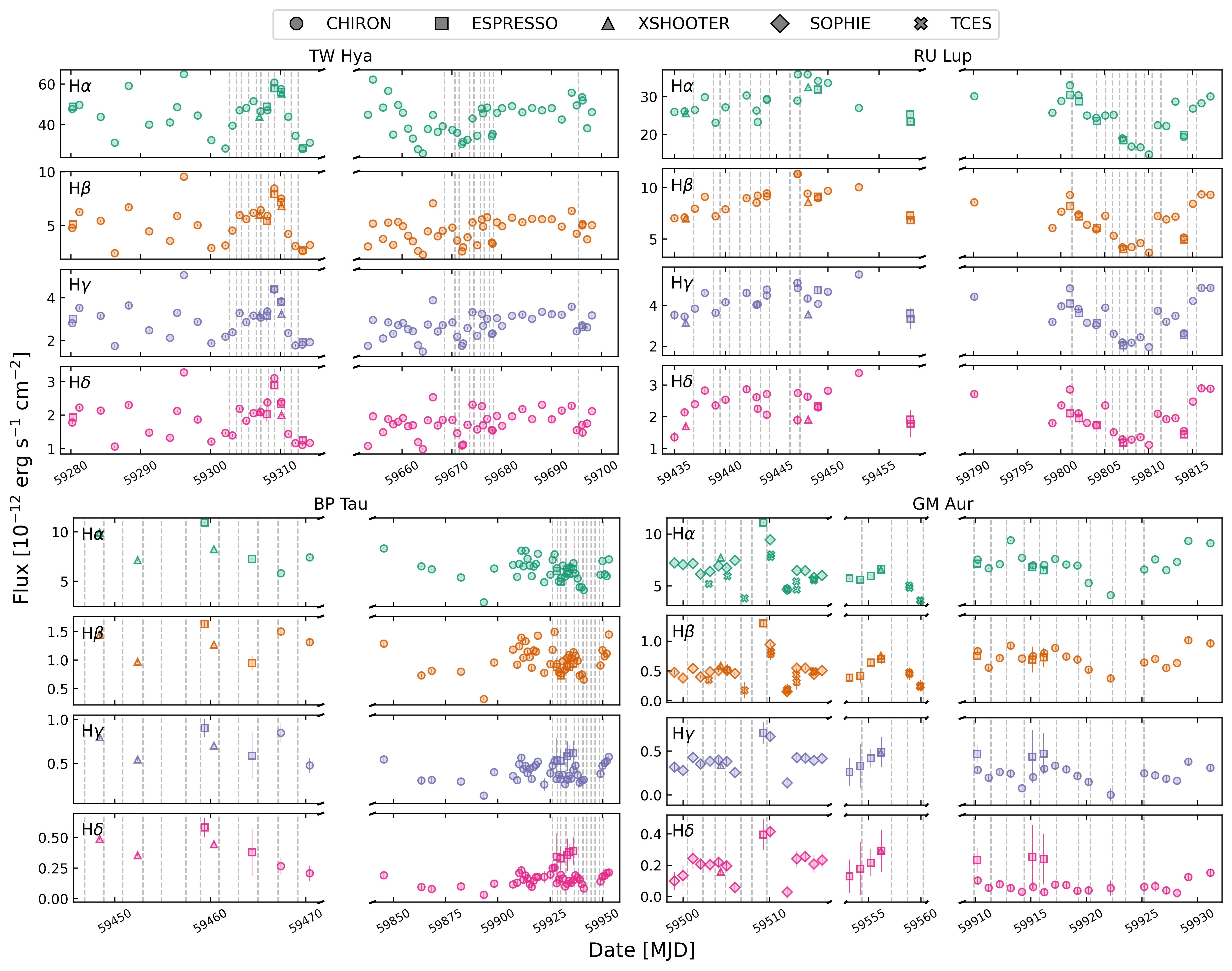}
    \caption{Flux versus time for \ha, \hb, \hy, and \hd (top to bottom) for TW~Hya (top left), RU~Lup (top right), BP~Tau (bottom left) and GM~Aur (bottom right). Marker shape denotes the instrument used, with CHIRON, ESPRESSO, XSHOOTER, SOPHIE, and TCES as circles, squares, triangles, diamonds, and crosses, respectively. Line flux is in units of 10$^{-12}$ erg/s/cm$^2$. Dashed grey lines are times of the $HST$ observations from Paper I. Note that some observations that are far removed in time are not included here. Note the broken axis for GM~Aur, E1.}
    \label{fig: Balmer Fluxes vs Time}
\end{figure*}

\input{Optical_Lines}

We focus here on 11 emission lines (see details in Table \ref{tab: Emission Line Info}), all of which have previously been correlated with accretion \citep{Alcala2017}. To extract the line, we estimate the continuum as a 3rd degree polynomial fit to a region of $\pm$200 {\AA}, ignoring the central line region of width described in Table~\ref{tab: Emission Line Info}. We calculate the line flux as the total integrated flux under the continuum-subtracted spectrum. Due to some uncertainty in both the flux calibration and determination of the continuum, we add an additional 10\% uncertainty in quadrature to the fluxes for all lines except \ha and \hb, where their continuum is more well-defined. Calculated fluxes are listed in Appendix A (Tables \ref{tab: Line Fluxes TW Hya}--\ref{tab: Line Fluxes GM Aur}), while fluxes versus time for the Balmer lines (\ha, \hb, \hy, and \hd) are shown in Figure~\ref{fig: Balmer Fluxes vs Time}. Profiles for the three brightest Balmer lines, \ha, \hb, and \hy are shown in Figure \ref{fig: Balmer Profiles}, with observations from E1 in red and E2 in blue. Fluxes versus time for the remaining lines are shown in Figures~\ref{fig: Line Fluxes Other, TW Hya and RU Lup}--\ref{fig: Line Fluxes Other, BP Tau and GM Aur} in Appendix~\ref{appendix: Optical Line Luminosities}.

\subsubsection{Balmer Lines} \label{sec: Results - Balmer Lines}

Here we focus on \ha, \hb, \hy, and \hd (though we do not discuss \hd). These are the brightest lines in our study and are known to closely trace the accretion flows \citep{Muzerolle2001, Bouvier2023}. In some cases they also trace outflows/winds, which manifest as absorption features and can complicate the line shape \citep{Dupree2014, Bouvier2023}. Variability in the emission strength is indicative of changes in the accretion flow geometry and/or rate, while changes in the absorption features are largely indicative of changes in the wind and are often stochastic.

TW~Hya's emission lines are characterized by bright, asymmetric profiles (Figure \ref{fig: Balmer Profiles}, top row). The asymmetry is primarily due to strong, blue-shifted absorption from --100 to --50~km s$^{-1}$. This absorption reaches below the continuum in \hb and \hy and is strongest in E2. There is also a narrow, low-velocity red-shifted emission component present in some observations in E2. Peak-to-peak, \ha, \hb, and \hy fluxes vary by factors of 2.5, 4.1, and 3.4, respectively.

As seen in our shock modeling and photometry from Papers I and II, RU~Lup's emission lines (Figure \ref{fig: Balmer Profiles}, second row) are brightest in E1 and reach their lowest point during our $HST$ monitoring in E2. \ha and \hb are largely symmetric in both epochs, though some blue-shifted absorption appears stronger in E2. \hy is largely asymmetric, with stronger emission on its red side. Additionally, the red and blue wings in \hy are different from one another and from the other targets: the blue wing appears slightly shallower than the other targets while the red wing drops steeply below $\sim$200 km s$^{-1}$, but slowly afterwards. Peak-to-peak, \ha, \hb, and \hy fluxes vary by factors of 2.4, 3.1, and 2.8, respectively. 

BP~Tau (Figure \ref{fig: Balmer Profiles}, third row) exhibits the most consistent line shapes in our sample, varying primarily in intensity. All three lines show red-shifted absorption with symmetric red/blue wings and little to no obvious blue-shifted absorption. The red absorption appears to be fairly broad and/or weak, as BP~Tau does not show strong, sharp absorption features like TW~Hya and GM~Aur. Peak-to-peak, \ha, \hb, and \hy fluxes vary by factors of 3.8, 5.1, and 6.7, respectively.

The Balmer lines in GM~Aur (Figure \ref{fig: Balmer Profiles}, last row) show the strongest variability and most complex morphology. The shape of \ha is dominated by strong, variable blue-shifted absorption from -200 to -50 km s$^{-1}$ in both E1 and E2. \hb and \hy also show very strong absorption in both epochs, but it is much more pronounced in E1. While the red wing in \hy is never unambiguously in absorption, it does appear to be more depressed than the other lines/targets, suggestive of matter inflow. Peak-to-peak, \ha, \hb, and \hy fluxes vary by factors of 3.1, 8.5, 201.4, respectively. The dramatic variability in \hy is due to a particularly low observation where \hy is largely washed out by noise. Ignoring this observation, the peak-to-peak variability in \hy is 9.0.

While we cannot rule out the possibility that extinction plays some role in the different variability amplitudes of the Balmer lines, it is unlikely to be the primary cause. It is more likely that these differences originate in the accretion flows and outflows. The absorption features (from outflows) can be quite strong, but are stochastic and not directly correlated with accretion. Additionally, each Balmer line has a different sensitivity to the flow’s temperature and optical depth \citep{Muzerolle2001}, leading to different variability between each line since they probe different parts of the accretion flow. Previous accretion shock modeling shows that the accretion occurs along columns of different radial density profiles \citep[e.g.][]{Ingleby2013, RE19, Nature, Pittman2022}. Material of different densities has different contributions to the Balmer lines and so the accretion rates derived from these lines may be different from one another or from that of the total flow, as measured by the accretion shock model.

\subsubsection{Non-Balmer Lines} \label{sec: Results - Non-Balmer Lines}

Beyond the three bright Balmer lines described above, we also focus on 7 additional He lines present in most of our optical spectra. These are listed in Table \ref{tab: Emission Line Info}. In general, these lines exhibit similar variability when compared to the Balmer lines, though with additional scatter. In GM~Aur, some of these lines are not detected, resulting in near- or sub-zero fluxes. 

\begin{figure*}
    \centering
    \includegraphics[width=0.975\textwidth]{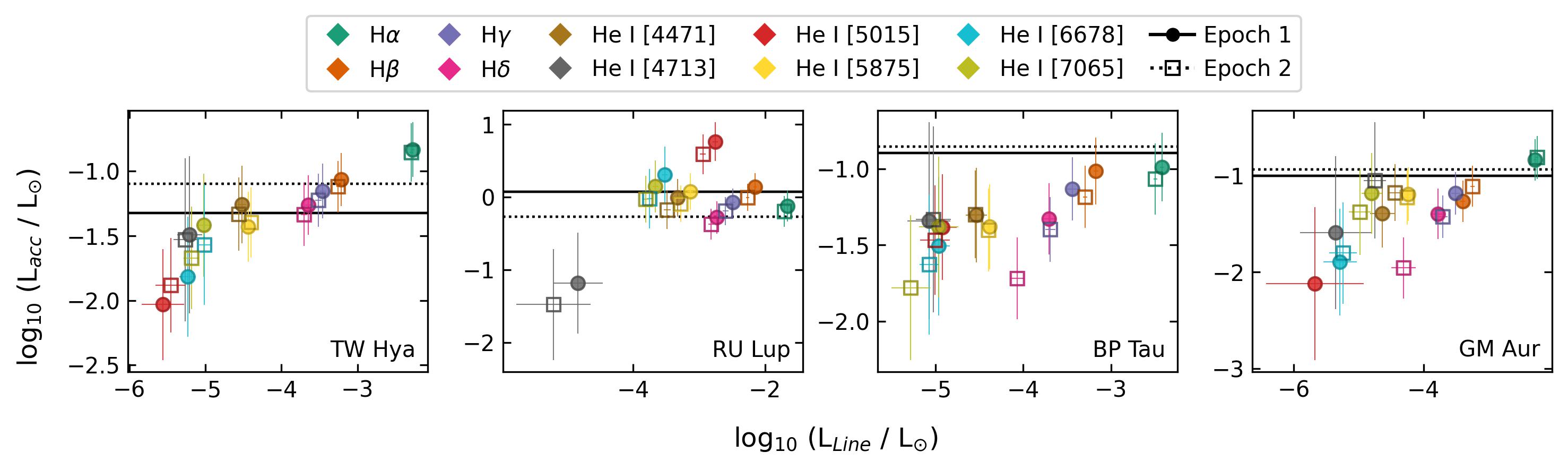}
    \caption{Median line luminosities converted to accretion luminosities using the empirical relationships of \citet{Alcala2017}. Filled circles denote \lacc from Epoch 1, while open squares denote Epoch 2. Lines are the median \lacc from the shock modeling from Paper I, where the solid/dotted lines are E1/E2. Marker colors denote different lines; see the legend.}
    \label{fig: L_Line to L_acc}
\end{figure*}

\citet{Alcala2017} simultaneously and self-consistently estimate the accretion luminosities and line luminosities of 92 YSOs and derive relationships between these parameters. For every line they study, they find strong correlations, suggesting a connection between optical line luminosity and accretion luminosity. We use these same relationships to estimate mean accretion luminosities in our sample. These results are shown in Figure \ref{fig: L_Line to L_acc}.

Besides \hd, the Balmer lines, which tend to be bright and robustly detected in all observations, provide the most consistent \lacc as compared to our shock modeling from Paper I. On average, \ha, \hb, and \hy are consistent to within 0.5 dex (a factor of $\sim$3), with \ha typically being closest. That said, there is little consistency as to whether an epoch with higher median accretion also sees brighter median emission lines; only in RU~Lup are the lines dimmer in conjunction with the lower accretion rate in Epoch 2. Some He lines (in particular He I$_{4471}$, He I$_{5875}$, and He I$_{7065}$) are generally equally as consistent as the Balmer lines. The other lines can deviate by up 1.5 dex (a factor of $\sim$30) with large uncertainties. Like the Balmer lines, fluxes of the He lines, and therefore their use with the \citet{Alcala2017} relationships, are sensitive to the adopted value of extinction.

\subsection{Optical Veiling}

The strong continuum emission produced by the accretion process at optical wavelengths can be comparable to the underlying photospheric emission. This can act to ``fill-in" absorption lines in a process called \textit{veiling}. Estimating the veiling is important not only for the accretion shock model (see Paper I), but is closely tied to accretion \citep[e.g.][]{Ingleby2013, Sousa2023, Herczeg2023, Nelissen2023}. These veiling measurements, along with contemporaneous $V$-band photometry from Paper II, are used to scale WTTS spectra as accurate photospheric templates in Paper I.

We estimate veiling in our optical spectra following the basic method of \citet{Hartigan1989}. First, after convolving our de-reddened (assuming constant A$_V$, see Paper I) target and template spectra to a uniform wavelength grid, we split them into 75~{\AA} intervals, removing intervals with missing data or that overlap with bright emission lines like \ha and \hb. Then for each interval, we perform continuum normalization by fitting a 3rd degree polynomial to non-outlying points. After continuum normalization, we align the target and template spectra using cross-correlation. Next, we add some fraction of the template continuum (a flat, featureless spectrum) to the target spectrum and calculate the Root Mean Square Error (RMSE) between the target and template spectra. The added continuum varies between 0--300\% of the template continuum level. The fractional veiling in each 75~{\AA} bin is the added continuum that minimizes the dispersion between the target and template spectra, divided by the template continuum level. Finally, we fit a 3rd degree polynomial to all the fractional veilings, and r$_V$ is the value of this fitted curve at $\lambda$=5500~{\AA}. We apply a Monte Carlo approach to this procedure, repeating the veiling calculation 500 times on a noise-added spectrum. This acts to minimize noise and outliers. We do not consider rotational broadening in our calculation of r$_V$, as some others do \citep[e.g.,][]{Frasca2016, Manara2020}. However, tests with rotational broadening showed r$_V$s about 0.1 larger, in turn resulting in accretion rates about 2.1\%, 6.3\%, and 2.9\% higher for TW~Hya, BP~Tau, and GM~Aur, respectively, which does not significantly impact any of the results reported here.  

\begin{figure*}[h]
    \centering
    \includegraphics[width=0.8\textwidth]{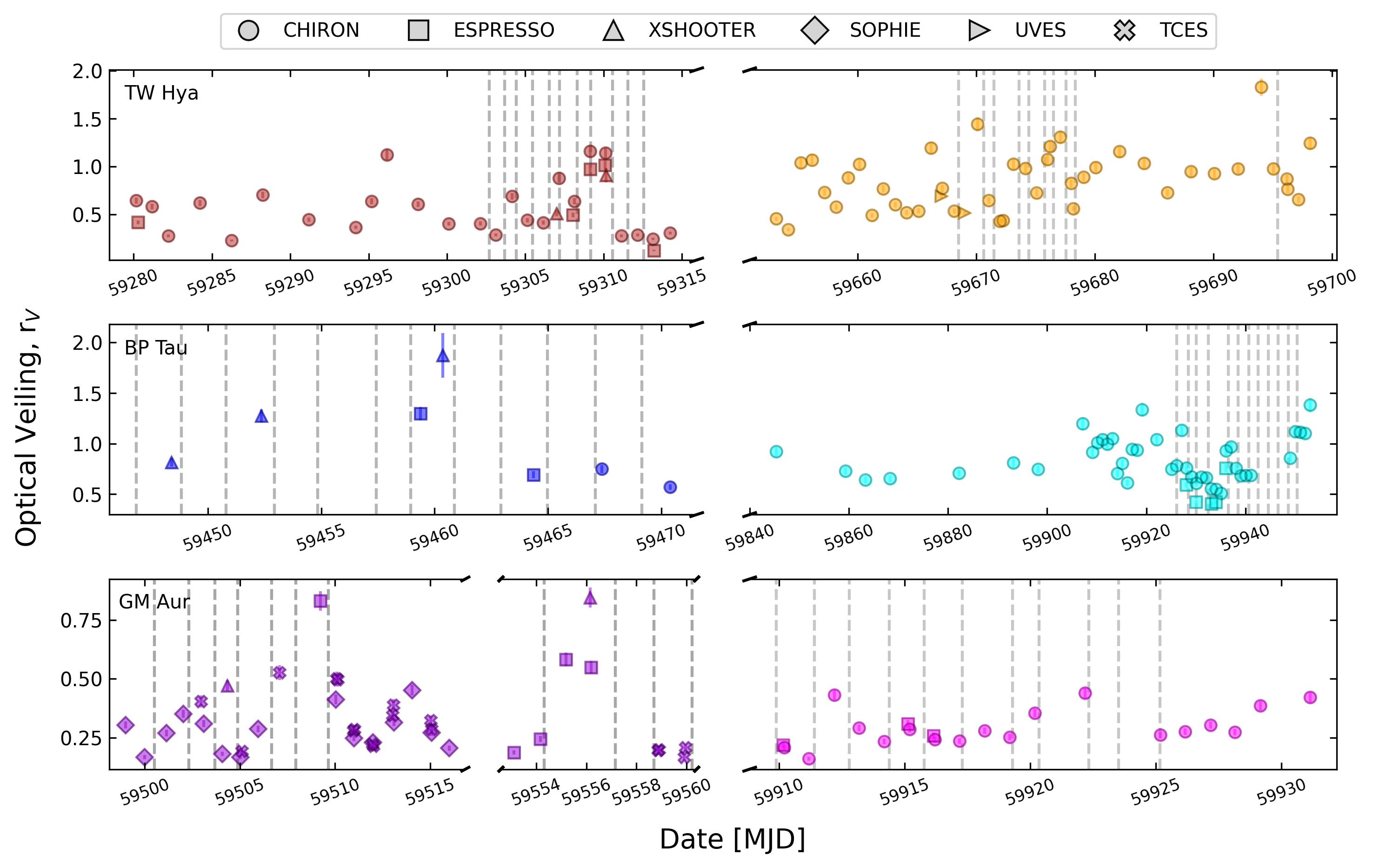}
    \caption{Optical veiling (r$_V$) estimates for TW~Hya (red/orange, top row), BP~Tau (blue/cyan, middle row), and GM~Aur (purple/pink, bottom row). We do not estimate r$_V$ for RU~Lup due to the abundance of strong emission lines.}
    \label{fig: Veilings}
\end{figure*}

Veiling estimates for TW~Hya, BP~Tau, and GM~Aur are shown in Figure \ref{fig: Veilings} and Appendix~A (Tables \ref{tab: Flow Model Results TW Hya}--\ref{tab: Flow Model Results GM Aur}). Note that because of an abundance of emission lines in its optical spectra (due to its high accretion rate), we are unable to determine r$_V$ for RU~Lup. r$_V$ in TW~Hya varies between 0.12--1.83, with generally higher values in E2. BP~Tau is typically more strongly veiled than TW~Hya, with r$_V$ between 0.40--1.87. GM~Aur is the least veiled in our sample, with r$_V$ typically between 0.16--0.84. 

\subsection{Accretion Flow Modeling} \label{sec: Results - Flow Model}

We utilize the accretion flow models of \citet{Hartmann1994} and \citet{Muzerolle1998, Muzerolle2001}. These models assume an axisymmetric accretion flow from the surrounding gas disk. This flow follows along the magnetic field structure, which is assumed to be dipolar and aligned with the stellar rotation axis. The flow is parameterized by the disk truncation radius (R$_{in}$), the width of the flow (W$_{r}$), the mass accretion rate (\mdot), the maximum temperature of the flow ($T_{max}$) and the viewing inclination angle ($i$). After preliminary tests using a large, coarse grid for \ha, we create a smaller, finer grid of models for each target, with ranges of parameters shown in Table \ref{tab: Flow Model Grid}. In nearly all cases of fitting \ha, the resulting parameters are far from the parameter bounds. Only in a few cases for TW~Hya, BP~Tau, and GM~Aur does our fitting suggest that the flow width, \Wr, is smaller than the lower bound of our grid, typically 0.1 \rstar. \hb and \hy also prefer \Wr below the lower bound of our grid. We do not expand our grid to smaller \Wr than 0.1 \rstar, as even 0.1 \rstar is smaller than expected from 3D MHD modeling \citep[e.g., ][]{Zhu2023}. Additionally, such thin accretion flows would map to correspondingly small filling factors \citep{Muzerolle2001}, but our shock modeling from Paper I reveals modest filling factors up to 40\%.

\input{Flow_Model_Grid_Table.tex}

To ensure consistent wavelength calibration for each fitted profile, we perform further wavelength calibration for \ha, \hb, and \hy. We select $\sim$100 {\AA}-wide regions to the left and right of the line center, making sure not to include the extended wings. We then performed cross-correlation on these regions against a template BT-SETTL \citep{btsettl} photosphere model spectrum and applied the highest-power wavelength shift to the observed line spectrum. We found that the intrinsic wavelength calibration is generally consistent to within 0.3 {\AA} for our CHIRON, X-Shooter, UVES, and TCES spectra, but was often off by up to 2 {\AA} for our ESPRESSO spectra. Here we use all observations, as the flow model does not require flux-calibrated spectra. 

When performing the flow model fitting, because of the complex line shapes, generally due to blue-shifted absorption which the model does not account for, we exclude several regions in the spectra from fitting. For \ha, this is -100 to 0 km s$^{-1}$ for TW~Hya, -100 to +50 km s$^{-1}$ for RU~Lup, no mask for BP~Tau, and -200 to +50 km s$^{-1}$ for GM~Aur. For \hb and \hy, we only mask between $\pm$50 km s$^{-1}$ for RU~Lup. In some cases, spurious emission/absorption features can appear within our fitted regions that deviate significantly from any fitted model, which can affect the final fit, but in general these masked regions contained persistent features that could not be fitted by any model.

\begin{figure*}
    \centering
    \includegraphics[width=0.97\textwidth]{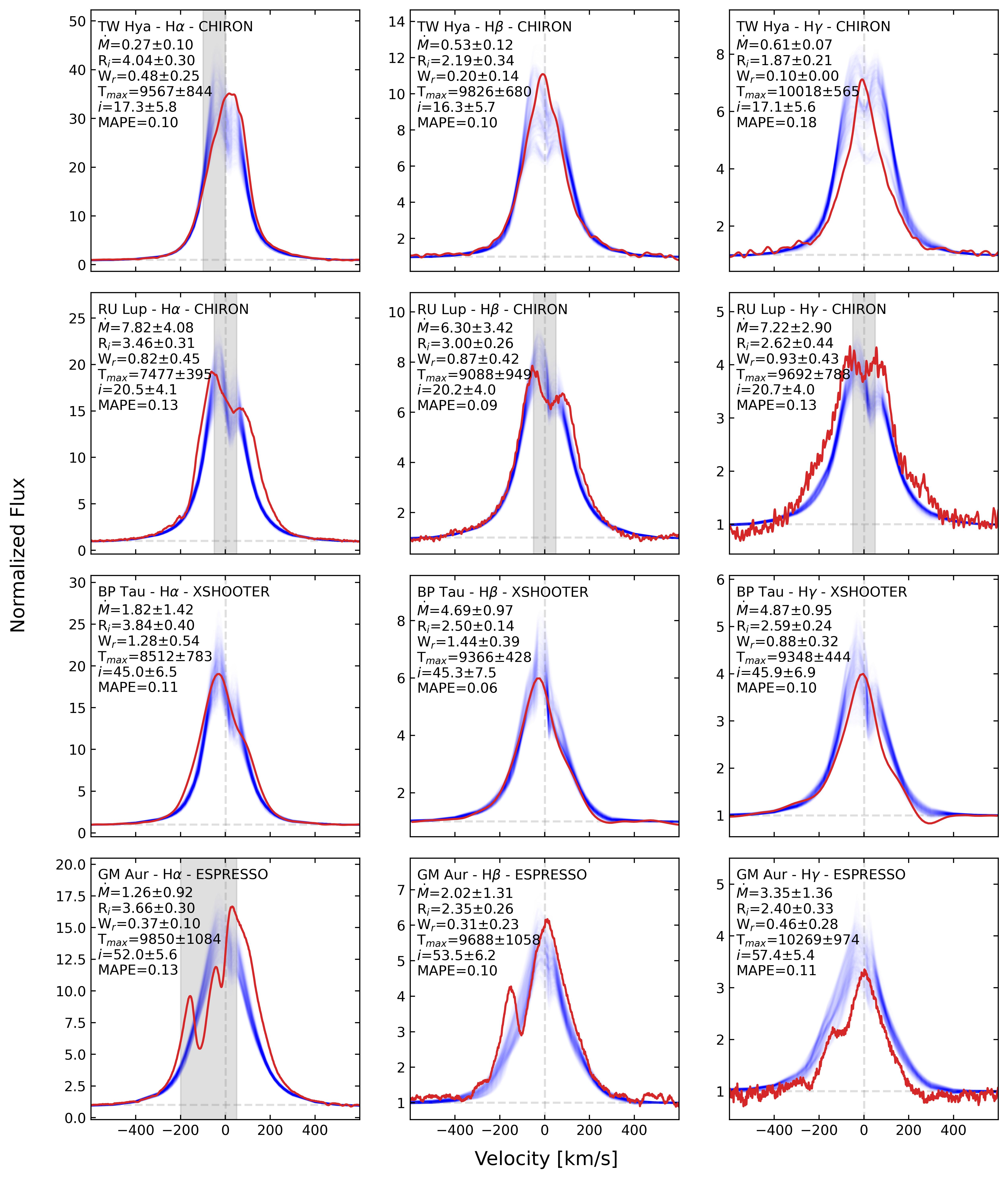}
    \caption{Examples of flow model fits for the best-fit observation for each target. Top to bottom: TW~Hya, RU~Lup, BP~Tau, GM~Aur. Left to right: \ha, \hb, \hy. Solid red line is observed spectrum. Blue lines are 500 best-fit models. Grey regions denote regions not fit by the flow model. Examples of the worst-fit observations can be found in Figure \ref{fig: Flow Model Examples Bad}.}
    \label{fig: Flow Model Examples Good}
\end{figure*}

Aside from these masked regions, we fit our grid of models between +/-500 km s$^{-1}$, roughly consistent with the free-fall velocity of each target. For each line, we then fit our grid of models and select the 500 best-fitting models according to the Mean Absolute Percent Error (MAPE) between the model and the observed spectrum. Using these 500 best-fit models, we then calculate weighted averages for the output parameters, where the weights are the inverse MAPE values for each model. Uncertainties are taken as the weighted standard deviation of those models. Example fits for well-fitting observation for each target are shown in Figure \ref{fig: Flow Model Examples Good}, while fits for poorly-fit observations are shown in Figure \ref{fig: Flow Model Examples Bad} in Appendix \ref{appendix: Flow Model Results}.

\begin{figure*}
    \centering
    \includegraphics[width=0.975\textwidth]{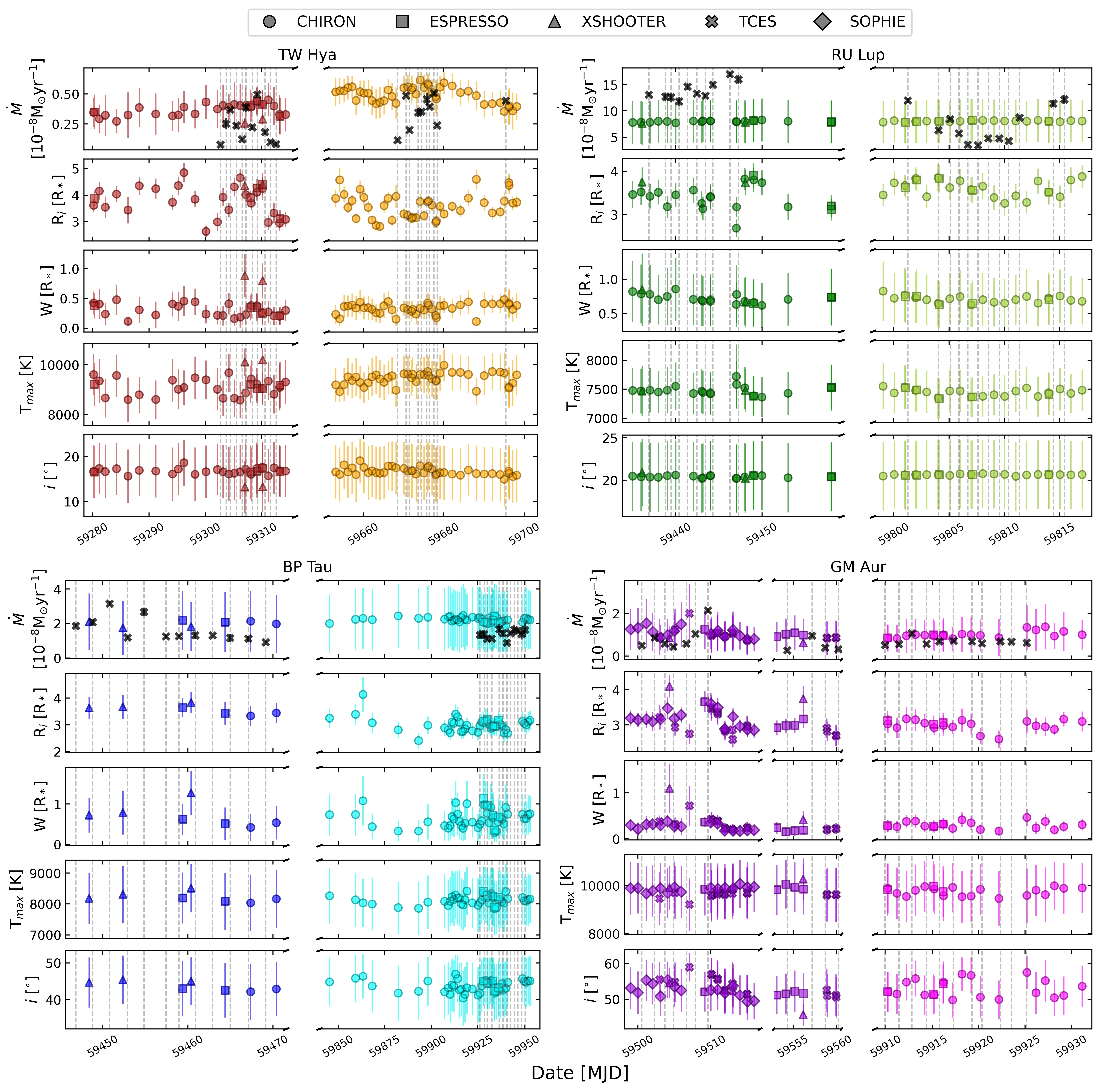}
    \caption{\ha flow model results for TW~Hya (top left), RU~Lup (top right), BP~Tau (bottom left), and GM~Aur (bottom right). For each target, left/right are E1/E2 while top to bottom are \mdot, \rin, \Wr, \Tmax, and $i$. Marker shape refers to the instrument (see legend). Black crosses in top row of each panel are accretion rates from shock modeling from Paper I.}
    \label{fig: Flow Results}
\end{figure*}

In principle, all lines could be fit simultaneously for each observation, but we elect to fit each line separately in order to gauge how the model responds to the variability in each line. The accretion rates from \ha exhibited the closest agreement to the accretion rates derived from shock modeling in Paper I. To that end, the results of our modeling for just \ha are shown in Figure \ref{fig: Flow Results}, while Table~\ref{tab: Average Flow Parameters} shows the median values of our flow modeling for each object and line. Tables~\ref{tab: Flow Model Results TW Hya}--\ref{tab: Flow Model Results GM Aur} show the results for all \ha observations. 


\input{Average_Flow_Model_Params}

\subsubsection{Modeling Results for TW~Hya} \label{sec: Flow Model - TW Hya}

The median accretion rates estimated by the flow model for TW~Hya are 0.4/0.5\accrateunits, 0.5/0.6 \accrateunits, and 0.6/0.6\accrateunits in E1/E2 for \ha, \hb, and \hy, respectively. The median accretion rates from the shock model (see Paper I) are 0.22/0.37\accrateunits for E1/E2. \hb and \hy produce accretion rates at or near the upper boundaries set by our grid.

\ha most closely recovers the shock model accretion rates in E1, while \hb does in E2 (though \hb is highly inconsistent with \mdot derived in Paper I in E1). In general, though, all 3 lines recover the median accretion rate consistent to within about a factor of 2 in both epochs. The variability in \mdot is lower in the flow models, with a standard deviation between all \ha observations of 0.06\accrateunits for the flow model compared to $\sim$0.13\accrateunits for the shock model.

\rin is comparably low for both \hb and \hy ($\sim$1.5--2.8 \rstar), but is higher for \ha (2.6--4.8 \rstar). For all three lines, \rin is lower in E2. \Wr is very low in \hb and \hy, in most cases preferring the lower bound of our grid (0.1 \rstar), but for \ha the median is 0.33 \rstar. \Tmax and $i$ are lowest for \ha\ and highest for \hy.

The flow model appears to fit \ha and \hy equally well (average MAPEs are 0.14 and 0.16) in E1 and for some of the later visits in E2, but the earlier visits in E2 with particularly strong blue-shifted emission are poorly fit. Thus, observations between MJD=59653$\sim$59680 should be considered carefully.

\subsubsection{Modeling Results for RU~Lup} \label{sec: Flow Model - RU Lup}

The median accretion rates estimated by the flow model for RU~Lup are 8.0/8.1\accrateunits, 7.6/8.4\accrateunits, and 8.2/10.0\accrateunits in E1/E2 for \ha, \hb, and \hy, respectively. The median accretion rates from the shock model (see Paper I) are 13.2/6.1\accrateunits for E1/E2. The standard deviation of the accretion rates for all observations (reflective of the accretion variability) is about 1\accrateunits, lower than the $\sim$2\accrateunits for the shock model. Thus, the flow model is unable to recover the accretion rate specific to either epoch, but does recover accretion rates within about a factor of 2. It also does not recover the variability in the accretion rate, producing much less variability overall.

For \rin, we find 3.4/3.6 \rstar, 3.3/3.6 \rstar, and 2.8/3.9 \rstar in E1/E2 for \ha, \hb, and \hy. Overall, this is lower than the typically assumed 5 \rstar, but like TW~Hya, \hb and \hy yield lower values. \Wr is about 0.7, 1.0, and 1.2 for \ha, \hb, and \hy, with little difference between epochs. The flow appears to be cooler than TW~Hya, but shows the same behavior where temperature increases towards \hy. For all three lines, we find inclinations between 20--23\textdegree, slightly higher than the assumed inclination of RU~Lup {18.8\textdegree}.

Overall, \hb is fit better than either \ha or \hy, with median MAPE values of 0.12 for \hb, but 0.16 and 0.18 for \ha and \hy across all observations. In all cases, though, the flow model underestimates the blue-shifted emission and is unable to recover the broad blue wing.

\subsubsection{Modeling Results for BP~Tau} \label{sec: Flow Model - BP Tau}

The median accretion rates estimated by the flow model for BP~Tau are 2.0/2.2\accrateunits, 4.1/4.3\accrateunits, and 4.5/4.6\accrateunits in E1/E2 for \ha, \hb, and \hy, respectively. The median accretion rates from the shock model (see Paper I) are 1.3/1.4\accrateunits for E1/E2. Unlike TW~Hya and RU~Lup, the flow model over-predicts the accretion rate for all 3 lines in both epochs, though again \ha is the closest. With the exception of E1 (whose variability may be skewed by 2 high visits), the variability predicted by the flow model is higher than that of the shock model. In either case, both models suggest the accretion variability in BP~Tau is the lowest in our sample.

The truncation radius, \rin, is similar to that of RU~Lup: 2.7--4.1 in \ha and $\sim$1.6--2.9 in \hb and \hy, again smaller than 5 \rstar. The flow width, \Wr, is comparable across the three lines, from 0.3--1.5 \rstar. The flow temperature, \Tmax, is coolest in \ha ($\sim$8120 K) and similar in \hb and \hy ($\sim$9200 K), higher than \ha. The inclinations predicted by all lines vary between 41\textdegree--45\textdegree, higher than the true inclination of BP~Tau of $\sim$38\textdegree.

The flow model fits \ha well, but like the other targets, is not as broad as the observations. Like \ha, the blue-shifted wing in \hb is narrower in the model. Additionally, the red wing is often poorly fit, as there is often enhanced emission near 100 km s$^{-1}$, though when this emission feature is absent, the red wing is fit well. The flow model also predicts red-shifted absorption that is absent in the data. For \hy, the blue wing and central peak are often underestimated by the model, though this may be due to poor continuum subtraction. Like \hb, the model predicts red-shifted absorption that is absent in the data.

\subsubsection{Modeling Results for GM~Aur} \label{sec: Flow Model - GM Aur}

The median accretion rates estimated by the flow model for GM~Aur are 1.04/1.01\accrateunits, 2.2/2.9\accrateunits, 3.6/2.4\accrateunits in E1/E2 for \ha, \hb, and \hy, respectively. The median accretion rates from the shock model (see Paper I) are 0.58/0.67\accrateunits for E1/E2.

For \rin, we find 3.1/3.0 \rstar, 1.9/2.1 \rstar, and 2.0/2.4 \rstar in E1/E2 for \ha, \hb, and \hy. For \Wr we find 0.3/0.3 \rstar, 0.3/0.5 \rstar, and 0.4/0.5 \rstar. For \Tmax we find 9780/9750 K, 9600/9850 K, and 10160/9840 K. For $i$ we find 52.7/53.1\textdegree, 53.7/54.5\textdegree, and 54.9/54.2\textdegree. \Wr, \Tmax, and $i$ are fairly consistent between each line, though they all tend to increase from \ha to \hy. \rin is less consistent, with the highest values coming from \ha like in the other targets.

Our \ha observations of GM~Aur show that there are consistent blue- and red-shifted wings that the flow model cannot recover. Low velocity red-shifted emission between about 0--150 km s$^{-1}$ is typically fit by the model, however. The blue wing in \hb is often fit among the best in our sample, though some observations in E2 exhibit very strong blue-shifted absorption that severely contaminates the model. The red wing is also generally fitted well, though there are cases of enhanced red-shifted emission not recovered by the flow model. GM~Aur shows some red-shifted absorption that is recovered by the model. \hy is especially noisy in GM~Aur and is not fitted well by the model. It also appears to suffer from poor continuum subtraction, like BP~Tau.

%% file: Optical_Lines.tex
\begin{deluxetable}{c c c c c c}[htp]
\setlength{\tabcolsep}{10pt}
\tablecaption{Optical emission lines  \label{tab: Emission Line Info}}
\centering
\tablehead{
\colhead{Line} & \colhead{$\lambda_0$ [\AA]} & \colhead{Width [\AA / km s$^{-1}$]}
}
\startdata
\ha & 6562.79 & 30 / 1370.41 \\
\hb & 4861.29 & 15 / 925.03 \\
\hy & 4340.47 & 10 / 690.69 \\
\hd & 4101.7 & 8 / 584.71 \\
He {\sc i}$_{4387}$ & 4387.9 & 3 / 204.97 \\
He {\sc i}$_{4471}$ & 4471.5 & 3 / 201.13 \\
He {\sc i}$_{4713}$ & 4713.1 & 1.5 / 95.41 \\
He {\sc i}$_{5015}$ & 5015.7 & 5 / 298.85 \\
He {\sc i}$_{5875}$ & 5875.6 & 3.5 / 178.58 \\
He {\sc i}$_{6678}$ & 6678.2 & 3 / 134.67 \\
He {\sc i}$_{7065}$ & 7065.2 & 3 / 127.30 \\
\enddata
\end{deluxetable}

%% file: Flow_Model_Grid_Table.tex
\begin{deluxetable*}{c c c c c c}[htp]
\setlength{\tabcolsep}{10pt}
\tablecaption{Accretion Flow Model Grid Parameters \label{tab: Flow Model Grid}}
\centering
\tablehead{
\colhead{Object} & \colhead{\rin} & \colhead{\Wr} & \colhead{\mdot} & \colhead{\Tmax} & \colhead{Incl.} \\
\colhead{} & \colhead{[\rsun]} & \colhead{[\rsun]} & \colhead{[1\accrateunits]} & \colhead{[K]} & \colhead{[\textdegree]}
}
\startdata
TW Hya & 1.5--6.0 (0.3) & 0.1--1.9 (0.3) & 0.01-0.7 (0.03) & 7500--10700 (400) & 3--30 (3) \\
RU Lup & 1.4--6.2 (0.3) & 0.2--2.4 (0.2) & 2.25-20.0 (0.75) & 6000-10800 (300) & 5-45 (5) \\
BP Tau & 1.5--6.0 (0.3) & 0.1--2.2 (0.3) & 0.2--6.0 (0.025) & 7000--9800 (400) & 25--57 (4) \\
GM Aur & 1.2--6.0 (0.4) & 0.1--2.1 (0.2) & 0.1--0.98 (0.04), 1.0--5.5 (0.4) & 7500--11000 (400) & 35--67 (4) \\
\enddata
\tablecomments{Hyphenated numbers denote the range of each parameter, while parenthesis denote step size.}
\end{deluxetable*}

%% file: Average_Flow_Model_Params.tex
\begin{deluxetable*}{c c c | c c c c c c} \label{tab: Average Flow Parameters}
\tablecaption{Median flow model results}
\tablehead{
\colhead{Object} & \colhead{Line} & \colhead{Epoch} & \colhead{\mdot} & \colhead{\rin} & \colhead{\Wr} & \colhead{\Tmax} & \colhead{Incl.} & \colhead{MAPE} \\
\colhead{} & \colhead{} & \colhead{} & \colhead{[1\accrateunits]} & \colhead{[\rstar]} & \colhead{[\rstar]} & \colhead{[K]} & \colhead{[\textdegree]} & \colhead{}
}
\startdata
TW Hya & \ha & 1 & 0.37 (0.05) & 3.84 (0.55) & 0.34 (0.16) & 9173 (398) & 16.6 (1.1) & 0.14 (0.03) \\
  &  & 2 & 0.49 (0.07) & 3.58 (0.45) & 0.33 (0.09) & 9450 (239) & 16.7 (0.8) & 0.15 (0.03) \\
\hline
TW Hya & \hb & 1 & 0.53 (0.03) & 2.00 (0.27) & 0.19 (0.11) & 9751 (98) & 16.2 (0.9) & 0.13 (0.03) \\
  &  & 2 & 0.55 (0.03) & 1.70 (0.17) & 0.12 (0.04) & 9803 (119) & 16.1 (0.7) & 0.18 (0.06) \\
\hline
TW Hya & \hy & 1 & 0.59 (0.02) & 1.77 (0.12) & 0.11 (0.01) & 9923 (89) & 16.9 (0.8) & 0.19 (0.02) \\
  &  & 2 & 0.58 (0.03) & 1.68 (0.12) & 0.11 (0.01) & 9843 (150) & 17.3 (0.6) & 0.22 (0.04) \\
\hline
\hline
RU Lup & \ha & 1 & 7.98 (0.15) & 3.43 (0.29) & 0.72 (0.07) & 7476 (75) & 20.4 (0.2) & 0.15 (0.02) \\
  &  & 2 & 8.06 (0.11) & 3.61 (0.18) & 0.71 (0.05) & 7441 (62) & 20.6 (0.1) & 0.16 (0.01) \\
\hline
RU Lup & \hb & 1 & 7.63 (0.86) & 3.30 (0.35) & 0.90 (0.24) & 8734 (217) & 20.5 (0.3) & 0.11 (0.01) \\
  &  & 2 & 8.38 (0.87) & 3.57 (0.33) & 1.14 (0.33) & 8420 (93) & 21.4 (0.4) & 0.14 (0.02) \\
\hline
RU Lup & \hy & 1 & 8.25 (0.93) & 2.83 (0.36) & 1.10 (0.27) & 9394 (228) & 21.0 (0.3) & 0.19 (0.03) \\
  &  & 2 & 10.03 (1.49) & 3.90 (0.72) & 1.29 (0.39) & 8761 (184) & 21.5 (0.4) & 0.16 (0.04) \\
\hline
\hline
BP Tau & \ha & 1 & 2.01 (0.15) & 3.57 (0.16) & 0.70 (0.26) & 8212 (146) & 43.7 (1.2) & 0.13 (0.01) \\
  &  & 2 & 2.22 (0.14) & 3.01 (0.25) & 0.65 (0.22) & 8127 (146) & 43.7 (1.5) & 0.11 (0.02) \\
\hline
BP Tau & \hb & 1 & 4.07 (0.44) & 2.42 (0.14) & 1.03 (0.30) & 9080 (189) & 42.0 (2.6) & 0.08 (0.02) \\
  &  & 2 & 4.31 (0.36) & 2.19 (0.19) & 1.06 (0.32) & 9240 (229) & 41.9 (2.0) & 0.09 (0.02) \\
\hline
BP Tau & \hy & 1 & 4.46 (0.77) & 2.57 (0.30) & 0.82 (0.41) & 9166 (443) & 45.4 (2.1) & 0.19 (0.12) \\
  &  & 2 & 4.62 (0.51) & 2.29 (0.28) & 0.76 (0.28) & 9287 (387) & 43.3 (2.3) & 0.14 (0.02) \\
\hline
\hline
GM Aur & \ha & 1 & 1.04 (0.26) & 3.11 (0.33) & 0.30 (0.17) & 9781 (186) & 52.7 (2.4) & 0.13 (0.01) \\
  &  & 2 & 1.01 (0.15) & 3.01 (0.15) & 0.30 (0.07) & 9753 (157) & 53.1 (2.4) & 0.12 (0.01) \\
\hline
GM Aur & \hb & 1 & 2.23 (1.18) & 1.90 (0.21) & 0.30 (0.20) & 9604 (521) & 53.7 (1.8) & 0.13 (0.03) \\
  &  & 2 & 2.85 (0.67) & 2.12 (0.24) & 0.53 (0.20) & 9853 (266) & 54.5 (1.5) & 0.11 (0.04) \\
\hline
GM Aur & \hy & 1 & 3.56 (0.59) & 2.00 (0.42) & 0.43 (0.22) & 10159 (151) & 54.9 (2.7) & 0.17 (0.03) \\
  &  & 2 & 2.45 (1.08) & 2.37 (0.56) & 0.44 (0.21) & 9838 (359) & 54.2 (3.4) & 0.17 (0.04) \\
\hline
\hline
\enddata
\tablecomments{Values are the median value for the given parameter in E1/E2. Values in parentheses are the standard deviation of the given parameter in E1/E2.}
\end{deluxetable*}

%% file: Discussion.tex
\section{Discussion} \label{sec: Discussion}

Below we compare the flow model results for the three different Balmer lines to each other, compare these results to accretion rates estimated in Paper I using accretion shock modeling, and discuss the inner truncation radii (\rin) and inclinations ($i$) predicted by the flow model.

\subsection{\texorpdfstring{$\mathrm{H}\alpha\ \mathrm{vs}\ \mathrm{H}\beta\ \mathrm{vs}\ \mathrm{H}\gamma$}{Ha vs HB vs Hy}} \label{sec: Ha vs Hb vs Hy}

We have modeled three different Balmer lines using our accretion flow model and here we discuss the similarities and differences between those results. First, the top 500 models for each line tended to be equally well fitting. The median MAPEs for all objects and observations were 13.6, 12.4, and 18.0\% for \ha, \hb, and \hy respectively, meaning the quality of the fits was consistent to within about 5.6\%. This suggests that the typical quality of the fits does not vary significantly between the lines. We also see no notable trends in the uncertainties produced by each line, suggesting that each line is equally uncertain for all parameters. The results for each line do differ in their median output parameters, though. 

For example, \hb and \hy, in nearly all cases, yield the highest accretion rates. These high accretion rates are on average $\sim$79\% higher than those predicted by \ha, which already produces rates higher than predicted by our shock modeling (Paper I). They are also at or above the highest accretion rates estimated for these targets (see Paper I, Section 2). In many cases, especially in TW~Hya and BP~Tau, these rates are at or near the boundaries of our grid, so an even higher \mdot may be preferred by these lines. \hb and \hy also produce, on average, the lowest truncation radii (\rin) and the highest flow temperatures (\Tmax). \rin is often between (or even lower than) 2.0--2.5 \rstar, which is well below the typically assumed 5 \rstar or any observation estimates of \rin for these targets.

Considering their high transition energies, it may be that the upper lines (\hb, \hy) trace more compact, hotter flows that lay inside of larger, more diffuse flows more closely traced by \ha. This more compact flow geometry may then produce higher accretion rates if a roughly constant energy flux in the flow is assumed. From \citet{Calvet1998}, the energy flux is $F\propto1/(1-1/R_{in})$. Thus, for a constant energy flux in the accretion flow, \mdot and \rin are inversely proportional, reflective of our results. This layered flow structure was employed in the flow modeling of \citet{Thanathibodee2019} to explain the complex Balmer profiles seen in low accretors. Recent 3D MHD modeling by \citet{Zhu2024} shows precisely this onion-like structure, where multiple, layered flows result in multiple hotspots and ``hidden" flows. It may thus be important to incorporate a self-attenuating, multi-layered flow structure in future accretion flow modeling.

Despite these differences, the correlation coefficient, $r$, between the accretion rates output by each line are very high: $r_{\dot{M}, H\alpha-H\beta}$=0.92, $r_{\dot{M}, H\alpha-H\gamma}$=0.92, and $r_{\dot{M}, H\beta-H\gamma}$=0.95. This suggests that the same physical effects that induce variability are manifesting in each line, though our model appears to interpret them differently.

\subsection{Accretion Rates: Shock vs Flow} \label{sec: Shock vs Flow}

Because of the potential issues in applying the flow model to \hb and \hy described above, and because \ha produces \mdot and \rin most consistent with previous work, we consider results using only \ha for the flow model for the following discussion. This is in contrast to other studies \citep[e.g., ][]{Natta2004, Herczeg2008, Alcala2014, Herczeg2023}, who recommend against using \ha to estimate \mdot due to the large spread in the \lacc--L$_{H\alpha}$ relationship and to the many processes that contribute to its emission profile. These previous studies have typically used either the total \ha luminosity (L$_{H\alpha}$) or the full width at 10\% of the line peak (W$_{H\alpha,\ 10\%}$) to compare against \lacc or \mdot. Both L$_{H\alpha}$ and W$_{H\alpha,\ 10\%}$ are subject to blue-shifted absorption from an ejected wind, which we show here can be quite strong and variable, which could produce the additional scatter observed. Our flow modeling mitigates the effect of wind absorption by masking the blue regions most susceptible to absorption, which should allow \ha to be a more reliable tracer of accretion in CTTSs.

\begin{figure}
    \centering
    \includegraphics[width=0.44\textwidth]{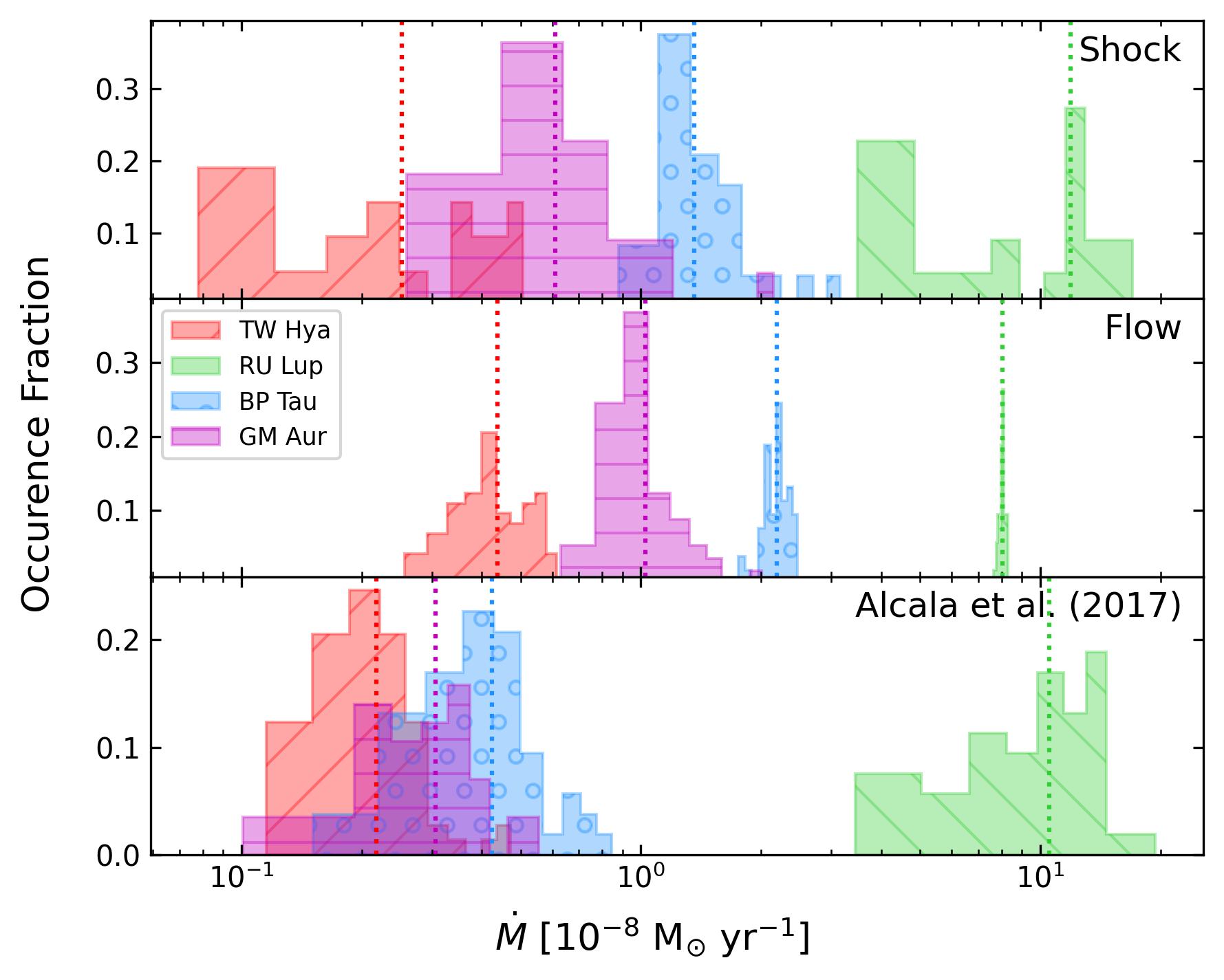}
    \includegraphics[width=0.46\textwidth]{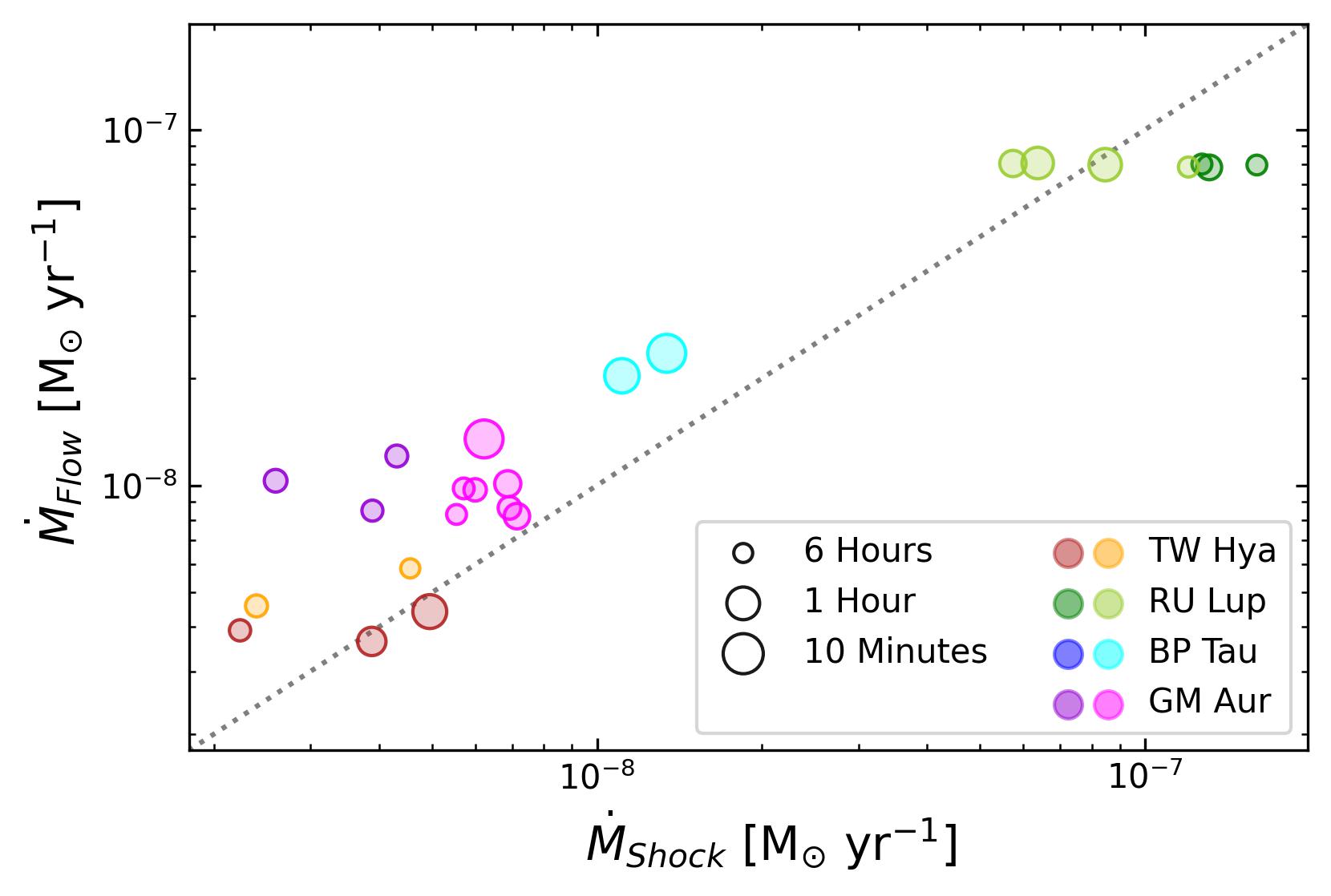}
    \caption{\textit{Left}: Histograms of accretion rates from the shock model from Paper I (top row), flow model from \ha from this work (middle row), and using relationships from \citet{Alcala2017}. Colored dotted lines show the median of each distribution. \textit{Right}: Accretion rates from the flow model (\ha) vs the shock model, where data are contemporaneous to within 6 hours. Larger points represent observations obtained closer in time. For both plots, colors (yellow/red, green, blue, and pink/purple) denote the different targets (TW Hya, RU Lup, BP Tau, and GM Aur).}
    \label{fig: Shock vs Flow}
\end{figure}

Figure \ref{fig: Shock vs Flow} compares the accretion rates derived from our two models: an accretion shock model \citep{Calvet1998, RE19} from Paper I and an accretion flow model \citep{Muzerolle1998} used here. The top panel shows the distribution of accretion rates for each target, while the bottom panel directly compares contemporaneous accretion rate measurements.

With the exception of RU Lup, the flow model produces median accretion rates that are 51\%, 47\%, and 46\% higher than the shock model for TW~Hya, BP~Tau, and GM~Aur, respectively, showing that the flow model can recover accretion rates that generally describe the system (i.e., within a factor of 2). When looking at the contemporaneous rates, though, these differences become 40\%, 79\%, and 103\%, showing that changes in the accretion are not consistently recovered between the shock and flow models. The median rate predicted by the flow model for RU Lup is about 21\% less than the shock model and 23\% less for contemporaneous observations. Additionally, the flow model produces significantly less variability than the shock model, both in terms of variance and range. 

\citet{Nature} use the same shock and flow (using only \ha) models to estimate the accretion rate of GM~Aur across $\sim$1 week of $HST$ observations and $\sim$3 weeks of daily CHIRON observations. They also found that, on average, the flow model over-predicted the shock model by 42\% and by 30\% for contemporaneous data. \citet{Thanathibodee2018} also use the same flow model and a single-column version of our shock model to estimate accretion rates of low accretors, where \mdot $\lessapprox 10^{-9} M_{\odot} yr^{-1}$. Here, the shock model generally produced higher accretion rates than the flow model ($\sim$27\%), though they note that their \mdot$_{\mathrm{Shock}}$ are likely upper limits. 

Our findings, along with previous use of these models, suggest that the flow model is able to recover accretion rates in line with quiescent accretion in these systems (as predicted by our shock modeling in Paper I), but cannot accurately reproduce accretion variability, and in some cases it produces effectively no variability. Some of these results could be due to the built-in assumption of the flow model: that it is axisymmetric and that mass flows from the disk onto the star in both the upper and lower hemispheres. In the shock model, we only measure accretion luminosity from what we see on the visible side of the star. The flow model assumes that mass also flows in on the other ``unseen'' side. Additionally, the flow model assumes a ring of emission on the stellar surface, so by construction it cannot reproduce variability due to non-axisymmetry of the hotspots themselves.

The accretion rates derived here are also generally in line with previous studies. Using empirical relationships from \citet{Alcala2017}, \citet{Gangi2022} estimate the accretion rate of BP~Tau at 0.56\accrateunits in January 2020. This is consistent with our estimates using those same relationships, which underestimate the accretion rate as compared to our shock modeling n BP~Tau specifically. \citet{Gangi2022} also find the accretion rate of GM~Aur to be 0.51\accrateunits, which is at the higher end of our estimates using those empirical relationships, suggesting that GM Aur may have been in an enhanced accretion state during their observations. \citet{Armeniinprep} estimate the accretion rate of RU Lup also using the empirical relationships from \citet{Alcala2017} (though only using the He I$_{5875}$  line), finding average rates between 7–40\accrateunits. These accretion rates are roughly consistent with, though slightly higher than, our findings using any method. \citet{Armeniinprep} also see that the accretion rate in RU Lup in 2022 is about half that of 2021.

The top panel of Figure~\ref{fig: Shock vs Flow} also shows a comparison of accretion rates calculated using the 11 emission lines from Section~\ref{sec: Analysis and Results} and the empirical relationships of \citet{Alcala2017} to those calculated using the shock model (Paper I) and the flow model. The empirical relationships produce similar accretion rates as compared to the shock model in TW~Hya and RU~Lup: $\sim$23.8\% lower in TW~Hya and 4.7\% higher in RU~Lup. Accretion rates in BP~Tau and GM~Aur and 71.9\% and 56.1\% lower using the \citet{Alcala2017} relationships. This further reinforces that while these empirical relationships can provide accurate accretion rates in some cases, they fail in others and should be used sparingly in studies of variability.

\subsection{Truncation Radius and Stellar Inclination}

The flow model appears capable of recovering inclinations and truncation radii that are consistent with current theory and previous studies. The flow model assumes aligned rotation and magnetic axes, which is often not the case \citep{Donati2007, Donati2008, Donati2010, Donati2011}, so there is additional uncertainty on our predicted inclinations. These axes tend to be misaligned by up to 20\textdegree, making the flow model's predictions consistent with previous estimates.

Using high spatial resolution Br$_{\gamma}$ spectra, TW~Hya's inner truncation radius (\rin) has been estimated between 3.5--12.35 \rstar \citep{Gravity2020, Gravity2023}, depending on the treatment of the data and assumed stellar parameters. Our median estimate of 3.7$\pm$0.5 \rstar is consistent with these studies and is only slightly smaller than the typically assumed 5 \rstar. Those studies also fitted for inclination and found $i$=14$^{+6}_{-14}$\textdegree, consistent both with our estimate of $\sim$17\textdegree\ and with the assumed value of 7\textdegree. RU~Lup's \rin was also estimated by \citet{Gravity2023} using \Bry. They found \rin=3.3--6.6 \rstar from two distinct observations, consistent with our finding of 3.5$\pm$0.3 \rstar. They find $i$ between 12--24\textdegree, also consistent with our findings ($\sim$20.5\textdegree) and previous studies. Our estimate of \rin for BP~Tau (\rin=3.2$\pm$0.4 \rstar) is consistent with 3D magnetospheric accretion models \citep{Long2011}, which place \rin between 3.6--7 \rstar, depending on the assumed stellar parameters, especially the accretion rate. We estimate median $i$ of 43.7$\pm$1.3\textdegree, higher and inconsistent with our assumed value of 38.2\textdegree. And finally, our flow model places \rin between 2.6--4.1 \rstar in GM~Aur, though no other study can place this into further context. The flow model recovers inclinations (53$\pm$2.4\textdegree) consistent with previous work.

%% file: Conclusion.tex
\section{Summary} \label{sec: Conclusion}

We conducted a comprehensive multi-wavelength, multi-epoch monitoring campaign of 4 CTTSs (TW~Hya, RU~Lup, BP~Tau, and GM~Aur), including $HST$ UV spectra (see Paper I), contemporaneous optical light curves (see Paper II), and contemporaneous optical spectra, which is the focus of this paper. Here we model the \ha, \hb, and \hy emission line profiles using a magnetospheric accretion flow model to estimate the accretion characteristics of these CTTSs over many stellar rotations in both 2021 and 2022.  We also measure optical line luminosities and compare with empirical relationships. Our main findings are as follows:

\begin{enumerate}

    \item Averaged across multiple observations, our accretion flow model reproduces accretion rates that are consistent to within a factor of 2 when compared to the shock modeling using $HST$ UV spectra from Paper I. Additionally, it typically recovers inner truncation radii and stellar inclinations consistent with previous studies.

    \item While the accretion rates from all three Balmer lines are highly correlated and show similar variability trends, \ha produces accretion rates most similar to those obtained using accretion shock modeling from Paper I. \hb and \hy systematically produce the highest accretion rates, the smallest truncation radii, and the highest flow temperatures when compared to \ha. Thus, we suggest that when the variable absorption components are properly mitigated, \ha (among the Balmer lines) is the most reliable tracer of the accretion rate in CTTSs, but still fails to recover variability.

    \item Our results suggest that the magnetic truncation radius is typically between 2.5--5~\rstar, and is between 3--4 \rstar in most cases.

    \item Like UV emission lines in Paper I and photometry in Paper II, optical emission lines are useful for estimating accretion rates in CTTSs but should be used with caution. Using empirical relationships, several lines yield consistent accretion luminosities as accretion shock modeling, but the variability trends differ, suggesting that optical emission lines are a poor direct tracer of accretion variability.
  
\end{enumerate}

Our results here reinforce that UV spectra remain the best tool to understand the accretion rate in CTTSs, particularly variability. However, optical spectra can provide useful information about the accretion such as the average accretion rate and magnetospheric geometry. Further, this small sample shows the diversity in CTTSs, that variability characteristics and line morphologies can differ considerably between targets. 

It is also important to emphasize that emission from CTTSs is highly variable at all wavelengths from the UV to the NIR and beyond. Future studies of individual CTTSs should incorporate simultaneous, multi-wavelength spectra from the UV to the NIR to obtain the most accurate picture of accretion.

%% file: Acknowledgements.tex
\begin{acknowledgments}
This material is based upon work supported by the National Science Foundation under Grant Number AST-2108446. This work is supported by \hst AR-16129 from the Space Telescope Science Institute, which is operated by AURA, Inc. This work has been supported by Deutsche Forschungsgemeinschaft (DFG) in the framework of the YTTHACA Project (469334657) under the project code  EI 409/20-1. Funded by the European Union (ERC, WANDA, 101039452). Views and opinions expressed are however those of the author(s) only and do not necessarily reflect those of the European Union or the European Research Council Executive Agency. Neither the European Union nor the granting authority can be held responsible for them. This work was also supported by the NKFIH excellence grant TKP2021-NKTA-64.

We thank Todd Henry and the CHIRON team for their prompt scheduling of the CHIRON observations. This work benefited from discussions with the ODYSSEUS team (\url{https://sites.bu.edu/odysseus/}); see \cite{Espaillat2022} for an overview of the ODYSSEUS survey. 

\end{acknowledgments}

%% file: Appendix.tex
\appendix

\section{Flow Model Results} \label{appendix: Flow Model Results}

Tables~\ref{tab: Flow Model Results TW Hya}--\ref{tab: Flow Model Results GM Aur} present the results of our accretion flow modeling and optical veiling for TW~Hya, RU~Lup, BP~Tau, and GM~Aur.

\input{Flow_Model_Results_Table}

\begin{figure*}
    \centering
    \includegraphics[width=0.97\textwidth]{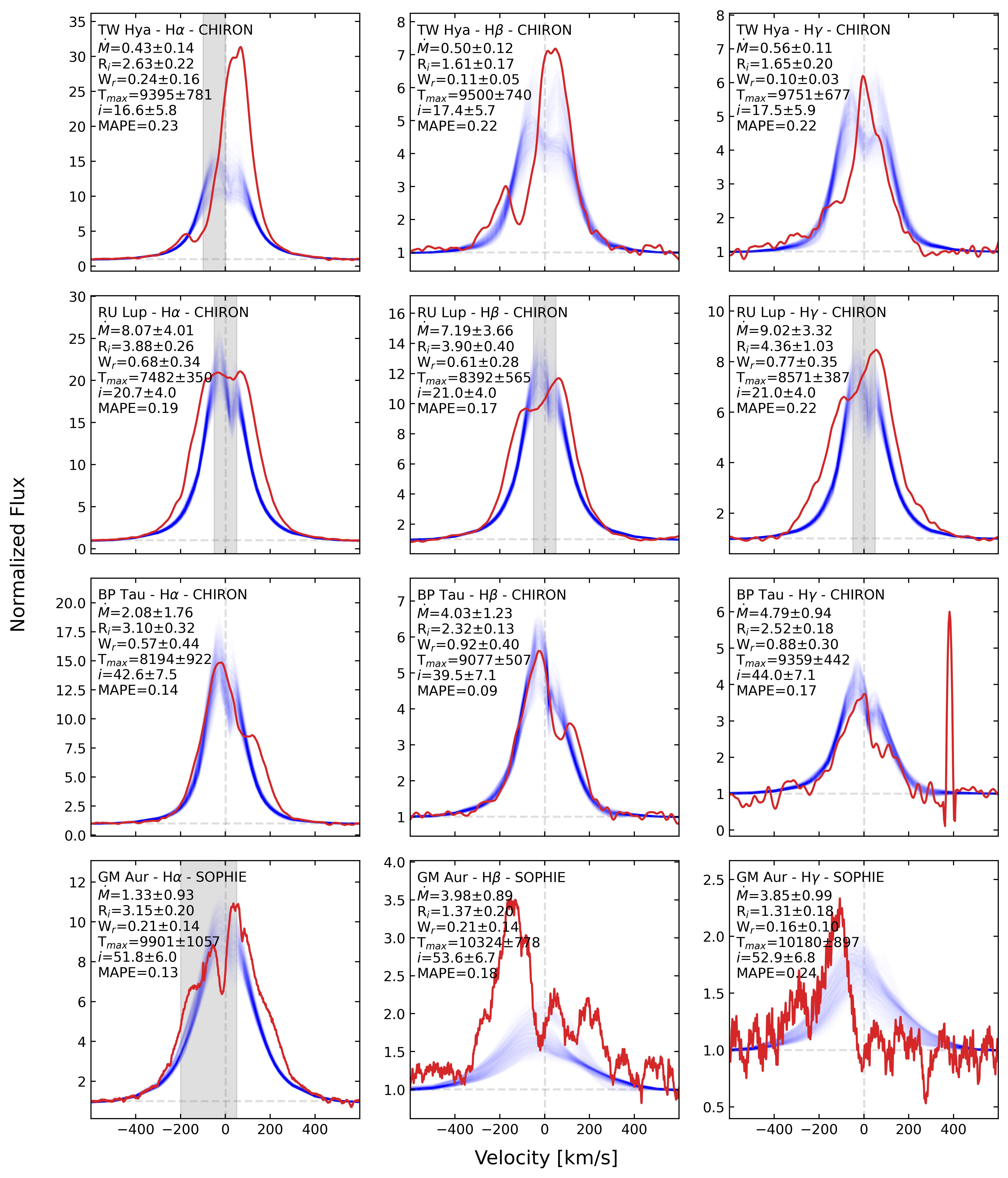}
    \caption{Examples of flow model fits for the worst-fit observation for each target. Top to bottom: TW~Hya, RU~Lup, BP~Tau, GM~Aur. Left to right: \ha, \hb, \hy. Solid red line is observed spectrum. Blue lines are 500 best-fit models. Grey regions denote regions not fit by the flow model. Examples of best-fit observations can be found in Figure \ref{fig: Flow Model Examples Good}.}
    \label{fig: Flow Model Examples Bad}
\end{figure*}

\section{Optical Line Luminosities} \label{appendix: Optical Line Luminosities}

Tables \ref{tab: Line Fluxes TW Hya}--\ref{tab: Line Fluxes GM Aur} present the measured lines luminosities for all 14 lines we focus on. Figure \ref{fig: Line Fluxes Other, TW Hya and RU Lup} and \ref{fig: Line Fluxes Other, BP Tau and GM Aur} shows the line luminosities for the remaining 7 lines over time for TW~Hya/RU~Lup and BP~Tau/GM~Aur, respectively. \ha, \hb, and \hy are shown in Figure \ref{fig: Balmer Fluxes vs Time} in Section \ref{sec: Results - Balmer Lines}.

\input{Line_Luminosities}

\begin{figure*}
    \centering
    \includegraphics[width=0.975\textwidth]{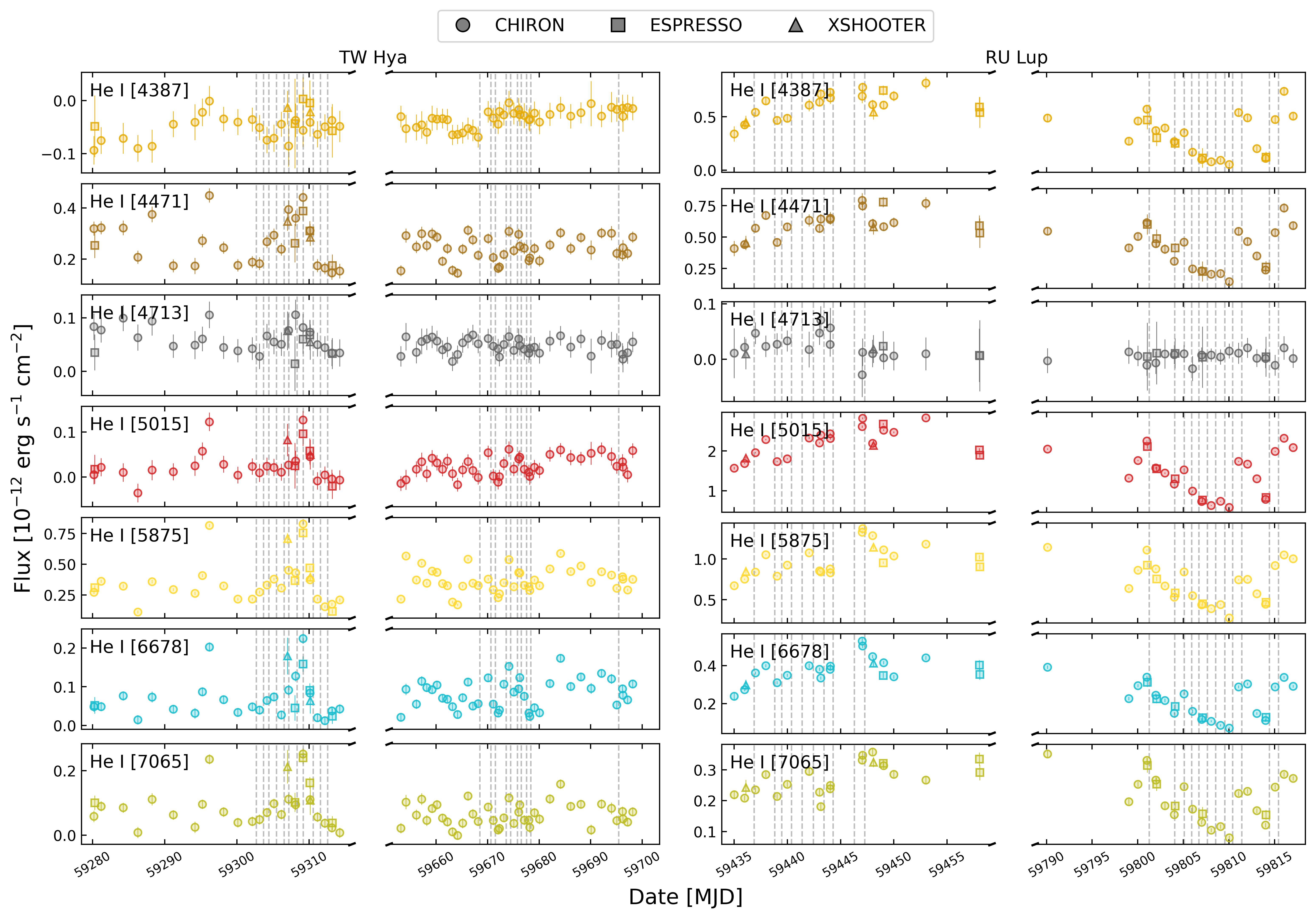}
    \caption{Flux versus time for the remaining 11 lines for TW~Hya (left) and RU~Lup (right). Marker shape denotes the instrument used, with CHIRON, ESPRESSO, XSHOOTER, SOPHIE, and \coude as circles, squares, triangles, diamonds, and crosses, respectively. Line flux is in units of 10${-12}$ erg s$^{-1}$ cm$^{-2}$. Dashed grey lines are times of the $HST$ observations from Paper I.}
    \label{fig: Line Fluxes Other, TW Hya and RU Lup}
\end{figure*}

\begin{figure*}
    \centering
    \includegraphics[width=0.975\textwidth]{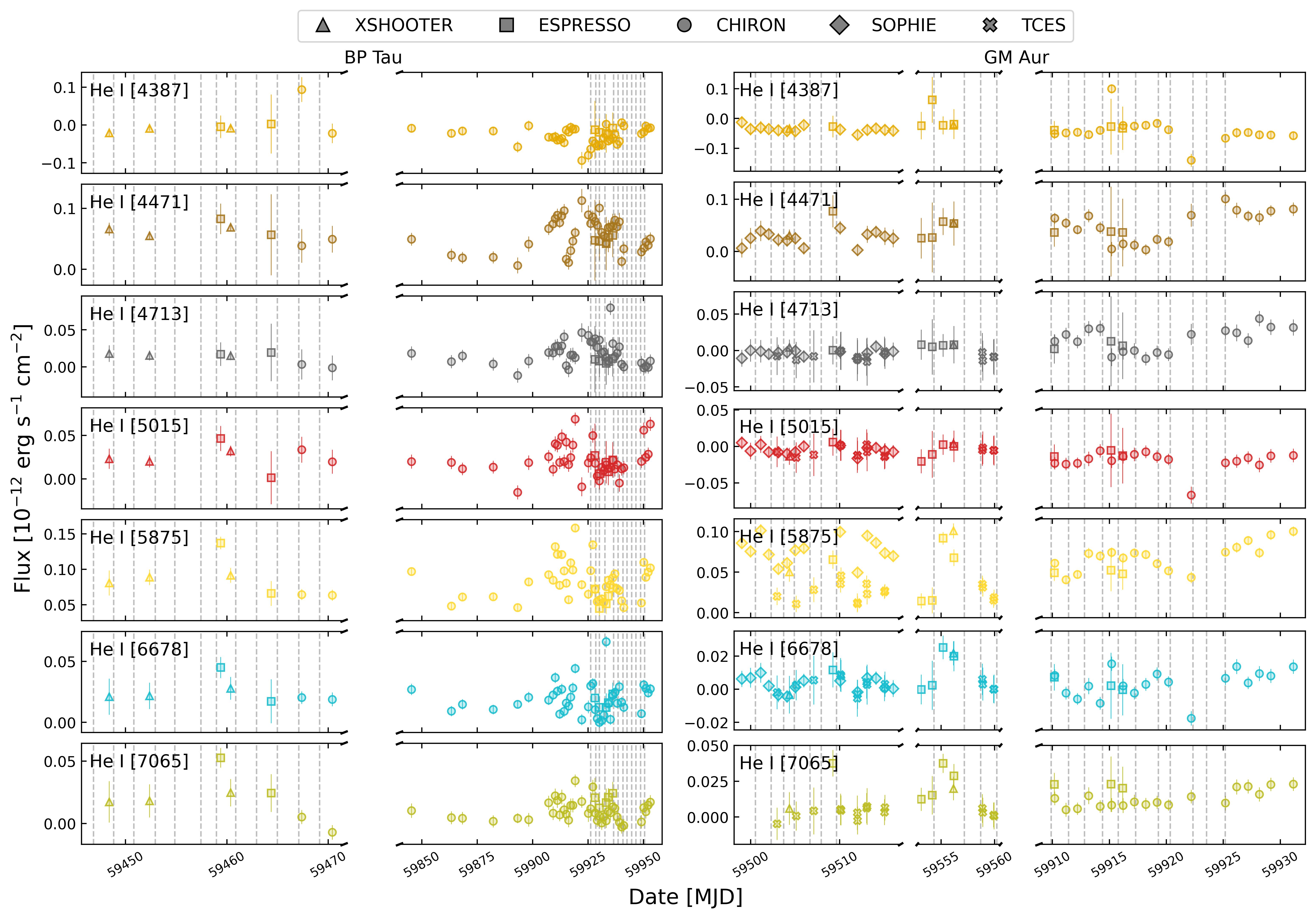}
    \caption{Same as Figure \ref{fig: Line Fluxes Other, TW Hya and RU Lup} but for BP~Tau and GM~Aur.}
    \label{fig: Line Fluxes Other, BP Tau and GM Aur}
\end{figure*}

%% file: Flow_Model_Results_Table.tex
\startlongtable
\begin{deluxetable*}{c c c c c c c c c c} \label{tab: Flow Model Results TW Hya}
\tabletypesize{\footnotesize}
\tablecaption{Accretion flow model results and veilings for TW Hya}
\tablehead{
\colhead{Object} & \colhead{MJD} & \colhead{Instrument} & \colhead{r$_V$} & \colhead{\mdot} & \colhead{\rin} & \colhead{\Wr} & \colhead{\Tmax} & \colhead{Incl.} & \colhead{MAPE} \\
\colhead{} & \colhead{} & \colhead{} & \colhead{} & \colhead{[10$^{8}$ M$_{\odot}$yr$^{-1}$]} & \colhead{[\rstar]} & \colhead{[\rstar]} & \colhead{[K]} & \colhead{[\textdegree]} & \colhead{}
}
\startdata
TW Hya & 59241.30 & C & 0.23$\pm$0.01 & 0.4$\pm$0.2 & 2.7$\pm$0.2 & 0.2$\pm$0.1 & 9195.1$\pm$902.1 & 15.7$\pm$6.0 & 0.11 \\
TW Hya & 59280.17 & C & 0.64$\pm$0.03 & 0.3$\pm$0.1 & 3.6$\pm$0.3 & 0.4$\pm$0.2 & 9606.3$\pm$820.1 & 16.7$\pm$5.9 & 0.14 \\
TW Hya & 59280.29 & E & 0.42$\pm$0.02 & 0.4$\pm$0.1 & 3.9$\pm$0.3 & 0.4$\pm$0.2 & 9216.2$\pm$864.9 & 16.6$\pm$5.8 & 0.12 \\
TW Hya & 59281.18 & C & 0.58$\pm$0.03 & 0.3$\pm$0.1 & 4.2$\pm$0.3 & 0.4$\pm$0.3 & 9352.1$\pm$876.4 & 17.4$\pm$5.7 & 0.14 \\
TW Hya & 59282.23 & C & 0.28$\pm$0.01 & 0.3$\pm$0.2 & 3.5$\pm$0.4 & 0.2$\pm$0.2 & 8666.6$\pm$778.4 & 16.7$\pm$5.9 & 0.11 \\
TW Hya & 59284.24 & C & 0.62$\pm$0.03 & 0.3$\pm$0.1 & 4.0$\pm$0.3 & 0.5$\pm$0.3 & 9567.3$\pm$844.5 & 17.3$\pm$5.8 & 0.10 \\
TW Hya & 59286.26 & C & 0.23$\pm$0.01 & 0.3$\pm$0.2 & 3.4$\pm$0.5 & 0.12$\pm$0.07 & 8596.7$\pm$894.3 & 15.6$\pm$6.0 & 0.13 \\
TW Hya & 59288.25 & C & 0.70$\pm$0.04 & 0.4$\pm$0.2 & 4.4$\pm$0.3 & 0.3$\pm$0.2 & 8793.6$\pm$718.3 & 16.9$\pm$5.8 & 0.10 \\
TW Hya & 59291.18 & C & 0.45$\pm$0.02 & 0.3$\pm$0.2 & 4.2$\pm$0.4 & 0.2$\pm$0.2 & 8605.1$\pm$771.2 & 16.8$\pm$5.8 & 0.11 \\
TW Hya & 59294.17 & C & 0.37$\pm$0.02 & 0.3$\pm$0.1 & 3.7$\pm$0.3 & 0.4$\pm$0.2 & 9382.7$\pm$857.5 & 16.1$\pm$5.9 & 0.10 \\
TW Hya & 59295.20 & C & 0.64$\pm$0.03 & 0.3$\pm$0.1 & 4.4$\pm$0.3 & 0.4$\pm$0.2 & 9017.4$\pm$798.3 & 17.2$\pm$5.7 & 0.09 \\
TW Hya & 59296.18 & C & 1.12$\pm$0.06 & 0.4$\pm$0.1 & 4.8$\pm$0.4 & 0.5$\pm$0.3 & 9085.1$\pm$687.2 & 18.6$\pm$5.1 & 0.12 \\
TW Hya & 59298.15 & C & 0.61$\pm$0.03 & 0.3$\pm$0.1 & 3.8$\pm$0.3 & 0.4$\pm$0.2 & 9477.4$\pm$848.3 & 16.1$\pm$6.0 & 0.12 \\
TW Hya & 59300.11 & C & 0.40$\pm$0.02 & 0.4$\pm$0.1 & 2.6$\pm$0.2 & 0.2$\pm$0.2 & 9395.9$\pm$781.8 & 16.6$\pm$5.8 & 0.23 \\
TW Hya & 59302.13 & C & 0.41$\pm$0.02 & 0.4$\pm$0.2 & 3.0$\pm$0.3 & 0.2$\pm$0.2 & 9015.0$\pm$845.9 & 17.1$\pm$5.7 & 0.15 \\
TW Hya & 59303.12 & C & 0.29$\pm$0.01 & 0.4$\pm$0.2 & 3.9$\pm$0.3 & 0.2$\pm$0.2 & 8660.8$\pm$710.7 & 16.6$\pm$5.9 & 0.15 \\
TW Hya & 59304.16 & C & 0.69$\pm$0.03 & 0.4$\pm$0.1 & 3.4$\pm$0.3 & 0.4$\pm$0.2 & 9679.0$\pm$759.0 & 16.1$\pm$6.0 & 0.18 \\
TW Hya & 59305.12 & C & 0.44$\pm$0.02 & 0.4$\pm$0.2 & 4.3$\pm$0.4 & 0.2$\pm$0.1 & 8660.1$\pm$749.9 & 16.3$\pm$6.0 & 0.13 \\
TW Hya & 59306.15 & C & 0.42$\pm$0.02 & 0.4$\pm$0.2 & 4.7$\pm$0.3 & 0.2$\pm$0.2 & 8582.3$\pm$729.0 & 16.6$\pm$6.0 & 0.13 \\
TW Hya & 59307.00 & X & 0.51$\pm$0.03 & 0.26$\pm$0.04 & 4.4$\pm$0.3 & 0.9$\pm$0.4 & 10109.9$\pm$544.8 & 13.2$\pm$5.8 & 0.14 \\
TW Hya & 59307.16 & C & 0.88$\pm$0.04 & 0.4$\pm$0.1 & 4.0$\pm$0.3 & 0.2$\pm$0.2 & 8867.1$\pm$674.1 & 17.2$\pm$5.7 & 0.14 \\
TW Hya & 59308.04 & E & 0.49$\pm$0.02 & 0.4$\pm$0.1 & 3.9$\pm$0.3 & 0.3$\pm$0.2 & 9218.4$\pm$747.2 & 16.6$\pm$5.9 & 0.13 \\
TW Hya & 59308.13 & C & 0.64$\pm$0.03 & 0.4$\pm$0.1 & 3.7$\pm$0.2 & 0.4$\pm$0.2 & 9433.1$\pm$774.4 & 16.0$\pm$6.0 & 0.13 \\
TW Hya & 59309.14 & C & 1.16$\pm$0.06 & 0.4$\pm$0.1 & 4.1$\pm$0.3 & 0.4$\pm$0.2 & 9223.7$\pm$660.3 & 17.4$\pm$5.7 & 0.13 \\
TW Hya & 59309.14 & E & 0.97$\pm$0.05 & 0.4$\pm$0.1 & 4.3$\pm$0.3 & 0.3$\pm$0.2 & 9052.6$\pm$579.9 & 16.8$\pm$5.9 & 0.13 \\
TW Hya & 59310.09 & E & 1.01$\pm$0.05 & 0.4$\pm$0.1 & 4.4$\pm$0.4 & 0.3$\pm$0.2 & 9045.6$\pm$625.5 & 17.4$\pm$5.7 & 0.17 \\
TW Hya & 59310.13 & C & 1.14$\pm$0.06 & 0.4$\pm$0.1 & 4.3$\pm$0.4 & 0.3$\pm$0.2 & 9049.8$\pm$651.7 & 17.7$\pm$5.7 & 0.17 \\
TW Hya & 59310.15 & X & 0.91$\pm$0.07 & 0.29$\pm$0.04 & 4.4$\pm$0.3 & 0.8$\pm$0.3 & 10197.8$\pm$490.2 & 13.3$\pm$5.6 & 0.18 \\
TW Hya & 59311.14 & C & 0.28$\pm$0.01 & 0.5$\pm$0.1 & 3.0$\pm$0.3 & 0.3$\pm$0.2 & 9344.9$\pm$759.8 & 15.7$\pm$5.9 & 0.23 \\
TW Hya & 59312.16 & C & 0.29$\pm$0.01 & 0.4$\pm$0.2 & 3.3$\pm$0.4 & 0.2$\pm$0.2 & 8818.2$\pm$773.3 & 17.5$\pm$5.7 & 0.18 \\
TW Hya & 59313.15 & C & 0.24$\pm$0.01 & 0.3$\pm$0.2 & 2.9$\pm$0.3 & 0.2$\pm$0.1 & 9090.8$\pm$918.8 & 16.9$\pm$5.6 & 0.13 \\
TW Hya & 59313.22 & E & 0.12$\pm$0.01 & 0.3$\pm$0.2 & 3.1$\pm$0.3 & 0.2$\pm$0.2 & 9184.6$\pm$933.3 & 16.6$\pm$5.7 & 0.13 \\
TW Hya & 59314.23 & C & 0.31$\pm$0.02 & 0.3$\pm$0.1 & 3.1$\pm$0.3 & 0.3$\pm$0.2 & 9310.3$\pm$883.3 & 16.8$\pm$5.7 & 0.14 \\
TW Hya & 59653.15 & C & 0.46$\pm$0.02 & 0.5$\pm$0.1 & 3.9$\pm$0.3 & 0.2$\pm$0.2 & 9194.1$\pm$665.5 & 16.6$\pm$5.9 & 0.16 \\
TW Hya & 59654.15 & C & 0.34$\pm$0.02 & 0.5$\pm$0.1 & 4.6$\pm$0.5 & 0.2$\pm$0.1 & 8918.2$\pm$412.8 & 16.0$\pm$6.0 & 0.16 \\
TW Hya & 59655.22 & C & 1.04$\pm$0.05 & 0.5$\pm$0.1 & 4.0$\pm$0.3 & 0.4$\pm$0.2 & 9203.6$\pm$657.1 & 18.1$\pm$5.4 & 0.13 \\
TW Hya & 59656.17 & C & 1.07$\pm$0.05 & 0.6$\pm$0.1 & 3.5$\pm$0.2 & 0.4$\pm$0.2 & 9515.0$\pm$750.6 & 16.6$\pm$5.8 & 0.14 \\
TW Hya & 59657.21 & C & 0.73$\pm$0.04 & 0.56$\pm$0.09 & 3.8$\pm$0.2 & 0.4$\pm$0.2 & 9367.1$\pm$693.1 & 17.5$\pm$5.7 & 0.16 \\
TW Hya & 59658.19 & C & 0.58$\pm$0.03 & 0.4$\pm$0.1 & 3.1$\pm$0.2 & 0.3$\pm$0.2 & 9503.4$\pm$766.7 & 16.0$\pm$6.0 & 0.11 \\
TW Hya & 59659.20 & C & 0.88$\pm$0.04 & 0.52$\pm$0.09 & 4.2$\pm$0.3 & 0.4$\pm$0.2 & 9290.4$\pm$653.3 & 19.1$\pm$4.9 & 0.15 \\
TW Hya & 59660.15 & C & 1.02$\pm$0.05 & 0.5$\pm$0.1 & 3.9$\pm$0.3 & 0.3$\pm$0.2 & 9088.7$\pm$607.6 & 17.6$\pm$5.7 & 0.14 \\
TW Hya & 59661.22 & C & 0.49$\pm$0.02 & 0.5$\pm$0.1 & 3.5$\pm$0.2 & 0.3$\pm$0.2 & 9276.8$\pm$694.3 & 16.4$\pm$6.0 & 0.15 \\
TW Hya & 59662.16 & C & 0.77$\pm$0.04 & 0.4$\pm$0.1 & 3.1$\pm$0.2 & 0.3$\pm$0.2 & 9513.3$\pm$756.8 & 16.8$\pm$5.8 & 0.13 \\
TW Hya & 59663.18 & C & 0.60$\pm$0.03 & 0.4$\pm$0.1 & 2.9$\pm$0.2 & 0.3$\pm$0.1 & 9591.7$\pm$764.5 & 16.4$\pm$5.8 & 0.14 \\
TW Hya & 59664.14 & C & 0.52$\pm$0.03 & 0.4$\pm$0.1 & 2.8$\pm$0.2 & 0.2$\pm$0.2 & 9429.6$\pm$762.2 & 16.3$\pm$5.9 & 0.15 \\
TW Hya & 59665.16 & C & 0.53$\pm$0.03 & 0.4$\pm$0.1 & 3.6$\pm$0.3 & 0.4$\pm$0.2 & 9296.4$\pm$714.5 & 16.4$\pm$5.9 & 0.12 \\
TW Hya & 59666.19 & C & 1.19$\pm$0.06 & 0.53$\pm$0.09 & 3.9$\pm$0.2 & 0.4$\pm$0.2 & 9479.2$\pm$661.4 & 17.9$\pm$5.3 & 0.13 \\
TW Hya & 59667.14 & C & 0.77$\pm$0.04 & 0.5$\pm$0.1 & 3.1$\pm$0.2 & 0.3$\pm$0.1 & 9556.4$\pm$747.5 & 17.9$\pm$5.5 & 0.20 \\
TW Hya & 59668.14 & C & 0.54$\pm$0.03 & 0.4$\pm$0.1 & 4.0$\pm$0.4 & 0.2$\pm$0.1 & 8979.4$\pm$654.9 & 17.5$\pm$5.8 & 0.21 \\
TW Hya & 59670.10 & C & 1.44$\pm$0.07 & 0.6$\pm$0.1 & 3.3$\pm$0.2 & 0.3$\pm$0.1 & 9637.4$\pm$717.0 & 17.4$\pm$5.8 & 0.17 \\
TW Hya & 59671.06 & C & 0.65$\pm$0.03 & 0.6$\pm$0.1 & 3.2$\pm$0.2 & 0.3$\pm$0.1 & 9628.3$\pm$719.1 & 16.8$\pm$5.9 & 0.18 \\
TW Hya & 59672.01 & C & 0.43$\pm$0.02 & 0.5$\pm$0.1 & 3.1$\pm$0.2 & 0.3$\pm$0.2 & 9476.3$\pm$719.7 & 16.8$\pm$6.0 & 0.16 \\
TW Hya & 59672.26 & C & 0.44$\pm$0.02 & 0.6$\pm$0.1 & 3.2$\pm$0.2 & 0.3$\pm$0.1 & 9619.7$\pm$711.1 & 16.9$\pm$6.0 & 0.16 \\
TW Hya & 59673.13 & C & 1.02$\pm$0.05 & 0.5$\pm$0.1 & 3.1$\pm$0.2 & 0.2$\pm$0.2 & 9319.3$\pm$702.7 & 16.1$\pm$6.1 & 0.15 \\
TW Hya & 59674.14 & C & 0.98$\pm$0.05 & 0.61$\pm$0.07 & 3.8$\pm$0.3 & 0.3$\pm$0.1 & 9609.0$\pm$658.4 & 18.0$\pm$5.3 & 0.17 \\
TW Hya & 59675.09 & C & 0.73$\pm$0.04 & 0.6$\pm$0.1 & 3.2$\pm$0.2 & 0.3$\pm$0.2 & 9605.5$\pm$725.2 & 16.4$\pm$6.1 & 0.14 \\
TW Hya & 59676.01 & C & 1.07$\pm$0.05 & 0.58$\pm$0.08 & 3.7$\pm$0.2 & 0.4$\pm$0.2 & 9728.1$\pm$681.0 & 17.5$\pm$5.5 & 0.16 \\
TW Hya & 59676.23 & C & 1.21$\pm$0.06 & 0.57$\pm$0.09 & 3.8$\pm$0.2 & 0.4$\pm$0.2 & 9563.5$\pm$742.5 & 17.9$\pm$5.3 & 0.16 \\
TW Hya & 59677.07 & C & 1.31$\pm$0.07 & 0.5$\pm$0.1 & 3.6$\pm$0.2 & 0.3$\pm$0.2 & 9428.3$\pm$723.9 & 17.2$\pm$5.6 & 0.17 \\
TW Hya & 59678.01 & C & 0.82$\pm$0.04 & 0.5$\pm$0.1 & 3.0$\pm$0.2 & 0.2$\pm$0.2 & 9360.5$\pm$760.0 & 15.9$\pm$5.9 & 0.15 \\
TW Hya & 59678.18 & C & 0.56$\pm$0.03 & 0.5$\pm$0.1 & 3.0$\pm$0.3 & 0.2$\pm$0.1 & 9355.3$\pm$762.9 & 16.2$\pm$6.0 & 0.14 \\
TW Hya & 59679.05 & C & 0.89$\pm$0.04 & 0.6$\pm$0.1 & 3.6$\pm$0.2 & 0.4$\pm$0.2 & 9664.7$\pm$719.7 & 16.4$\pm$5.9 & 0.18 \\
TW Hya & 59680.07 & C & 0.99$\pm$0.05 & 0.59$\pm$0.09 & 3.3$\pm$0.2 & 0.4$\pm$0.1 & 9980.8$\pm$633.8 & 16.4$\pm$6.0 & 0.18 \\
TW Hya & 59682.09 & C & 1.16$\pm$0.06 & 0.6$\pm$0.1 & 3.6$\pm$0.2 & 0.3$\pm$0.2 & 9693.3$\pm$724.6 & 16.1$\pm$5.8 & 0.19 \\
TW Hya & 59684.15 & C & 1.03$\pm$0.05 & 0.5$\pm$0.1 & 3.4$\pm$0.2 & 0.4$\pm$0.2 & 9691.6$\pm$702.9 & 15.9$\pm$6.0 & 0.19 \\
TW Hya & 59686.12 & C & 0.73$\pm$0.04 & 0.52$\pm$0.09 & 3.9$\pm$0.3 & 0.4$\pm$0.2 & 9609.6$\pm$729.2 & 17.0$\pm$5.8 & 0.16 \\
TW Hya & 59688.09 & C & 0.95$\pm$0.05 & 0.5$\pm$0.1 & 4.6$\pm$0.4 & 0.11$\pm$0.06 & 8936.7$\pm$533.5 & 16.2$\pm$6.1 & 0.14 \\
TW Hya & 59690.06 & C & 0.93$\pm$0.05 & 0.4$\pm$0.1 & 3.7$\pm$0.3 & 0.4$\pm$0.2 & 9564.6$\pm$785.0 & 16.1$\pm$6.0 & 0.15 \\
TW Hya & 59692.07 & C & 0.97$\pm$0.05 & 0.4$\pm$0.1 & 3.3$\pm$0.2 & 0.4$\pm$0.2 & 9724.5$\pm$735.1 & 16.2$\pm$6.0 & 0.15 \\
TW Hya & 59694.03 & C & 1.83$\pm$0.09 & 0.4$\pm$0.1 & 3.4$\pm$0.2 & 0.4$\pm$0.2 & 9707.3$\pm$733.7 & 16.0$\pm$6.0 & 0.14 \\
TW Hya & 59695.05 & C & 0.98$\pm$0.05 & 0.4$\pm$0.1 & 3.8$\pm$0.3 & 0.5$\pm$0.2 & 9662.6$\pm$830.9 & 15.0$\pm$6.0 & 0.13 \\
TW Hya & 59696.18 & C & 0.87$\pm$0.04 & 0.4$\pm$0.1 & 4.5$\pm$0.3 & 0.4$\pm$0.2 & 9108.0$\pm$787.3 & 16.7$\pm$5.9 & 0.08 \\
TW Hya & 59696.24 & C & 0.76$\pm$0.04 & 0.4$\pm$0.1 & 4.4$\pm$0.3 & 0.4$\pm$0.2 & 9064.6$\pm$814.4 & 16.2$\pm$6.0 & 0.08 \\
TW Hya & 59697.15 & C & 0.65$\pm$0.03 & 0.4$\pm$0.2 & 3.7$\pm$0.3 & 0.3$\pm$0.2 & 9263.3$\pm$910.5 & 15.6$\pm$5.8 & 0.09 \\
TW Hya & 59698.14 & C & 1.24$\pm$0.06 & 0.4$\pm$0.1 & 3.7$\pm$0.2 & 0.4$\pm$0.2 & 9594.0$\pm$787.8 & 15.9$\pm$5.9 & 0.09 \\
\enddata
\tablecomments{Instrument names are abbreviated as: C: CHIRON, E: ESPRESSO: X: XSHOOTER, S: SOPHIE, U: UVES, T: TCES}
\end{deluxetable*}

\startlongtable
\begin{deluxetable*}{c c c c c c c c c} \label{tab: Flow Model Results RU Lup}
\tabletypesize{\footnotesize}
\tablecaption{Accretion flow model results for RU Lup}
\tablehead{
\colhead{Object} & \colhead{MJD} & \colhead{Instrument} & \colhead{\mdot} & \colhead{\rin} & \colhead{\Wr} & \colhead{\Tmax} & \colhead{Incl.} & \colhead{MAPE} \\
\colhead{} & \colhead{} & \colhead{} & \colhead{[10$^{8}$ M$_{\odot}$yr$^{-1}$]} & \colhead{[\rstar]} & \colhead{[\rstar]} & \colhead{[K]} & \colhead{[\textdegree]} & \colhead{}
}
\startdata
RU Lup & 59264.37 & C & 8.3$\pm$4.0 & 3.4$\pm$0.3 & 0.6$\pm$0.3 & 7368.8$\pm$343.6 & 20.4$\pm$4.1 & 0.16 \\
RU Lup & 59318.26 & C & 8.2$\pm$4.1 & 3.3$\pm$0.3 & 0.6$\pm$0.3 & 7336.0$\pm$337.1 & 20.3$\pm$4.1 & 0.20 \\
RU Lup & 59395.05 & C & 8.2$\pm$4.0 & 3.7$\pm$0.3 & 0.7$\pm$0.4 & 7434.5$\pm$350.3 & 20.5$\pm$4.1 & 0.16 \\
RU Lup & 59434.99 & C & 7.8$\pm$4.1 & 3.5$\pm$0.3 & 0.8$\pm$0.4 & 7477.3$\pm$395.1 & 20.5$\pm$4.1 & 0.13 \\
RU Lup & 59435.98 & C & 7.9$\pm$4.1 & 3.5$\pm$0.3 & 0.8$\pm$0.4 & 7468.7$\pm$387.1 & 20.4$\pm$4.1 & 0.14 \\
RU Lup & 59436.10 & X & 7.6$\pm$4.1 & 3.7$\pm$0.3 & 0.8$\pm$0.5 & 7467.8$\pm$429.2 & 20.9$\pm$4.0 & 0.14 \\
RU Lup & 59437.00 & C & 7.8$\pm$4.0 & 3.4$\pm$0.3 & 0.8$\pm$0.4 & 7482.0$\pm$380.8 & 20.4$\pm$4.1 & 0.14 \\
RU Lup & 59437.97 & C & 8.1$\pm$4.0 & 3.5$\pm$0.3 & 0.7$\pm$0.4 & 7450.0$\pm$358.0 & 20.4$\pm$4.1 & 0.16 \\
RU Lup & 59439.02 & C & 8.0$\pm$4.0 & 3.2$\pm$0.3 & 0.7$\pm$0.4 & 7483.2$\pm$395.2 & 20.5$\pm$4.1 & 0.13 \\
RU Lup & 59439.99 & C & 7.7$\pm$4.1 & 3.5$\pm$0.3 & 0.9$\pm$0.5 & 7551.5$\pm$410.7 & 20.6$\pm$4.1 & 0.14 \\
RU Lup & 59442.03 & C & 8.1$\pm$4.1 & 3.6$\pm$0.3 & 0.7$\pm$0.4 & 7425.3$\pm$359.5 & 20.5$\pm$4.1 & 0.15 \\
RU Lup & 59443.02 & C & 8.0$\pm$4.0 & 3.3$\pm$0.3 & 0.7$\pm$0.4 & 7464.7$\pm$377.2 & 20.2$\pm$4.1 & 0.14 \\
RU Lup & 59443.14 & C & 8.1$\pm$4.1 & 3.1$\pm$0.3 & 0.7$\pm$0.4 & 7443.3$\pm$371.7 & 20.3$\pm$4.1 & 0.15 \\
RU Lup & 59444.02 & C & 8.1$\pm$4.0 & 3.4$\pm$0.3 & 0.7$\pm$0.4 & 7415.9$\pm$356.2 & 20.5$\pm$4.1 & 0.16 \\
RU Lup & 59444.03 & C & 8.1$\pm$4.0 & 3.4$\pm$0.3 & 0.7$\pm$0.4 & 7438.4$\pm$365.9 & 20.6$\pm$4.1 & 0.15 \\
RU Lup & 59447.02 & C & 8.0$\pm$4.1 & 2.7$\pm$0.2 & 0.6$\pm$0.3 & 7719.6$\pm$552.8 & 20.2$\pm$4.0 & 0.17 \\
RU Lup & 59447.05 & C & 7.9$\pm$4.0 & 3.2$\pm$0.2 & 0.8$\pm$0.4 & 7581.4$\pm$405.1 & 20.2$\pm$4.1 & 0.15 \\
RU Lup & 59448.00 & C & 8.0$\pm$4.0 & 3.8$\pm$0.2 & 0.7$\pm$0.3 & 7523.0$\pm$363.4 & 20.4$\pm$4.1 & 0.16 \\
RU Lup & 59448.07 & X & 7.8$\pm$4.0 & 3.7$\pm$0.3 & 0.7$\pm$0.4 & 7478.7$\pm$364.2 & 20.2$\pm$4.0 & 0.20 \\
RU Lup & 59449.00 & E & 8.2$\pm$4.1 & 3.9$\pm$0.3 & 0.6$\pm$0.3 & 7381.6$\pm$345.2 & 20.5$\pm$4.1 & 0.18 \\
RU Lup & 59449.04 & C & 8.2$\pm$4.0 & 3.8$\pm$0.3 & 0.7$\pm$0.3 & 7385.6$\pm$335.1 & 20.6$\pm$4.0 & 0.18 \\
RU Lup & 59449.98 & C & 8.3$\pm$4.1 & 3.7$\pm$0.3 & 0.6$\pm$0.3 & 7362.8$\pm$341.5 & 20.5$\pm$4.1 & 0.17 \\
RU Lup & 59453.01 & C & 8.0$\pm$4.1 & 3.2$\pm$0.3 & 0.7$\pm$0.4 & 7429.6$\pm$372.3 & 20.3$\pm$4.1 & 0.14 \\
RU Lup & 59458.02 & E & 7.9$\pm$4.0 & 3.2$\pm$0.3 & 0.7$\pm$0.4 & 7537.0$\pm$386.3 & 20.4$\pm$4.1 & 0.13 \\
RU Lup & 59458.06 & E & 8.0$\pm$4.0 & 3.1$\pm$0.3 & 0.7$\pm$0.4 & 7524.6$\pm$396.0 & 20.4$\pm$4.1 & 0.13 \\
RU Lup & 59676.20 & C & 8.0$\pm$4.1 & 3.6$\pm$0.3 & 0.7$\pm$0.4 & 7384.6$\pm$356.1 & 20.4$\pm$4.0 & 0.15 \\
RU Lup & 59683.17 & C & 8.1$\pm$4.0 & 3.4$\pm$0.3 & 0.8$\pm$0.4 & 7407.1$\pm$372.0 & 20.5$\pm$4.1 & 0.15 \\
RU Lup & 59690.25 & C & 8.0$\pm$4.1 & 3.4$\pm$0.3 & 0.7$\pm$0.4 & 7436.9$\pm$373.7 & 20.3$\pm$4.1 & 0.15 \\
RU Lup & 59699.19 & C & 8.2$\pm$4.0 & 3.3$\pm$0.3 & 0.7$\pm$0.4 & 7407.6$\pm$348.0 & 20.3$\pm$4.1 & 0.15 \\
RU Lup & 59715.17 & C & 8.1$\pm$4.1 & 3.5$\pm$0.3 & 0.7$\pm$0.4 & 7418.0$\pm$356.7 & 20.5$\pm$4.1 & 0.16 \\
RU Lup & 59723.14 & C & 8.0$\pm$4.0 & 3.1$\pm$0.2 & 0.7$\pm$0.4 & 7495.1$\pm$384.0 & 20.5$\pm$4.1 & 0.15 \\
RU Lup & 59736.11 & C & 7.8$\pm$4.0 & 3.7$\pm$0.2 & 0.8$\pm$0.4 & 7538.9$\pm$379.5 & 20.6$\pm$4.1 & 0.16 \\
RU Lup & 59790.11 & C & 7.9$\pm$4.0 & 3.6$\pm$0.3 & 0.9$\pm$0.4 & 7565.2$\pm$396.6 & 20.6$\pm$4.1 & 0.14 \\
RU Lup & 59799.03 & C & 7.9$\pm$4.0 & 3.5$\pm$0.2 & 0.8$\pm$0.4 & 7550.9$\pm$395.1 & 20.5$\pm$4.1 & 0.15 \\
RU Lup & 59800.03 & C & 8.2$\pm$4.0 & 3.7$\pm$0.3 & 0.7$\pm$0.4 & 7434.4$\pm$355.0 & 20.7$\pm$4.0 & 0.16 \\
RU Lup & 59801.00 & C & 7.9$\pm$4.0 & 3.7$\pm$0.2 & 0.8$\pm$0.4 & 7537.4$\pm$374.4 & 20.6$\pm$4.0 & 0.17 \\
RU Lup & 59801.05 & E & 7.8$\pm$4.0 & 3.6$\pm$0.2 & 0.8$\pm$0.4 & 7524.4$\pm$371.7 & 20.6$\pm$4.0 & 0.17 \\
RU Lup & 59801.98 & C & 8.0$\pm$4.0 & 3.8$\pm$0.3 & 0.7$\pm$0.4 & 7477.8$\pm$362.6 & 20.7$\pm$4.0 & 0.16 \\
RU Lup & 59802.07 & E & 8.0$\pm$4.0 & 3.8$\pm$0.3 & 0.8$\pm$0.4 & 7487.0$\pm$372.8 & 20.6$\pm$4.0 & 0.16 \\
RU Lup & 59802.97 & C & 8.1$\pm$4.1 & 3.4$\pm$0.3 & 0.7$\pm$0.4 & 7451.8$\pm$359.2 & 20.7$\pm$4.1 & 0.16 \\
RU Lup & 59804.00 & C & 8.1$\pm$4.1 & 3.8$\pm$0.3 & 0.6$\pm$0.3 & 7349.0$\pm$354.2 & 20.6$\pm$4.1 & 0.16 \\
RU Lup & 59804.10 & E & 8.0$\pm$4.0 & 3.8$\pm$0.3 & 0.6$\pm$0.3 & 7337.4$\pm$345.7 & 20.7$\pm$4.0 & 0.17 \\
RU Lup & 59805.06 & C & 8.0$\pm$4.0 & 3.6$\pm$0.3 & 0.7$\pm$0.4 & 7469.4$\pm$370.3 & 20.6$\pm$4.0 & 0.15 \\
RU Lup & 59806.00 & C & 8.0$\pm$4.0 & 3.8$\pm$0.3 & 0.7$\pm$0.4 & 7469.5$\pm$365.0 & 20.7$\pm$4.0 & 0.15 \\
RU Lup & 59807.00 & C & 8.2$\pm$4.1 & 3.6$\pm$0.3 & 0.6$\pm$0.3 & 7362.5$\pm$340.2 & 20.7$\pm$4.0 & 0.17 \\
RU Lup & 59807.10 & E & 8.2$\pm$4.1 & 3.6$\pm$0.3 & 0.6$\pm$0.3 & 7367.3$\pm$344.6 & 20.7$\pm$4.0 & 0.16 \\
RU Lup & 59808.05 & C & 8.2$\pm$4.1 & 3.7$\pm$0.3 & 0.7$\pm$0.4 & 7376.8$\pm$356.0 & 20.8$\pm$4.0 & 0.16 \\
RU Lup & 59809.07 & C & 8.2$\pm$4.0 & 3.4$\pm$0.3 & 0.7$\pm$0.3 & 7401.8$\pm$353.1 & 20.7$\pm$4.1 & 0.16 \\
RU Lup & 59810.03 & C & 8.1$\pm$4.1 & 3.3$\pm$0.3 & 0.7$\pm$0.4 & 7374.9$\pm$363.8 & 20.7$\pm$4.1 & 0.16 \\
RU Lup & 59811.05 & C & 8.1$\pm$4.1 & 3.4$\pm$0.3 & 0.7$\pm$0.4 & 7467.2$\pm$364.8 & 20.5$\pm$4.1 & 0.15 \\
RU Lup & 59812.02 & C & 8.0$\pm$3.9 & 3.3$\pm$0.3 & 0.7$\pm$0.4 & 7520.5$\pm$367.2 & 20.6$\pm$4.1 & 0.15 \\
RU Lup & 59813.05 & C & 8.2$\pm$4.0 & 3.8$\pm$0.3 & 0.7$\pm$0.3 & 7372.6$\pm$344.2 & 20.7$\pm$4.0 & 0.19 \\
RU Lup & 59814.00 & C & 8.0$\pm$4.1 & 3.5$\pm$0.3 & 0.7$\pm$0.4 & 7437.7$\pm$363.7 & 20.7$\pm$4.1 & 0.16 \\
RU Lup & 59814.04 & E & 8.1$\pm$4.1 & 3.5$\pm$0.3 & 0.7$\pm$0.4 & 7413.8$\pm$360.2 & 20.6$\pm$4.0 & 0.17 \\
RU Lup & 59815.05 & C & 7.9$\pm$4.0 & 3.4$\pm$0.3 & 0.8$\pm$0.4 & 7502.5$\pm$367.8 & 20.7$\pm$4.1 & 0.17 \\
RU Lup & 59816.02 & C & 8.2$\pm$4.0 & 3.8$\pm$0.3 & 0.7$\pm$0.4 & 7426.4$\pm$337.3 & 20.6$\pm$4.0 & 0.17 \\
RU Lup & 59817.04 & C & 8.1$\pm$4.0 & 3.9$\pm$0.3 & 0.7$\pm$0.3 & 7482.1$\pm$350.2 & 20.7$\pm$4.0 & 0.19 \\
\enddata
\tablecomments{Instrument names are abbreviated as: C: CHIRON, E: ESPRESSO: X: XSHOOTER, S: SOPHIE, U: UVES, T: TCES}
\end{deluxetable*}

\startlongtable
\begin{deluxetable*}{c c c c c c c c c c} \label{tab: Flow Model Results BP Tau}
\tabletypesize{\footnotesize}
\tablecaption{Accretion flow model results and veilings for BP Tau}
\tablehead{
\colhead{Object} & \colhead{MJD} & \colhead{Instrument} & \colhead{r$_V$} & \colhead{\mdot} & \colhead{\rin} & \colhead{\Wr} & \colhead{\Tmax} & \colhead{Incl.} & \colhead{MAPE} \\
\colhead{} & \colhead{} & \colhead{} & \colhead{} & \colhead{[10$^{8}$ M$_{\odot}$yr$^{-1}$]} & \colhead{[\rstar]} & \colhead{[\rstar]} & \colhead{[K]} & \colhead{[\textdegree]} & \colhead{}
}
\startdata
BP Tau & 59448.39 & X & 0.81$\pm$0.04 & 2.1$\pm$1.6 & 3.6$\pm$0.4 & 0.7$\pm$0.4 & 8181.4$\pm$819.5 & 44.7$\pm$6.8 & 0.13 \\
BP Tau & 59452.35 & X & 1.27$\pm$0.06 & 1.8$\pm$1.6 & 3.7$\pm$0.4 & 0.8$\pm$0.5 & 8311.2$\pm$913.6 & 45.5$\pm$6.5 & 0.11 \\
BP Tau & 59459.38 & E & 1.30$\pm$0.06 & 2.2$\pm$1.7 & 3.6$\pm$0.4 & 0.6$\pm$0.4 & 8195.0$\pm$821.8 & 43.0$\pm$7.5 & 0.13 \\
BP Tau & 59460.37 & X & 1.87$\pm$0.22 & 1.8$\pm$1.4 & 3.8$\pm$0.4 & 1.3$\pm$0.5 & 8512.9$\pm$783.3 & 45.0$\pm$6.5 & 0.11 \\
BP Tau & 59464.36 & E & 0.69$\pm$0.03 & 2.1$\pm$1.7 & 3.4$\pm$0.4 & 0.5$\pm$0.4 & 8087.9$\pm$919.2 & 42.6$\pm$7.4 & 0.13 \\
BP Tau & 59467.38 & C & 0.75$\pm$0.04 & 2.1$\pm$1.8 & 3.3$\pm$0.4 & 0.4$\pm$0.3 & 8035.9$\pm$899.1 & 42.2$\pm$7.6 & 0.14 \\
BP Tau & 59470.38 & C & 0.57$\pm$0.03 & 2.0$\pm$1.7 & 3.4$\pm$0.4 & 0.5$\pm$0.4 & 8165.6$\pm$925.4 & 42.9$\pm$7.3 & 0.14 \\
BP Tau & 59845.35 & C & 0.92$\pm$0.05 & 2.0$\pm$1.7 & 3.2$\pm$0.4 & 0.7$\pm$0.5 & 8266.1$\pm$900.4 & 44.8$\pm$6.6 & 0.11 \\
BP Tau & 59859.33 & C & 0.73$\pm$0.04 & 2.2$\pm$1.8 & 3.4$\pm$0.4 & 0.7$\pm$0.5 & 8134.4$\pm$866.9 & 45.8$\pm$6.4 & 0.11 \\
BP Tau & 59863.31 & C & 0.64$\pm$0.03 & 2.3$\pm$1.8 & 4.1$\pm$0.6 & 1.1$\pm$0.6 & 8035.8$\pm$782.3 & 46.4$\pm$6.0 & 0.11 \\
BP Tau & 59868.29 & C & 0.66$\pm$0.03 & 2.2$\pm$1.8 & 3.1$\pm$0.4 & 0.4$\pm$0.3 & 7998.3$\pm$847.8 & 43.7$\pm$7.2 & 0.11 \\
BP Tau & 59882.25 & C & 0.71$\pm$0.04 & 2.4$\pm$1.9 & 2.8$\pm$0.3 & 0.3$\pm$0.3 & 7881.4$\pm$842.4 & 41.8$\pm$7.5 & 0.14 \\
BP Tau & 59893.21 & C & 0.81$\pm$0.04 & 2.3$\pm$1.8 & 2.4$\pm$0.3 & 0.3$\pm$0.3 & 7860.0$\pm$850.6 & 42.3$\pm$7.7 & 0.09 \\
BP Tau & 59898.19 & C & 0.75$\pm$0.04 & 2.4$\pm$1.8 & 3.0$\pm$0.3 & 0.6$\pm$0.4 & 8058.6$\pm$862.5 & 45.1$\pm$6.6 & 0.11 \\
BP Tau & 59907.19 & C & 1.20$\pm$0.06 & 2.2$\pm$1.8 & 2.9$\pm$0.3 & 0.5$\pm$0.4 & 8084.9$\pm$878.4 & 42.2$\pm$7.4 & 0.14 \\
BP Tau & 59909.17 & C & 0.91$\pm$0.05 & 2.4$\pm$1.8 & 2.8$\pm$0.3 & 0.4$\pm$0.3 & 7949.8$\pm$842.6 & 43.2$\pm$7.1 & 0.11 \\
BP Tau & 59910.17 & C & 1.01$\pm$0.05 & 2.4$\pm$1.7 & 2.7$\pm$0.3 & 0.5$\pm$0.3 & 8087.6$\pm$785.0 & 41.5$\pm$7.5 & 0.13 \\
BP Tau & 59911.18 & C & 1.04$\pm$0.05 & 2.1$\pm$1.7 & 3.1$\pm$0.3 & 0.7$\pm$0.5 & 8232.7$\pm$845.4 & 44.0$\pm$7.0 & 0.13 \\
BP Tau & 59912.15 & C & 1.00$\pm$0.05 & 2.2$\pm$1.8 & 3.1$\pm$0.3 & 0.6$\pm$0.4 & 8145.1$\pm$860.1 & 43.7$\pm$7.0 & 0.11 \\
BP Tau & 59913.17 & C & 1.05$\pm$0.05 & 2.2$\pm$1.6 & 3.4$\pm$0.4 & 1.0$\pm$0.5 & 8289.3$\pm$774.4 & 46.9$\pm$5.3 & 0.11 \\
BP Tau & 59914.18 & C & 0.70$\pm$0.04 & 2.2$\pm$1.7 & 3.3$\pm$0.4 & 0.8$\pm$0.5 & 8188.4$\pm$870.7 & 45.7$\pm$6.3 & 0.11 \\
BP Tau & 59915.18 & C & 0.80$\pm$0.04 & 2.1$\pm$1.8 & 3.1$\pm$0.3 & 0.6$\pm$0.4 & 8194.3$\pm$922.6 & 42.6$\pm$7.5 & 0.14 \\
BP Tau & 59916.19 & C & 0.61$\pm$0.03 & 2.4$\pm$1.8 & 2.7$\pm$0.3 & 0.5$\pm$0.4 & 8007.4$\pm$829.9 & 42.9$\pm$7.3 & 0.14 \\
BP Tau & 59917.16 & C & 0.94$\pm$0.05 & 2.5$\pm$1.8 & 2.8$\pm$0.3 & 0.4$\pm$0.3 & 7966.7$\pm$822.2 & 40.5$\pm$7.5 & 0.15 \\
BP Tau & 59918.16 & C & 0.94$\pm$0.05 & 2.2$\pm$1.8 & 2.9$\pm$0.3 & 0.5$\pm$0.4 & 8137.9$\pm$889.5 & 41.4$\pm$7.5 & 0.12 \\
BP Tau & 59919.15 & C & 1.33$\pm$0.07 & 2.2$\pm$1.5 & 3.0$\pm$0.3 & 1.0$\pm$0.6 & 8416.5$\pm$771.9 & 43.2$\pm$7.0 & 0.13 \\
BP Tau & 59922.14 & C & 1.04$\pm$0.05 & 2.4$\pm$1.8 & 2.8$\pm$0.3 & 0.5$\pm$0.4 & 8035.6$\pm$809.6 & 42.9$\pm$7.2 & 0.10 \\
BP Tau & 59925.16 & C & 0.75$\pm$0.04 & 2.3$\pm$1.8 & 2.7$\pm$0.3 & 0.4$\pm$0.3 & 7957.2$\pm$866.6 & 41.2$\pm$7.5 & 0.13 \\
BP Tau & 59926.14 & C & 0.78$\pm$0.04 & 2.4$\pm$1.8 & 3.0$\pm$0.3 & 0.5$\pm$0.3 & 8050.8$\pm$813.9 & 41.7$\pm$7.4 & 0.12 \\
BP Tau & 59927.15 & C & 1.13$\pm$0.06 & 2.3$\pm$1.6 & 3.1$\pm$0.3 & 0.6$\pm$0.4 & 8126.2$\pm$767.3 & 43.8$\pm$7.2 & 0.12 \\
BP Tau & 59928.11 & E & 0.59$\pm$0.03 & 2.1$\pm$1.6 & 3.2$\pm$0.3 & 1.1$\pm$0.6 & 8408.5$\pm$802.9 & 45.1$\pm$6.4 & 0.10 \\
BP Tau & 59928.13 & C & 0.76$\pm$0.04 & 2.1$\pm$1.6 & 3.1$\pm$0.3 & 1.0$\pm$0.6 & 8359.5$\pm$860.2 & 44.8$\pm$6.7 & 0.10 \\
BP Tau & 59929.14 & C & 0.67$\pm$0.03 & 2.1$\pm$1.7 & 3.2$\pm$0.5 & 1.0$\pm$0.6 & 8291.1$\pm$900.4 & 45.3$\pm$6.3 & 0.10 \\
BP Tau & 59930.07 & E & 0.42$\pm$0.02 & 2.1$\pm$1.7 & 3.2$\pm$0.4 & 1.0$\pm$0.6 & 8278.4$\pm$894.6 & 44.6$\pm$6.8 & 0.09 \\
BP Tau & 59930.13 & C & 0.61$\pm$0.03 & 2.0$\pm$1.7 & 3.1$\pm$0.4 & 0.8$\pm$0.5 & 8201.3$\pm$922.6 & 44.3$\pm$7.1 & 0.09 \\
BP Tau & 59931.13 & C & 0.67$\pm$0.03 & 2.3$\pm$1.8 & 3.0$\pm$0.4 & 0.5$\pm$0.4 & 8028.5$\pm$892.2 & 44.4$\pm$7.0 & 0.12 \\
BP Tau & 59932.13 & C & 0.66$\pm$0.03 & 2.1$\pm$1.6 & 2.9$\pm$0.3 & 0.9$\pm$0.6 & 8335.7$\pm$840.9 & 45.2$\pm$6.3 & 0.12 \\
BP Tau & 59933.13 & C & 0.55$\pm$0.03 & 2.2$\pm$1.8 & 2.9$\pm$0.3 & 0.7$\pm$0.5 & 8206.4$\pm$931.8 & 44.3$\pm$6.8 & 0.12 \\
BP Tau & 59933.17 & E & 0.40$\pm$0.02 & 2.2$\pm$1.8 & 3.0$\pm$0.3 & 0.7$\pm$0.6 & 8252.3$\pm$905.3 & 45.1$\pm$6.5 & 0.12 \\
BP Tau & 59934.07 & E & 0.42$\pm$0.02 & 2.4$\pm$1.9 & 3.0$\pm$0.3 & 0.4$\pm$0.3 & 7892.1$\pm$869.7 & 42.4$\pm$7.5 & 0.11 \\
BP Tau & 59934.12 & C & 0.55$\pm$0.03 & 2.4$\pm$1.9 & 2.9$\pm$0.3 & 0.3$\pm$0.2 & 7842.1$\pm$842.6 & 41.8$\pm$7.5 & 0.12 \\
BP Tau & 59935.11 & C & 0.51$\pm$0.03 & 2.4$\pm$1.8 & 2.9$\pm$0.3 & 0.3$\pm$0.3 & 7885.2$\pm$851.5 & 42.1$\pm$7.5 & 0.12 \\
BP Tau & 59936.08 & E & 0.76$\pm$0.04 & 2.1$\pm$1.6 & 3.2$\pm$0.3 & 0.7$\pm$0.5 & 8227.9$\pm$851.6 & 43.2$\pm$7.3 & 0.11 \\
BP Tau & 59936.11 & C & 0.93$\pm$0.05 & 2.1$\pm$1.8 & 3.1$\pm$0.3 & 0.6$\pm$0.4 & 8138.8$\pm$889.2 & 42.8$\pm$7.5 & 0.11 \\
BP Tau & 59937.12 & C & 0.97$\pm$0.05 & 2.0$\pm$1.7 & 3.0$\pm$0.4 & 0.7$\pm$0.5 & 8156.8$\pm$910.0 & 42.5$\pm$7.4 & 0.08 \\
BP Tau & 59938.12 & C & 0.76$\pm$0.04 & 2.1$\pm$1.7 & 3.0$\pm$0.4 & 0.6$\pm$0.4 & 8080.4$\pm$871.4 & 44.4$\pm$6.9 & 0.10 \\
BP Tau & 59939.11 & C & 0.68$\pm$0.03 & 2.2$\pm$1.7 & 2.7$\pm$0.3 & 0.5$\pm$0.4 & 8028.8$\pm$926.5 & 42.8$\pm$7.5 & 0.09 \\
BP Tau & 59940.11 & C & 0.68$\pm$0.03 & 1.9$\pm$1.6 & 2.8$\pm$0.3 & 0.9$\pm$0.6 & 8400.3$\pm$901.7 & 43.1$\pm$7.3 & 0.10 \\
BP Tau & 59941.13 & C & 0.69$\pm$0.03 & 2.2$\pm$1.7 & 2.7$\pm$0.3 & 0.7$\pm$0.5 & 8173.9$\pm$857.2 & 44.7$\pm$6.5 & 0.11 \\
BP Tau & 59949.08 & C & 0.86$\pm$0.04 & 2.1$\pm$1.6 & 3.1$\pm$0.4 & 0.7$\pm$0.5 & 8207.5$\pm$854.7 & 45.7$\pm$6.4 & 0.13 \\
BP Tau & 59950.08 & C & 1.12$\pm$0.06 & 2.3$\pm$1.7 & 3.1$\pm$0.3 & 0.6$\pm$0.4 & 8105.3$\pm$762.3 & 45.0$\pm$6.7 & 0.14 \\
BP Tau & 59951.08 & C & 1.11$\pm$0.06 & 2.3$\pm$1.8 & 3.0$\pm$0.3 & 0.6$\pm$0.4 & 8087.0$\pm$800.9 & 44.7$\pm$6.7 & 0.12 \\
BP Tau & 59952.08 & C & 1.10$\pm$0.06 & 2.2$\pm$1.7 & 3.1$\pm$0.4 & 0.7$\pm$0.5 & 8137.3$\pm$847.8 & 45.1$\pm$6.4 & 0.10 \\
BP Tau & 59953.07 & C & 1.38$\pm$0.07 & 2.2$\pm$1.7 & 3.2$\pm$0.4 & 0.8$\pm$0.4 & 8160.0$\pm$780.1 & 45.1$\pm$6.8 & 0.12 \\
\enddata
\tablecomments{Instrument names are abbreviated as: C: CHIRON, E: ESPRESSO: X: XSHOOTER, S: SOPHIE, U: UVES, T: TCES}
\end{deluxetable*}

\startlongtable
\begin{deluxetable*}{c c c c c c c c c c} \label{tab: Flow Model Results GM Aur}
\tabletypesize{\footnotesize}
\tablecaption{Accretion flow model results and veilings for GM Aur}
\tablehead{
\colhead{Object} & \colhead{MJD} & \colhead{Instrument} & \colhead{r$_V$} & \colhead{\mdot} & \colhead{\rin} & \colhead{\Wr} & \colhead{\Tmax} & \colhead{Incl.} & \colhead{MAPE} \\
\colhead{} & \colhead{} & \colhead{} & \colhead{} & \colhead{[10$^{8}$ M$_{\odot}$yr$^{-1}$]} & \colhead{[\rstar]} & \colhead{[\rstar]} & \colhead{[K]} & \colhead{[\textdegree]} & \colhead{}
}
\startdata
GM Aur & 59498.99 & S & 0.30$\pm$0.03 & 1.3$\pm$1.0 & 3.2$\pm$0.2 & 0.3$\pm$0.1 & 9891.0$\pm$1085.9 & 53.1$\pm$5.8 & 0.14 \\
GM Aur & 59499.98 & S & 0.17$\pm$0.02 & 1.3$\pm$0.9 & 3.1$\pm$0.2 & 0.2$\pm$0.1 & 9901.8$\pm$1057.6 & 51.8$\pm$6.0 & 0.13 \\
GM Aur & 59501.13 & S & 0.27$\pm$0.03 & 1.5$\pm$1.1 & 3.2$\pm$0.2 & 0.3$\pm$0.2 & 9674.5$\pm$1130.9 & 55.3$\pm$5.7 & 0.13 \\
GM Aur & 59502.03 & S & 0.35$\pm$0.04 & 1.1$\pm$0.9 & 3.1$\pm$0.3 & 0.3$\pm$0.1 & 9784.0$\pm$1063.2 & 54.3$\pm$5.8 & 0.12 \\
GM Aur & 59502.96 & T & 0.50$\pm$0.05 & 0.9$\pm$0.8 & 3.2$\pm$0.4 & 0.4$\pm$0.1 & 9465.4$\pm$1088.5 & 55.5$\pm$4.8 & 0.13 \\
GM Aur & 59503.11 & S & 0.31$\pm$0.03 & 0.9$\pm$0.6 & 3.2$\pm$0.3 & 0.3$\pm$0.1 & 9896.9$\pm$1041.6 & 50.8$\pm$5.4 & 0.12 \\
GM Aur & 59504.08 & S & 0.18$\pm$0.02 & 1.0$\pm$0.8 & 3.5$\pm$0.3 & 0.4$\pm$0.1 & 9690.5$\pm$1045.8 & 55.4$\pm$5.1 & 0.11 \\
GM Aur & 59504.34 & X & 0.47$\pm$0.05 & 0.8$\pm$0.5 & 4.1$\pm$0.3 & 1.1$\pm$0.5 & 9913.5$\pm$913.6 & 54.5$\pm$4.3 & 0.09 \\
GM Aur & 59505.02 & S & 0.17$\pm$0.02 & 1.2$\pm$0.9 & 3.2$\pm$0.2 & 0.3$\pm$0.1 & 9940.1$\pm$1070.1 & 53.6$\pm$5.7 & 0.12 \\
GM Aur & 59505.11 & T & 0.29$\pm$0.03 & 1.2$\pm$1.0 & 2.9$\pm$0.2 & 0.3$\pm$0.2 & 9710.6$\pm$1092.8 & 54.8$\pm$5.5 & 0.13 \\
GM Aur & 59505.96 & S & 0.29$\pm$0.03 & 1.5$\pm$1.1 & 3.3$\pm$0.2 & 0.3$\pm$0.2 & 9746.0$\pm$1075.8 & 52.4$\pm$6.0 & 0.12 \\
GM Aur & 59507.09 & T & 0.63$\pm$0.06 & 2.0$\pm$1.4 & 2.7$\pm$0.3 & 0.7$\pm$0.4 & 9219.0$\pm$1089.1 & 58.9$\pm$4.0 & 0.12 \\
GM Aur & 59509.23 & E & 0.83$\pm$0.08 & 1.3$\pm$0.9 & 3.7$\pm$0.3 & 0.4$\pm$0.1 & 9850.0$\pm$1084.4 & 52.0$\pm$5.6 & 0.13 \\
GM Aur & 59510.04 & S & 0.41$\pm$0.04 & 1.1$\pm$0.7 & 3.6$\pm$0.3 & 0.4$\pm$0.1 & 9904.6$\pm$1030.3 & 52.6$\pm$5.8 & 0.13 \\
GM Aur & 59510.09 & T & 0.60$\pm$0.06 & 1.0$\pm$0.9 & 3.5$\pm$0.3 & 0.4$\pm$0.1 & 9575.9$\pm$1066.4 & 56.9$\pm$4.7 & 0.14 \\
GM Aur & 59510.14 & T & 0.60$\pm$0.06 & 1.0$\pm$0.8 & 3.5$\pm$0.3 & 0.4$\pm$0.1 & 9610.6$\pm$1042.8 & 56.9$\pm$4.7 & 0.14 \\
GM Aur & 59510.97 & T & 0.38$\pm$0.04 & 1.1$\pm$0.9 & 3.4$\pm$0.3 & 0.4$\pm$0.1 & 9642.8$\pm$1066.4 & 55.8$\pm$5.1 & 0.15 \\
GM Aur & 59511.00 & S & 0.25$\pm$0.02 & 1.2$\pm$0.9 & 3.5$\pm$0.3 & 0.3$\pm$0.1 & 9864.2$\pm$1063.5 & 52.6$\pm$5.7 & 0.13 \\
GM Aur & 59511.01 & T & 0.38$\pm$0.04 & 1.1$\pm$0.8 & 3.3$\pm$0.3 & 0.4$\pm$0.1 & 9708.8$\pm$1056.8 & 55.5$\pm$5.3 & 0.15 \\
GM Aur & 59511.97 & T & 0.32$\pm$0.03 & 0.9$\pm$0.7 & 2.9$\pm$0.2 & 0.2$\pm$0.1 & 9633.4$\pm$1034.7 & 52.5$\pm$6.0 & 0.12 \\
GM Aur & 59511.99 & S & 0.23$\pm$0.02 & 0.9$\pm$0.7 & 2.9$\pm$0.2 & 0.2$\pm$0.1 & 9908.1$\pm$995.4 & 51.7$\pm$5.8 & 0.10 \\
GM Aur & 59512.01 & T & 0.31$\pm$0.03 & 0.9$\pm$0.7 & 2.8$\pm$0.3 & 0.2$\pm$0.1 & 9684.6$\pm$1058.1 & 52.4$\pm$6.3 & 0.13 \\
GM Aur & 59513.03 & T & 0.44$\pm$0.04 & 1.1$\pm$0.9 & 2.9$\pm$0.2 & 0.2$\pm$0.1 & 9674.9$\pm$1038.3 & 54.3$\pm$6.0 & 0.14 \\
GM Aur & 59513.08 & T & 0.49$\pm$0.05 & 1.1$\pm$0.8 & 2.6$\pm$0.2 & 0.2$\pm$0.1 & 9836.4$\pm$1090.8 & 52.2$\pm$6.4 & 0.13 \\
GM Aur & 59513.10 & S & 0.31$\pm$0.03 & 1.2$\pm$0.9 & 3.2$\pm$0.2 & 0.2$\pm$0.1 & 9884.9$\pm$1031.3 & 53.7$\pm$6.1 & 0.13 \\
GM Aur & 59514.07 & S & 0.45$\pm$0.05 & 1.0$\pm$0.7 & 3.0$\pm$0.2 & 0.2$\pm$0.1 & 10071.8$\pm$990.7 & 51.1$\pm$5.6 & 0.13 \\
GM Aur & 59515.04 & T & 0.42$\pm$0.04 & 0.8$\pm$0.7 & 3.0$\pm$0.3 & 0.2$\pm$0.1 & 9714.5$\pm$1070.3 & 51.4$\pm$5.4 & 0.15 \\
GM Aur & 59515.08 & T & 0.39$\pm$0.04 & 0.7$\pm$0.7 & 2.9$\pm$0.3 & 0.2$\pm$0.1 & 9679.2$\pm$1068.5 & 51.3$\pm$5.4 & 0.15 \\
GM Aur & 59515.09 & S & 0.27$\pm$0.03 & 0.8$\pm$0.6 & 2.9$\pm$0.2 & 0.2$\pm$0.1 & 9938.5$\pm$1043.4 & 49.3$\pm$5.2 & 0.13 \\
GM Aur & 59516.01 & S & 0.21$\pm$0.02 & 0.8$\pm$0.6 & 2.8$\pm$0.2 & 0.2$\pm$0.1 & 9943.8$\pm$1021.7 & 49.5$\pm$5.2 & 0.14 \\
GM Aur & 59553.13 & E & 0.19$\pm$0.02 & 0.9$\pm$0.7 & 2.9$\pm$0.2 & 0.2$\pm$0.1 & 9828.5$\pm$1033.0 & 51.2$\pm$5.4 & 0.11 \\
GM Aur & 59554.17 & E & 0.24$\pm$0.02 & 1.0$\pm$0.7 & 3.0$\pm$0.2 & 0.16$\pm$0.09 & 10057.2$\pm$1004.7 & 51.5$\pm$5.7 & 0.11 \\
GM Aur & 59555.19 & E & 0.58$\pm$0.06 & 1.1$\pm$0.8 & 3.0$\pm$0.2 & 0.2$\pm$0.1 & 9935.9$\pm$1041.7 & 52.2$\pm$5.9 & 0.12 \\
GM Aur & 59556.15 & X & 0.84$\pm$0.08 & 0.6$\pm$0.2 & 3.7$\pm$0.4 & 0.4$\pm$0.2 & 10276.0$\pm$862.1 & 45.6$\pm$3.0 & 0.10 \\
GM Aur & 59556.19 & E & 0.55$\pm$0.05 & 1.0$\pm$0.7 & 3.2$\pm$0.3 & 0.2$\pm$0.1 & 9863.9$\pm$1061.7 & 51.6$\pm$5.8 & 0.11 \\
GM Aur & 59558.86 & T & 0.30$\pm$0.03 & 0.9$\pm$0.8 & 2.9$\pm$0.3 & 0.2$\pm$0.1 & 9603.7$\pm$1101.6 & 52.5$\pm$6.0 & 0.13 \\
GM Aur & 59558.90 & T & 0.30$\pm$0.03 & 0.8$\pm$0.8 & 2.8$\pm$0.3 & 0.2$\pm$0.1 & 9641.9$\pm$1112.8 & 51.0$\pm$6.0 & 0.13 \\
GM Aur & 59559.91 & T & 0.27$\pm$0.03 & 0.9$\pm$0.8 & 2.7$\pm$0.3 & 0.2$\pm$0.1 & 9621.8$\pm$1134.0 & 51.1$\pm$5.9 & 0.13 \\
GM Aur & 59559.95 & T & 0.31$\pm$0.03 & 0.9$\pm$0.8 & 2.7$\pm$0.3 & 0.2$\pm$0.1 & 9624.2$\pm$1129.2 & 50.5$\pm$5.7 & 0.13 \\
GM Aur & 59910.17 & E & 0.22$\pm$0.02 & 0.9$\pm$0.6 & 3.1$\pm$0.3 & 0.3$\pm$0.1 & 9833.3$\pm$1075.8 & 52.1$\pm$5.6 & 0.12 \\
GM Aur & 59910.21 & C & 0.21$\pm$0.02 & 0.8$\pm$0.6 & 3.0$\pm$0.3 & 0.3$\pm$0.1 & 9887.2$\pm$1039.5 & 52.0$\pm$5.5 & 0.13 \\
GM Aur & 59911.19 & C & 0.16$\pm$0.02 & 0.8$\pm$0.6 & 2.9$\pm$0.3 & 0.3$\pm$0.1 & 9683.6$\pm$1069.2 & 51.5$\pm$5.6 & 0.12 \\
GM Aur & 59912.20 & C & 0.43$\pm$0.04 & 1.0$\pm$0.8 & 3.2$\pm$0.3 & 0.4$\pm$0.1 & 9548.8$\pm$1074.9 & 54.8$\pm$5.1 & 0.12 \\
GM Aur & 59913.18 & C & 0.29$\pm$0.03 & 1.0$\pm$0.7 & 3.1$\pm$0.3 & 0.4$\pm$0.1 & 9806.9$\pm$1067.1 & 55.8$\pm$5.0 & 0.13 \\
GM Aur & 59914.19 & C & 0.23$\pm$0.02 & 1.0$\pm$0.7 & 3.0$\pm$0.3 & 0.3$\pm$0.1 & 9964.5$\pm$1053.8 & 51.2$\pm$5.5 & 0.11 \\
GM Aur & 59915.13 & E & 0.31$\pm$0.03 & 1.0$\pm$0.7 & 3.0$\pm$0.2 & 0.3$\pm$0.1 & 9980.7$\pm$1050.0 & 51.4$\pm$5.6 & 0.11 \\
GM Aur & 59915.20 & C & 0.29$\pm$0.03 & 0.9$\pm$0.7 & 2.9$\pm$0.2 & 0.3$\pm$0.1 & 9859.6$\pm$1022.0 & 51.2$\pm$5.2 & 0.11 \\
GM Aur & 59916.16 & E & 0.26$\pm$0.03 & 1.0$\pm$0.7 & 3.1$\pm$0.3 & 0.3$\pm$0.1 & 9755.1$\pm$1049.1 & 54.3$\pm$5.5 & 0.11 \\
GM Aur & 59916.21 & C & 0.24$\pm$0.02 & 1.0$\pm$0.8 & 3.0$\pm$0.3 & 0.3$\pm$0.1 & 9581.3$\pm$1059.2 & 54.8$\pm$5.8 & 0.12 \\
GM Aur & 59917.19 & C & 0.24$\pm$0.02 & 0.8$\pm$0.6 & 2.9$\pm$0.2 & 0.2$\pm$0.1 & 9929.7$\pm$1009.8 & 49.8$\pm$4.9 & 0.13 \\
GM Aur & 59918.19 & C & 0.28$\pm$0.03 & 1.0$\pm$0.8 & 3.1$\pm$0.3 & 0.4$\pm$0.2 & 9539.3$\pm$1078.0 & 57.0$\pm$4.7 & 0.12 \\
GM Aur & 59919.18 & C & 0.25$\pm$0.03 & 1.0$\pm$0.8 & 3.0$\pm$0.3 & 0.3$\pm$0.2 & 9568.2$\pm$1072.8 & 56.7$\pm$5.1 & 0.12 \\
GM Aur & 59920.18 & C & 0.35$\pm$0.04 & 1.0$\pm$0.7 & 2.7$\pm$0.2 & 0.2$\pm$0.1 & 9844.8$\pm$1057.3 & 50.5$\pm$5.9 & 0.12 \\
GM Aur & 59922.18 & C & 0.44$\pm$0.04 & 0.9$\pm$0.9 & 2.6$\pm$0.2 & 0.2$\pm$0.1 & 9468.9$\pm$1131.0 & 49.9$\pm$5.4 & 0.11 \\
GM Aur & 59925.18 & C & 0.26$\pm$0.03 & 1.4$\pm$1.1 & 3.1$\pm$0.3 & 0.5$\pm$0.2 & 9591.9$\pm$1133.6 & 57.5$\pm$4.6 & 0.13 \\
GM Aur & 59926.17 & C & 0.28$\pm$0.03 & 1.2$\pm$1.0 & 3.0$\pm$0.2 & 0.2$\pm$0.1 & 9811.3$\pm$1060.7 & 51.7$\pm$5.9 & 0.13 \\
GM Aur & 59927.18 & C & 0.30$\pm$0.03 & 1.4$\pm$1.1 & 2.9$\pm$0.3 & 0.4$\pm$0.2 & 9627.2$\pm$1129.8 & 55.2$\pm$5.9 & 0.12 \\
GM Aur & 59928.15 & C & 0.27$\pm$0.03 & 0.9$\pm$0.6 & 2.9$\pm$0.2 & 0.2$\pm$0.1 & 9997.8$\pm$993.4 & 50.4$\pm$5.1 & 0.13 \\
GM Aur & 59929.16 & C & 0.39$\pm$0.04 & 1.2$\pm$0.9 & 3.2$\pm$0.2 & 0.3$\pm$0.1 & 9890.0$\pm$1054.8 & 51.0$\pm$5.3 & 0.14 \\
GM Aur & 59931.14 & C & 0.42$\pm$0.04 & 1.0$\pm$0.7 & 3.1$\pm$0.3 & 0.3$\pm$0.1 & 9906.9$\pm$1048.1 & 53.6$\pm$5.7 & 0.14 \\
\enddata
\tablecomments{Instrument names are abbreviated as: C: CHIRON, E: ESPRESSO: X: XSHOOTER, S: SOPHIE, U: UVES, T: TCES}
\end{deluxetable*}

%% file: Line_Luminosities.tex
\begin{longrotatetable}
\movetabledown=10mm
\begin{deluxetable*}{c c c | c c c c c c c c c c c} \label{tab: Line Fluxes TW Hya}
\tabletypesize{\scriptsize}
\tablecaption{Line luminosities for TW Hya}
\tablehead{
\colhead{Object} & \colhead{MJD} & \colhead{Inst.} & \colhead{H${\alpha}$} & \colhead{H${\beta}$} & \colhead{H${\gamma}$} & \colhead{H${\delta}$} & \colhead{He I${4387}$} & \colhead{He I${4471}$} & \colhead{He I${4713}$} & \colhead{He I${5015}$} & \colhead{He I${5875}$} & \colhead{He I${6678}$} & \colhead{He I${7065}$}
}
\startdata
TW Hya & 59241.30 & C & 283.0$\pm$1.0 & 27.5$\pm$3.0 & 23.4$\pm$2.7 & 13.7$\pm$1.6 & -1.7$\pm$0.5 & 2.9$\pm$0.5 & 1.4$\pm$0.4 & -0.5$\pm$0.3 & 2.0$\pm$0.3 & 0.2$\pm$0.2 & 0.3$\pm$0.2 \\
TW Hya & 59280.17 & C & 477.0$\pm$0.7 & 47.9$\pm$4.9 & 28.1$\pm$2.9 & 17.9$\pm$1.9 & -0.9$\pm$0.3 & 3.2$\pm$0.4 & 0.8$\pm$0.3 & 0.1$\pm$0.2 & 2.7$\pm$0.3 & 0.5$\pm$0.1 & 0.6$\pm$0.2 \\
TW Hya & 59280.29 & E & 489.0$\pm$1.2 & 51.2$\pm$5.3 & 30.1$\pm$3.6 & 19.4$\pm$2.4 & -0.5$\pm$0.6 & 2.5$\pm$0.6 & 0.4$\pm$0.3 & 0.2$\pm$0.3 & 3.1$\pm$0.4 & 0.5$\pm$0.2 & 1.0$\pm$0.2 \\
TW Hya & 59281.18 & C & 497.0$\pm$0.6 & 62.6$\pm$6.3 & 35.2$\pm$3.6 & 22.3$\pm$2.3 & -0.8$\pm$0.3 & 3.2$\pm$0.4 & 0.8$\pm$0.2 & 0.2$\pm$0.2 & 3.6$\pm$0.4 & 0.5$\pm$0.1 & 0.9$\pm$0.2 \\
TW Hya & 59284.24 & C & 438.0$\pm$0.7 & 54.5$\pm$5.5 & 31.6$\pm$3.3 & 21.4$\pm$2.2 & -0.7$\pm$0.3 & 3.2$\pm$0.4 & 1.0$\pm$0.3 & 0.1$\pm$0.2 & 3.2$\pm$0.3 & 0.8$\pm$0.1 & 0.9$\pm$0.2 \\
TW Hya & 59286.26 & C & 310.0$\pm$0.7 & 24.5$\pm$2.6 & 17.4$\pm$1.9 & 10.6$\pm$1.1 & -0.9$\pm$0.3 & 2.1$\pm$0.3 & 0.6$\pm$0.3 & -0.4$\pm$0.2 & 1.1$\pm$0.2 & 0.1$\pm$0.1 & 0.1$\pm$0.2 \\
TW Hya & 59288.25 & C & 592.0$\pm$0.8 & 67.1$\pm$6.8 & 36.4$\pm$3.8 & 23.1$\pm$2.4 & -0.9$\pm$0.3 & 3.8$\pm$0.5 & 0.9$\pm$0.3 & 0.2$\pm$0.2 & 3.6$\pm$0.4 & 0.7$\pm$0.2 & 1.1$\pm$0.2 \\
TW Hya & 59291.18 & C & 401.0$\pm$0.6 & 44.7$\pm$4.5 & 24.8$\pm$2.6 & 14.8$\pm$1.5 & -0.4$\pm$0.2 & 1.7$\pm$0.3 & 0.5$\pm$0.2 & 0.1$\pm$0.2 & 2.9$\pm$0.3 & 0.4$\pm$0.1 & 0.6$\pm$0.2 \\
TW Hya & 59294.17 & C & 411.0$\pm$0.8 & 36.0$\pm$3.7 & 21.2$\pm$2.4 & 13.3$\pm$1.5 & -0.4$\pm$0.3 & 1.7$\pm$0.4 & 0.5$\pm$0.3 & 0.2$\pm$0.2 & 2.6$\pm$0.3 & 0.3$\pm$0.1 & 0.2$\pm$0.2 \\
TW Hya & 59295.20 & C & 487.0$\pm$0.7 & 59.1$\pm$6.0 & 33.0$\pm$3.4 & 21.2$\pm$2.2 & -0.2$\pm$0.3 & 2.7$\pm$0.4 & 0.6$\pm$0.2 & 0.6$\pm$0.2 & 4.1$\pm$0.4 & 0.9$\pm$0.1 & 1.0$\pm$0.2 \\
TW Hya & 59296.18 & C & 649.0$\pm$0.7 & 95.6$\pm$9.6 & 50.7$\pm$5.1 & 32.8$\pm$3.3 & -0.0$\pm$0.3 & 4.5$\pm$0.5 & 1.1$\pm$0.3 & 1.2$\pm$0.2 & 8.2$\pm$0.8 & 2.0$\pm$0.2 & 2.4$\pm$0.3 \\
TW Hya & 59298.15 & C & 446.0$\pm$0.6 & 50.4$\pm$5.1 & 28.8$\pm$3.0 & 18.7$\pm$1.9 & -0.3$\pm$0.2 & 2.5$\pm$0.3 & 0.4$\pm$0.2 & 0.3$\pm$0.2 & 3.2$\pm$0.3 & 0.7$\pm$0.1 & 0.7$\pm$0.2 \\
TW Hya & 59300.11 & C & 324.0$\pm$0.7 & 29.1$\pm$3.0 & 18.7$\pm$2.0 & 12.2$\pm$1.3 & -0.4$\pm$0.2 & 1.8$\pm$0.3 & 0.4$\pm$0.2 & 0.0$\pm$0.2 & 2.2$\pm$0.2 & 0.3$\pm$0.1 & 0.4$\pm$0.2 \\
TW Hya & 59302.13 & C & 282.0$\pm$0.6 & 31.6$\pm$3.2 & 21.8$\pm$2.3 & 14.7$\pm$1.5 & -0.4$\pm$0.2 & 1.9$\pm$0.3 & 0.4$\pm$0.2 & 0.2$\pm$0.2 & 2.2$\pm$0.2 & 0.5$\pm$0.1 & 0.4$\pm$0.2 \\
TW Hya & 59303.12 & C & 395.0$\pm$0.7 & 45.5$\pm$4.6 & 23.8$\pm$2.5 & 14.0$\pm$1.5 & -0.5$\pm$0.2 & 1.8$\pm$0.3 & 0.3$\pm$0.2 & 0.1$\pm$0.2 & 2.7$\pm$0.3 & 0.4$\pm$0.1 & 0.5$\pm$0.2 \\
TW Hya & 59304.16 & C & 471.0$\pm$0.6 & 59.7$\pm$6.0 & 32.8$\pm$3.4 & 21.9$\pm$2.2 & -0.7$\pm$0.3 & 2.7$\pm$0.4 & 0.7$\pm$0.2 & 0.2$\pm$0.2 & 3.3$\pm$0.3 & 0.6$\pm$0.1 & 0.7$\pm$0.2 \\
TW Hya & 59305.12 & C & 482.0$\pm$0.7 & 56.2$\pm$5.7 & 28.6$\pm$3.0 & 18.4$\pm$1.9 & -0.7$\pm$0.3 & 2.9$\pm$0.4 & 0.6$\pm$0.2 & 0.2$\pm$0.2 & 3.8$\pm$0.4 & 0.7$\pm$0.1 & 1.0$\pm$0.2 \\
TW Hya & 59306.15 & C & 516.0$\pm$0.6 & 62.0$\pm$6.3 & 31.7$\pm$3.2 & 20.6$\pm$2.1 & -0.4$\pm$0.2 & 2.4$\pm$0.3 & 0.5$\pm$0.2 & 0.1$\pm$0.2 & 3.1$\pm$0.3 & 0.3$\pm$0.1 & 0.6$\pm$0.2 \\
TW Hya & 59307.00 & X & 440.0$\pm$2.8 & 60.3$\pm$6.2 & 31.9$\pm$3.3 & 21.0$\pm$2.1 & -0.1$\pm$0.3 & 3.5$\pm$0.5 & 0.8$\pm$0.4 & 0.8$\pm$0.4 & 7.1$\pm$0.8 & 1.8$\pm$0.5 & 2.1$\pm$0.6 \\
TW Hya & 59307.16 & C & 466.0$\pm$0.8 & 64.5$\pm$6.5 & 30.7$\pm$3.3 & 21.0$\pm$2.3 & -0.9$\pm$0.4 & 3.9$\pm$0.5 & 0.8$\pm$0.3 & 0.3$\pm$0.2 & 4.5$\pm$0.5 & 0.9$\pm$0.2 & 1.1$\pm$0.2 \\
TW Hya & 59308.04 & E & 489.0$\pm$1.8 & 54.5$\pm$5.8 & 31.7$\pm$4.2 & 20.3$\pm$3.0 & -0.4$\pm$0.8 & 2.6$\pm$0.8 & 0.1$\pm$0.5 & 0.3$\pm$0.5 & 3.7$\pm$0.5 & 0.5$\pm$0.3 & 1.0$\pm$0.3 \\
TW Hya & 59308.13 & C & 471.0$\pm$0.8 & 59.2$\pm$6.0 & 33.7$\pm$3.5 & 23.8$\pm$2.5 & -0.4$\pm$0.3 & 3.6$\pm$0.5 & 1.1$\pm$0.3 & 0.4$\pm$0.2 & 4.3$\pm$0.5 & 1.3$\pm$0.2 & 0.9$\pm$0.2 \\
TW Hya & 59309.14 & C & 608.0$\pm$0.7 & 84.7$\pm$8.5 & 43.9$\pm$4.5 & 31.0$\pm$3.2 & -0.6$\pm$0.3 & 4.4$\pm$0.5 & 0.8$\pm$0.3 & 1.3$\pm$0.2 & 8.3$\pm$0.8 & 2.2$\pm$0.3 & 2.5$\pm$0.3 \\
TW Hya & 59309.14 & E & 579.0$\pm$1.1 & 79.5$\pm$8.0 & 44.3$\pm$4.6 & 28.9$\pm$3.1 & 0.0$\pm$0.4 & 3.9$\pm$0.5 & 0.6$\pm$0.3 & 1.0$\pm$0.3 & 7.5$\pm$0.8 & 1.6$\pm$0.3 & 2.4$\pm$0.3 \\
TW Hya & 59310.09 & E & 561.0$\pm$1.0 & 71.8$\pm$7.3 & 37.9$\pm$4.0 & 23.4$\pm$2.6 & -0.0$\pm$0.4 & 3.1$\pm$0.5 & 0.7$\pm$0.3 & 0.6$\pm$0.3 & 4.7$\pm$0.5 & 0.9$\pm$0.2 & 1.6$\pm$0.2 \\
TW Hya & 59310.13 & C & 576.0$\pm$0.7 & 75.3$\pm$7.6 & 38.4$\pm$3.9 & 23.9$\pm$2.4 & -0.4$\pm$0.3 & 3.1$\pm$0.4 & 0.7$\pm$0.2 & 0.5$\pm$0.2 & 3.8$\pm$0.4 & 0.8$\pm$0.1 & 1.1$\pm$0.2 \\
TW Hya & 59310.15 & X & 554.0$\pm$1.7 & 68.5$\pm$7.0 & 32.5$\pm$3.3 & 20.1$\pm$2.0 & -0.2$\pm$0.3 & 2.9$\pm$0.4 & 0.6$\pm$0.3 & 0.5$\pm$0.3 & 4.0$\pm$0.5 & 0.6$\pm$0.3 & 1.1$\pm$0.5 \\
TW Hya & 59311.14 & C & 440.0$\pm$0.6 & 42.4$\pm$4.3 & 23.5$\pm$2.5 & 14.4$\pm$1.5 & -0.6$\pm$0.3 & 1.7$\pm$0.3 & 0.5$\pm$0.2 & -0.1$\pm$0.2 & 2.2$\pm$0.2 & 0.2$\pm$0.1 & 0.6$\pm$0.2 \\
TW Hya & 59312.16 & C & 345.0$\pm$0.6 & 31.1$\pm$3.2 & 17.6$\pm$1.9 & 11.7$\pm$1.2 & -0.5$\pm$0.2 & 1.7$\pm$0.3 & 0.4$\pm$0.2 & 0.0$\pm$0.2 & 1.6$\pm$0.2 & 0.1$\pm$0.1 & 0.4$\pm$0.2 \\
TW Hya & 59313.15 & C & 279.0$\pm$0.7 & 26.2$\pm$2.8 & 18.0$\pm$2.0 & 11.1$\pm$1.2 & -0.4$\pm$0.3 & 1.5$\pm$0.3 & 0.3$\pm$0.2 & -0.0$\pm$0.2 & 1.8$\pm$0.2 & 0.4$\pm$0.1 & 0.2$\pm$0.2 \\
TW Hya & 59313.22 & E & 287.0$\pm$1.1 & 27.3$\pm$2.9 & 19.2$\pm$2.6 & 12.5$\pm$1.8 & -0.6$\pm$0.5 & 1.7$\pm$0.5 & 0.3$\pm$0.3 & -0.2$\pm$0.3 & 1.2$\pm$0.2 & 0.2$\pm$0.2 & 0.4$\pm$0.2 \\
TW Hya & 59314.23 & C & 311.0$\pm$0.7 & 32.0$\pm$3.3 & 19.1$\pm$2.1 & 11.7$\pm$1.3 & -0.5$\pm$0.3 & 1.5$\pm$0.3 & 0.3$\pm$0.3 & -0.1$\pm$0.2 & 2.1$\pm$0.2 & 0.4$\pm$0.1 & 0.1$\pm$0.2 \\
TW Hya & 59653.15 & C & 450.0$\pm$0.6 & 30.7$\pm$3.1 & 17.5$\pm$1.9 & 10.8$\pm$1.1 & -0.3$\pm$0.2 & 1.5$\pm$0.3 & 0.3$\pm$0.2 & -0.1$\pm$0.2 & 2.2$\pm$0.2 & 0.2$\pm$0.1 & 0.2$\pm$0.2 \\
TW Hya & 59654.15 & C & 623.0$\pm$0.8 & 51.9$\pm$5.3 & 29.6$\pm$3.0 & 19.7$\pm$2.0 & -0.5$\pm$0.3 & 2.9$\pm$0.4 & 0.6$\pm$0.3 & -0.1$\pm$0.2 & 5.7$\pm$0.6 & 0.9$\pm$0.2 & 1.0$\pm$0.2 \\
TW Hya & 59656.17 & C & 484.0$\pm$0.7 & 37.6$\pm$3.8 & 21.0$\pm$2.2 & 14.9$\pm$1.6 & -0.5$\pm$0.3 & 2.5$\pm$0.3 & 0.4$\pm$0.2 & 0.2$\pm$0.2 & 3.7$\pm$0.4 & 0.6$\pm$0.1 & 0.6$\pm$0.2 \\
TW Hya & 59657.21 & C & 566.0$\pm$0.7 & 52.8$\pm$5.4 & 28.5$\pm$3.0 & 18.9$\pm$1.9 & -0.5$\pm$0.3 & 3.0$\pm$0.4 & 0.6$\pm$0.2 & 0.3$\pm$0.2 & 5.1$\pm$0.5 & 1.1$\pm$0.2 & 1.1$\pm$0.2 \\
TW Hya & 59658.19 & C & 351.0$\pm$0.7 & 32.2$\pm$3.3 & 23.1$\pm$2.4 & 17.3$\pm$1.8 & -0.6$\pm$0.2 & 2.5$\pm$0.3 & 0.6$\pm$0.2 & 0.1$\pm$0.2 & 3.5$\pm$0.4 & 1.0$\pm$0.2 & 0.5$\pm$0.2 \\
TW Hya & 59659.20 & C & 497.0$\pm$0.6 & 53.5$\pm$5.4 & 27.2$\pm$2.8 & 18.1$\pm$1.9 & -0.3$\pm$0.2 & 3.0$\pm$0.4 & 0.6$\pm$0.2 & 0.4$\pm$0.2 & 4.5$\pm$0.5 & 0.9$\pm$0.1 & 0.8$\pm$0.2 \\
TW Hya & 59660.15 & C & 459.0$\pm$0.6 & 49.5$\pm$5.0 & 28.2$\pm$2.9 & 19.2$\pm$2.0 & -0.3$\pm$0.2 & 2.9$\pm$0.4 & 0.6$\pm$0.2 & 0.3$\pm$0.2 & 4.3$\pm$0.4 & 1.1$\pm$0.1 & 0.9$\pm$0.2 \\
TW Hya & 59661.22 & C & 381.0$\pm$0.5 & 40.7$\pm$4.1 & 25.2$\pm$2.6 & 16.7$\pm$1.7 & -0.3$\pm$0.2 & 1.9$\pm$0.3 & 0.4$\pm$0.2 & 0.2$\pm$0.2 & 3.4$\pm$0.4 & 0.7$\pm$0.1 & 0.5$\pm$0.1 \\
TW Hya & 59662.16 & C & 332.0$\pm$0.6 & 35.6$\pm$3.6 & 24.3$\pm$2.5 & 17.0$\pm$1.7 & -0.4$\pm$0.2 & 2.4$\pm$0.3 & 0.5$\pm$0.2 & 0.4$\pm$0.2 & 3.3$\pm$0.3 & 0.7$\pm$0.1 & 0.4$\pm$0.2 \\
TW Hya & 59663.18 & C & 278.0$\pm$0.6 & 26.4$\pm$2.7 & 17.8$\pm$1.9 & 12.0$\pm$1.2 & -0.6$\pm$0.2 & 1.6$\pm$0.3 & 0.2$\pm$0.2 & 0.1$\pm$0.2 & 1.9$\pm$0.2 & 0.5$\pm$0.1 & 0.1$\pm$0.1 \\
TW Hya & 59664.14 & C & 258.0$\pm$0.5 & 23.1$\pm$2.4 & 14.8$\pm$1.6 & 9.9$\pm$1.0 & -0.6$\pm$0.2 & 1.4$\pm$0.2 & 0.3$\pm$0.2 & -0.2$\pm$0.2 & 1.7$\pm$0.2 & 0.3$\pm$0.1 & -0.0$\pm$0.1 \\
TW Hya & 59665.16 & C & 379.0$\pm$0.6 & 44.7$\pm$4.5 & 27.5$\pm$2.8 & 18.5$\pm$1.9 & -0.6$\pm$0.2 & 2.4$\pm$0.3 & 0.5$\pm$0.2 & 0.2$\pm$0.2 & 3.2$\pm$0.3 & 0.7$\pm$0.1 & 0.4$\pm$0.2 \\
TW Hya & 59666.19 & C & 448.0$\pm$0.6 & 70.9$\pm$7.1 & 38.8$\pm$3.9 & 25.4$\pm$2.6 & -0.5$\pm$0.2 & 3.1$\pm$0.4 & 0.6$\pm$0.2 & 0.3$\pm$0.2 & 5.4$\pm$0.6 & 1.1$\pm$0.2 & 1.2$\pm$0.2 \\
TW Hya & 59667.14 & C & 364.0$\pm$0.6 & 40.2$\pm$4.1 & 24.2$\pm$2.5 & 16.9$\pm$1.7 & -0.6$\pm$0.2 & 2.8$\pm$0.3 & 0.7$\pm$0.2 & 0.1$\pm$0.2 & 3.5$\pm$0.4 & 0.5$\pm$0.1 & 0.7$\pm$0.2 \\
TW Hya & 59668.14 & C & 392.0$\pm$0.6 & 45.3$\pm$4.6 & 27.2$\pm$2.8 & 18.6$\pm$1.9 & -0.7$\pm$0.2 & 2.1$\pm$0.3 & 0.5$\pm$0.2 & -0.0$\pm$0.2 & 3.3$\pm$0.3 & 0.6$\pm$0.1 & 0.4$\pm$0.1 \\
TW Hya & 59670.10 & C & 374.0$\pm$0.6 & 48.2$\pm$4.9 & 28.5$\pm$2.9 & 18.6$\pm$1.9 & -0.2$\pm$0.2 & 2.8$\pm$0.3 & 0.6$\pm$0.2 & 0.5$\pm$0.2 & 3.8$\pm$0.4 & 1.2$\pm$0.2 & 0.9$\pm$0.2 \\
TW Hya & 59671.06 & C & 360.0$\pm$0.6 & 36.2$\pm$3.7 & 21.8$\pm$2.3 & 14.6$\pm$1.5 & -0.3$\pm$0.2 & 2.1$\pm$0.3 & 0.5$\pm$0.2 & 0.0$\pm$0.2 & 2.9$\pm$0.3 & 0.6$\pm$0.1 & 0.5$\pm$0.2 \\
TW Hya & 59672.01 & C & 304.0$\pm$0.5 & 26.0$\pm$2.7 & 17.4$\pm$1.8 & 11.0$\pm$1.1 & -0.4$\pm$0.2 & 1.6$\pm$0.3 & 0.4$\pm$0.2 & -0.1$\pm$0.2 & 2.3$\pm$0.2 & 0.3$\pm$0.1 & 0.2$\pm$0.1 \\
TW Hya & 59672.26 & C & 315.0$\pm$0.6 & 29.9$\pm$3.1 & 18.6$\pm$1.9 & 11.3$\pm$1.2 & -0.2$\pm$0.2 & 1.7$\pm$0.3 & 0.3$\pm$0.2 & 0.0$\pm$0.2 & 2.6$\pm$0.3 & 0.4$\pm$0.1 & 0.2$\pm$0.1 \\
TW Hya & 59673.13 & C & 325.0$\pm$0.5 & 39.2$\pm$4.0 & 25.8$\pm$2.6 & 17.1$\pm$1.7 & -0.3$\pm$0.2 & 2.2$\pm$0.3 & 0.5$\pm$0.2 & 0.3$\pm$0.2 & 3.5$\pm$0.4 & 1.1$\pm$0.1 & 0.5$\pm$0.1 \\
TW Hya & 59674.14 & C & 431.0$\pm$0.6 & 53.1$\pm$5.4 & 33.3$\pm$3.4 & 23.1$\pm$2.3 & -0.0$\pm$0.2 & 3.1$\pm$0.4 & 0.7$\pm$0.2 & 0.6$\pm$0.2 & 5.4$\pm$0.5 & 1.5$\pm$0.2 & 1.2$\pm$0.2 \\
TW Hya & 59675.09 & C & 345.0$\pm$0.6 & 31.8$\pm$3.3 & 22.2$\pm$2.3 & 15.7$\pm$1.6 & -0.2$\pm$0.2 & 2.3$\pm$0.3 & 0.4$\pm$0.2 & 0.2$\pm$0.2 & 3.2$\pm$0.3 & 0.9$\pm$0.1 & 0.4$\pm$0.1 \\
TW Hya & 59676.01 & C & 479.0$\pm$0.6 & 55.9$\pm$5.6 & 32.6$\pm$3.3 & 22.7$\pm$2.3 & -0.2$\pm$0.2 & 3.0$\pm$0.4 & 0.6$\pm$0.2 & 0.4$\pm$0.2 & 4.3$\pm$0.4 & 0.9$\pm$0.1 & 0.7$\pm$0.2 \\
TW Hya & 59676.23 & C & 456.0$\pm$0.6 & 49.2$\pm$5.0 & 26.8$\pm$2.8 & 17.0$\pm$1.7 & -0.3$\pm$0.2 & 2.5$\pm$0.3 & 0.5$\pm$0.2 & 0.4$\pm$0.2 & 4.3$\pm$0.4 & 1.2$\pm$0.2 & 0.9$\pm$0.2 \\
TW Hya & 59677.07 & C & 485.0$\pm$0.6 & 57.8$\pm$5.8 & 30.2$\pm$3.1 & 18.9$\pm$1.9 & -0.3$\pm$0.2 & 2.4$\pm$0.3 & 0.4$\pm$0.2 & 0.2$\pm$0.2 & 3.3$\pm$0.3 & 0.8$\pm$0.1 & 0.5$\pm$0.2 \\
TW Hya & 59678.01 & C & 343.0$\pm$0.6 & 34.4$\pm$3.5 & 23.0$\pm$2.4 & 15.6$\pm$1.6 & -0.4$\pm$0.2 & 1.9$\pm$0.3 & 0.3$\pm$0.2 & 0.1$\pm$0.2 & 3.2$\pm$0.3 & 0.3$\pm$0.1 & 0.5$\pm$0.2 \\
TW Hya & 59678.18 & C & 354.0$\pm$0.6 & 33.2$\pm$3.4 & 23.2$\pm$2.4 & 15.5$\pm$1.6 & -0.4$\pm$0.2 & 2.1$\pm$0.3 & 0.4$\pm$0.2 & 0.0$\pm$0.2 & 2.9$\pm$0.3 & 0.2$\pm$0.1 & 0.2$\pm$0.2 \\
TW Hya & 59679.05 & C & 458.0$\pm$0.6 & 53.0$\pm$5.3 & 30.5$\pm$3.1 & 19.8$\pm$2.0 & -0.2$\pm$0.2 & 2.4$\pm$0.3 & 0.5$\pm$0.2 & 0.2$\pm$0.2 & 3.7$\pm$0.4 & 0.5$\pm$0.1 & 0.7$\pm$0.2 \\
TW Hya & 59680.07 & C & 482.0$\pm$0.6 & 49.5$\pm$5.0 & 26.9$\pm$2.8 & 16.8$\pm$1.7 & -0.4$\pm$0.2 & 1.9$\pm$0.3 & 0.3$\pm$0.2 & 0.1$\pm$0.2 & 3.3$\pm$0.3 & 0.3$\pm$0.1 & 0.5$\pm$0.2 \\
TW Hya & 59682.09 & C & 491.0$\pm$0.6 & 57.5$\pm$5.8 & 31.6$\pm$3.2 & 19.7$\pm$2.0 & -0.3$\pm$0.2 & 2.6$\pm$0.3 & 0.6$\pm$0.2 & 0.5$\pm$0.2 & 4.6$\pm$0.5 & 1.1$\pm$0.2 & 1.1$\pm$0.2 \\
TW Hya & 59684.15 & C & 461.0$\pm$0.6 & 53.2$\pm$5.4 & 32.1$\pm$3.3 & 21.8$\pm$2.2 & -0.1$\pm$0.2 & 3.0$\pm$0.4 & 0.7$\pm$0.2 & 0.6$\pm$0.2 & 5.9$\pm$0.6 & 1.7$\pm$0.2 & 1.6$\pm$0.2 \\
TW Hya & 59686.12 & C & 482.0$\pm$0.6 & 56.5$\pm$5.7 & 30.2$\pm$3.1 & 18.8$\pm$1.9 & -0.3$\pm$0.2 & 2.4$\pm$0.3 & 0.5$\pm$0.2 & 0.4$\pm$0.2 & 4.4$\pm$0.5 & 1.0$\pm$0.1 & 0.9$\pm$0.2 \\
TW Hya & 59688.09 & C & 471.0$\pm$0.6 & 56.1$\pm$5.6 & 33.4$\pm$3.4 & 23.1$\pm$2.3 & -0.2$\pm$0.2 & 2.8$\pm$0.3 & 0.6$\pm$0.2 & 0.4$\pm$0.2 & 4.9$\pm$0.5 & 1.2$\pm$0.2 & 1.0$\pm$0.2 \\
TW Hya & 59690.06 & C & 482.0$\pm$0.8 & 56.0$\pm$5.7 & 32.3$\pm$3.5 & 18.8$\pm$2.1 & -0.1$\pm$0.4 & 2.4$\pm$0.4 & 0.3$\pm$0.3 & 0.5$\pm$0.3 & 3.5$\pm$0.4 & 1.0$\pm$0.2 & 0.2$\pm$0.2 \\
TW Hya & 59692.07 & C & 426.0$\pm$0.6 & 49.1$\pm$5.0 & 32.0$\pm$3.3 & 21.4$\pm$2.2 & -0.3$\pm$0.2 & 3.0$\pm$0.4 & 0.6$\pm$0.2 & 0.6$\pm$0.2 & 4.4$\pm$0.5 & 1.4$\pm$0.2 & 1.0$\pm$0.2 \\
TW Hya & 59694.03 & C & 558.0$\pm$0.7 & 63.8$\pm$6.4 & 35.9$\pm$3.7 & 22.8$\pm$2.3 & -0.1$\pm$0.3 & 3.0$\pm$0.4 & 0.5$\pm$0.2 & 0.5$\pm$0.2 & 4.1$\pm$0.4 & 1.2$\pm$0.2 & 0.8$\pm$0.2 \\
TW Hya & 59695.05 & C & 495.0$\pm$0.6 & 42.5$\pm$4.3 & 24.3$\pm$2.5 & 15.5$\pm$1.6 & -0.2$\pm$0.2 & 2.2$\pm$0.3 & 0.5$\pm$0.2 & 0.2$\pm$0.2 & 3.0$\pm$0.3 & 0.5$\pm$0.1 & 0.5$\pm$0.2 \\
TW Hya & 59696.18 & C & 536.0$\pm$0.6 & 50.3$\pm$5.1 & 27.1$\pm$2.8 & 17.1$\pm$1.8 & -0.1$\pm$0.3 & 2.2$\pm$0.3 & 0.3$\pm$0.2 & 0.3$\pm$0.2 & 4.0$\pm$0.4 & 0.9$\pm$0.1 & 0.7$\pm$0.2 \\
TW Hya & 59696.24 & C & 519.0$\pm$0.7 & 51.7$\pm$5.2 & 26.2$\pm$2.8 & 14.8$\pm$1.6 & -0.3$\pm$0.3 & 2.5$\pm$0.4 & 0.2$\pm$0.2 & 0.2$\pm$0.2 & 3.8$\pm$0.4 & 0.8$\pm$0.1 & 0.5$\pm$0.2 \\
TW Hya & 59697.15 & C & 383.0$\pm$0.6 & 37.4$\pm$3.8 & 26.2$\pm$2.7 & 17.5$\pm$1.8 & -0.1$\pm$0.2 & 2.2$\pm$0.3 & 0.3$\pm$0.2 & 0.1$\pm$0.2 & 2.9$\pm$0.3 & 0.7$\pm$0.1 & 0.4$\pm$0.1 \\
TW Hya & 59698.14 & C & 462.0$\pm$0.6 & 50.5$\pm$5.1 & 31.8$\pm$3.2 & 21.2$\pm$2.2 & -0.1$\pm$0.2 & 2.9$\pm$0.4 & 0.6$\pm$0.2 & 0.6$\pm$0.2 & 3.8$\pm$0.4 & 1.1$\pm$0.1 & 0.7$\pm$0.2 \\
\enddata
\tablecomments{Fluxes are in units of 10$^{13}$ erg s$^{-1}$ cm$^{-2}$. Additional 10\% uncertainty due to continuum subtraction has been included in all uncertanties except \ha and \hb.Instrument names are abbreviated as: C: CHIRON, E: ESPRESSO, X: XSHOOTER, S: SOPHIE, U: UVES, T: TCES}
\end{deluxetable*}
\end{longrotatetable}

\begin{longrotatetable}
\movetabledown=10mm
\begin{deluxetable*}{c c c | c c c c c c c c c c c} \label{tab: Line Fluxes RU Lup}
\tabletypesize{\scriptsize}
\tablecaption{Line luminosities for RU Lup}
\tablehead{
\colhead{Object} & \colhead{MJD} & \colhead{Inst.} & \colhead{H${\alpha}$} & \colhead{H${\beta}$} & \colhead{H${\gamma}$} & \colhead{H${\delta}$} & \colhead{He I${4387}$} & \colhead{He I${4471}$} & \colhead{He I${4713}$} & \colhead{He I${5015}$} & \colhead{He I${5875}$} & \colhead{He I${6678}$} & \colhead{He I${7065}$}
}
\startdata
RU Lup & 59264.37 & C & 288.0$\pm$0.6 & 69.7$\pm$7.0 & 30.2$\pm$3.2 & 18.5$\pm$2.0 & 3.8$\pm$0.5 & 4.2$\pm$0.5 & 0.0$\pm$0.2 & 18.1$\pm$1.8 & 8.6$\pm$0.9 & 3.2$\pm$0.3 & 2.6$\pm$0.3 \\
RU Lup & 59318.26 & C & 248.0$\pm$0.5 & 67.8$\pm$6.8 & 30.7$\pm$3.3 & 19.1$\pm$2.0 & 4.4$\pm$0.6 & 5.3$\pm$0.6 & 0.3$\pm$0.2 & 18.2$\pm$1.8 & 8.9$\pm$0.9 & 3.1$\pm$0.3 & 2.7$\pm$0.3 \\
RU Lup & 59395.05 & C & 327.0$\pm$0.5 & 83.5$\pm$8.4 & 42.9$\pm$4.4 & 25.6$\pm$2.6 & 5.0$\pm$0.6 & 5.4$\pm$0.6 & 0.3$\pm$0.2 & 20.9$\pm$2.1 & 10.4$\pm$1.0 & 3.9$\pm$0.4 & 2.7$\pm$0.3 \\
RU Lup & 59434.99 & C & 260.0$\pm$1.0 & 70.2$\pm$7.2 & 35.3$\pm$4.2 & 13.6$\pm$2.1 & 3.4$\pm$0.8 & 4.1$\pm$0.7 & 0.1$\pm$0.5 & 15.7$\pm$1.6 & 6.7$\pm$0.7 & 2.4$\pm$0.3 & 2.2$\pm$0.3 \\
RU Lup & 59435.98 & C & 261.0$\pm$0.5 & 71.1$\pm$7.1 & 34.6$\pm$3.6 & 21.4$\pm$2.2 & 4.2$\pm$0.5 & 4.4$\pm$0.5 & 0.2$\pm$0.2 & 16.9$\pm$1.7 & 7.5$\pm$0.8 & 2.7$\pm$0.3 & 2.1$\pm$0.2 \\
RU Lup & 59436.10 & X & 256.0$\pm$1.8 & 70.1$\pm$7.1 & 31.6$\pm$3.6 & 17.0$\pm$1.9 & 4.5$\pm$0.8 & 4.5$\pm$0.7 & 0.1$\pm$0.3 & 18.2$\pm$1.8 & 8.5$\pm$0.9 & 3.0$\pm$0.4 & 2.4$\pm$0.3 \\
RU Lup & 59437.00 & C & 265.0$\pm$0.5 & 79.6$\pm$8.0 & 38.5$\pm$4.0 & 24.0$\pm$2.5 & 5.4$\pm$0.7 & 5.7$\pm$0.7 & 0.5$\pm$0.2 & 19.6$\pm$2.0 & 8.4$\pm$0.8 & 3.6$\pm$0.4 & 2.4$\pm$0.2 \\
RU Lup & 59437.97 & C & 298.0$\pm$0.5 & 91.1$\pm$9.1 & 46.2$\pm$4.8 & 28.3$\pm$2.9 & 6.5$\pm$0.8 & 6.7$\pm$0.8 & 0.2$\pm$0.2 & 22.9$\pm$2.3 & 10.5$\pm$1.1 & 4.0$\pm$0.4 & 2.9$\pm$0.3 \\
RU Lup & 59439.02 & C & 231.0$\pm$0.5 & 72.1$\pm$7.2 & 36.4$\pm$3.8 & 23.6$\pm$2.4 & 4.7$\pm$0.6 & 4.6$\pm$0.6 & 0.3$\pm$0.2 & 17.4$\pm$1.7 & 7.9$\pm$0.8 & 3.1$\pm$0.3 & 2.1$\pm$0.2 \\
RU Lup & 59439.99 & C & 271.0$\pm$0.5 & 78.9$\pm$7.9 & 41.6$\pm$4.3 & 25.4$\pm$2.6 & 4.9$\pm$0.6 & 5.8$\pm$0.7 & 0.3$\pm$0.2 & 18.0$\pm$1.8 & 9.3$\pm$0.9 & 3.5$\pm$0.4 & 2.5$\pm$0.3 \\
RU Lup & 59442.03 & C & 303.0$\pm$0.7 & 89.8$\pm$9.0 & 46.1$\pm$4.9 & 28.7$\pm$3.1 & 6.1$\pm$0.8 & 6.3$\pm$0.8 & 0.2$\pm$0.3 & 23.3$\pm$2.3 & 10.8$\pm$1.1 & 4.0$\pm$0.4 & 3.0$\pm$0.3 \\
RU Lup & 59443.02 & C & 263.0$\pm$0.5 & 85.3$\pm$8.6 & 40.3$\pm$4.2 & 26.2$\pm$2.7 & 6.4$\pm$0.8 & 5.7$\pm$0.7 & 0.5$\pm$0.2 & 22.0$\pm$2.2 & 8.6$\pm$0.9 & 3.8$\pm$0.4 & 2.3$\pm$0.2 \\
RU Lup & 59443.14 & C & 233.0$\pm$0.5 & 92.2$\pm$9.3 & 40.6$\pm$4.3 & 22.5$\pm$2.4 & 7.1$\pm$0.8 & 6.5$\pm$0.8 & 0.7$\pm$0.3 & 24.0$\pm$2.4 & 8.4$\pm$0.8 & 3.3$\pm$0.3 & 1.8$\pm$0.2 \\
RU Lup & 59444.02 & C & 294.0$\pm$0.5 & 91.4$\pm$9.2 & 44.9$\pm$4.7 & 27.2$\pm$2.8 & 7.3$\pm$0.8 & 6.4$\pm$0.7 & 0.3$\pm$0.2 & 24.3$\pm$2.4 & 8.8$\pm$0.9 & 3.8$\pm$0.4 & 2.4$\pm$0.3 \\
RU Lup & 59444.03 & C & 291.0$\pm$0.8 & 94.4$\pm$9.5 & 47.9$\pm$5.1 & 20.7$\pm$2.5 & 6.8$\pm$0.9 & 6.5$\pm$0.8 & 0.6$\pm$0.4 & 23.1$\pm$2.3 & 8.3$\pm$0.8 & 4.0$\pm$0.4 & 2.5$\pm$0.3 \\
RU Lup & 59447.02 & C & 289.0$\pm$0.9 & 113.0$\pm$11.4 & 51.1$\pm$5.5 & 19.0$\pm$2.5 & 6.9$\pm$1.0 & 7.9$\pm$1.0 & -0.3$\pm$0.4 & 26.1$\pm$2.6 & 13.3$\pm$1.3 & 5.3$\pm$0.5 & 3.3$\pm$0.4 \\
RU Lup & 59447.05 & C & 359.0$\pm$0.6 & 113.0$\pm$11.3 & 48.5$\pm$5.0 & 27.5$\pm$2.9 & 7.8$\pm$0.9 & 7.5$\pm$0.8 & 0.1$\pm$0.3 & 28.2$\pm$2.8 & 13.7$\pm$1.4 & 5.0$\pm$0.5 & 3.5$\pm$0.4 \\
RU Lup & 59448.00 & C & 359.0$\pm$0.6 & 94.1$\pm$9.4 & 43.5$\pm$4.5 & 26.4$\pm$2.8 & 6.2$\pm$0.8 & 6.1$\pm$0.7 & 0.1$\pm$0.2 & 22.0$\pm$2.2 & 12.9$\pm$1.3 & 4.5$\pm$0.5 & 3.6$\pm$0.4 \\
RU Lup & 59448.07 & X & 325.0$\pm$1.1 & 86.2$\pm$8.6 & 35.7$\pm$4.0 & 19.1$\pm$2.1 & 5.4$\pm$0.9 & 5.8$\pm$0.8 & 0.2$\pm$0.3 & 21.4$\pm$2.2 & 11.4$\pm$1.2 & 4.1$\pm$0.4 & 3.2$\pm$0.4 \\
RU Lup & 59449.00 & E & 319.0$\pm$0.8 & 91.3$\pm$9.2 & 47.6$\pm$5.0 & 23.4$\pm$2.6 & 7.5$\pm$0.9 & 7.8$\pm$0.9 & 0.2$\pm$0.3 & 26.7$\pm$2.7 & 9.6$\pm$1.0 & 3.5$\pm$0.4 & 3.2$\pm$0.3 \\
RU Lup & 59449.04 & C & 342.0$\pm$0.5 & 89.8$\pm$9.0 & 40.8$\pm$4.2 & 23.0$\pm$2.4 & 6.1$\pm$0.7 & 5.8$\pm$0.7 & 0.0$\pm$0.2 & 25.3$\pm$2.5 & 11.1$\pm$1.1 & 4.2$\pm$0.4 & 3.1$\pm$0.3 \\
RU Lup & 59449.98 & C & 336.0$\pm$0.6 & 96.8$\pm$9.7 & 46.8$\pm$4.9 & 28.2$\pm$3.0 & 6.9$\pm$0.8 & 6.2$\pm$0.7 & 0.1$\pm$0.3 & 24.7$\pm$2.5 & 10.4$\pm$1.0 & 3.4$\pm$0.4 & 2.9$\pm$0.3 \\
RU Lup & 59453.01 & C & 270.0$\pm$0.6 & 100.0$\pm$10.0 & 55.2$\pm$5.8 & 33.8$\pm$3.5 & 8.2$\pm$1.0 & 7.7$\pm$0.9 & 0.1$\pm$0.3 & 28.3$\pm$2.8 & 11.8$\pm$1.2 & 4.4$\pm$0.5 & 2.7$\pm$0.3 \\
RU Lup & 59458.02 & E & 253.0$\pm$1.3 & 72.9$\pm$7.5 & 36.1$\pm$4.9 & 19.2$\pm$3.2 & 5.9$\pm$1.1 & 5.9$\pm$1.0 & 0.1$\pm$0.5 & 20.3$\pm$2.1 & 10.2$\pm$1.0 & 4.0$\pm$0.5 & 3.3$\pm$0.4 \\
RU Lup & 59458.06 & E & 233.0$\pm$1.5 & 68.4$\pm$7.2 & 33.5$\pm$6.1 & 17.8$\pm$4.6 & 5.4$\pm$1.5 & 5.3$\pm$1.3 & 0.1$\pm$0.6 & 18.9$\pm$2.0 & 9.1$\pm$1.0 & 3.5$\pm$0.5 & 2.9$\pm$0.4 \\
RU Lup & 59676.20 & C & 333.0$\pm$0.6 & 80.4$\pm$8.1 & 37.8$\pm$3.9 & 21.7$\pm$2.3 & 5.7$\pm$0.7 & 5.8$\pm$0.7 & 0.2$\pm$0.2 & 22.6$\pm$2.3 & 9.4$\pm$0.9 & 4.0$\pm$0.4 & 2.7$\pm$0.3 \\
RU Lup & 59723.14 & C & 204.0$\pm$0.5 & 54.9$\pm$5.5 & 27.0$\pm$2.9 & 16.7$\pm$1.8 & 3.7$\pm$0.5 & 3.8$\pm$0.5 & 0.2$\pm$0.2 & 14.7$\pm$1.5 & 7.1$\pm$0.7 & 2.7$\pm$0.3 & 2.1$\pm$0.2 \\
RU Lup & 59736.11 & C & 328.0$\pm$0.5 & 80.3$\pm$8.1 & 38.5$\pm$4.0 & 22.5$\pm$2.3 & 4.8$\pm$0.6 & 4.9$\pm$0.6 & -0.1$\pm$0.2 & 20.4$\pm$2.0 & 10.6$\pm$1.1 & 3.8$\pm$0.4 & 3.2$\pm$0.3 \\
RU Lup & 59790.11 & C & 301.0$\pm$0.5 & 85.8$\pm$8.6 & 44.3$\pm$4.6 & 27.2$\pm$2.8 & 4.9$\pm$0.6 & 5.5$\pm$0.6 & -0.0$\pm$0.2 & 20.5$\pm$2.1 & 11.4$\pm$1.1 & 3.9$\pm$0.4 & 3.5$\pm$0.4 \\
RU Lup & 59799.03 & C & 257.0$\pm$0.5 & 60.8$\pm$6.1 & 31.9$\pm$3.3 & 18.0$\pm$1.9 & 2.7$\pm$0.4 & 4.1$\pm$0.5 & 0.1$\pm$0.2 & 13.2$\pm$1.3 & 6.4$\pm$0.6 & 2.3$\pm$0.2 & 2.0$\pm$0.2 \\
RU Lup & 59800.03 & C & 288.0$\pm$0.5 & 76.6$\pm$7.7 & 39.7$\pm$4.1 & 23.6$\pm$2.4 & 4.6$\pm$0.6 & 5.1$\pm$0.6 & 0.1$\pm$0.2 & 17.6$\pm$1.8 & 8.6$\pm$0.9 & 3.0$\pm$0.3 & 2.5$\pm$0.3 \\
RU Lup & 59801.00 & C & 330.0$\pm$0.5 & 92.7$\pm$9.3 & 48.5$\pm$5.0 & 28.6$\pm$2.9 & 5.7$\pm$0.7 & 6.2$\pm$0.7 & -0.1$\pm$0.2 & 22.5$\pm$2.3 & 11.1$\pm$1.1 & 3.4$\pm$0.3 & 3.3$\pm$0.3 \\
RU Lup & 59801.05 & E & 305.0$\pm$1.5 & 82.0$\pm$8.5 & 41.0$\pm$5.0 & 21.1$\pm$3.0 & 4.7$\pm$1.0 & 6.0$\pm$1.0 & 0.0$\pm$0.6 & 21.1$\pm$2.2 & 9.2$\pm$1.0 & 3.1$\pm$0.4 & 3.1$\pm$0.4 \\
RU Lup & 59801.98 & C & 304.0$\pm$0.5 & 73.9$\pm$7.4 & 38.3$\pm$4.0 & 21.1$\pm$2.2 & 3.7$\pm$0.5 & 4.5$\pm$0.5 & -0.1$\pm$0.2 & 15.8$\pm$1.6 & 8.8$\pm$0.9 & 2.5$\pm$0.3 & 2.7$\pm$0.3 \\
RU Lup & 59802.07 & E & 287.0$\pm$1.3 & 72.0$\pm$7.5 & 36.3$\pm$4.4 & 19.5$\pm$2.7 & 3.0$\pm$0.8 & 4.9$\pm$0.9 & 0.1$\pm$0.6 & 15.6$\pm$1.6 & 7.6$\pm$0.8 & 2.2$\pm$0.3 & 2.5$\pm$0.3 \\
RU Lup & 59802.97 & C & 250.0$\pm$0.6 & 64.0$\pm$6.5 & 31.6$\pm$3.4 & 18.1$\pm$2.0 & 4.0$\pm$0.6 & 4.0$\pm$0.6 & 0.1$\pm$0.3 & 14.5$\pm$1.5 & 6.7$\pm$0.7 & 2.2$\pm$0.2 & 1.8$\pm$0.2 \\
RU Lup & 59804.00 & C & 244.0$\pm$0.4 & 59.6$\pm$6.0 & 30.4$\pm$3.1 & 17.4$\pm$1.8 & 2.7$\pm$0.4 & 3.1$\pm$0.4 & 0.1$\pm$0.2 & 11.7$\pm$1.2 & 5.3$\pm$0.5 & 1.5$\pm$0.2 & 1.6$\pm$0.2 \\
RU Lup & 59804.10 & E & 236.0$\pm$0.7 & 61.0$\pm$6.2 & 31.4$\pm$3.5 & 17.4$\pm$2.1 & 2.5$\pm$0.5 & 4.2$\pm$0.6 & 0.1$\pm$0.3 & 13.0$\pm$1.3 & 5.8$\pm$0.6 & 1.9$\pm$0.2 & 1.8$\pm$0.2 \\
RU Lup & 59805.06 & C & 250.0$\pm$0.4 & 72.6$\pm$7.3 & 39.0$\pm$4.0 & 23.6$\pm$2.4 & 3.5$\pm$0.4 & 4.6$\pm$0.5 & 0.1$\pm$0.2 & 15.3$\pm$1.5 & 8.4$\pm$0.8 & 2.5$\pm$0.3 & 2.5$\pm$0.3 \\
RU Lup & 59806.00 & C & 252.0$\pm$0.5 & 53.3$\pm$5.4 & 26.1$\pm$2.8 & 15.1$\pm$1.6 & 1.7$\pm$0.4 & 2.5$\pm$0.4 & -0.2$\pm$0.2 & 10.0$\pm$1.0 & 5.5$\pm$0.6 & 1.6$\pm$0.2 & 1.7$\pm$0.2 \\
RU Lup & 59807.00 & C & 189.0$\pm$0.4 & 42.1$\pm$4.2 & 21.9$\pm$2.3 & 12.9$\pm$1.3 & 1.0$\pm$0.2 & 2.3$\pm$0.3 & 0.1$\pm$0.1 & 7.3$\pm$0.7 & 4.4$\pm$0.4 & 1.2$\pm$0.1 & 1.3$\pm$0.1 \\
RU Lup & 59807.10 & E & 184.0$\pm$1.3 & 40.3$\pm$4.5 & 20.4$\pm$3.6 & 11.8$\pm$2.6 & 1.1$\pm$0.9 & 2.3$\pm$0.8 & 0.0$\pm$0.6 & 7.7$\pm$0.9 & 4.5$\pm$0.5 & 1.3$\pm$0.3 & 1.6$\pm$0.3 \\
RU Lup & 59808.05 & C & 168.0$\pm$0.3 & 42.3$\pm$4.2 & 21.8$\pm$2.2 & 12.8$\pm$1.3 & 0.8$\pm$0.2 & 2.1$\pm$0.3 & 0.1$\pm$0.1 & 6.4$\pm$0.6 & 3.9$\pm$0.4 & 1.1$\pm$0.1 & 1.1$\pm$0.1 \\
RU Lup & 59809.07 & C & 166.0$\pm$0.3 & 46.3$\pm$4.7 & 24.5$\pm$2.5 & 13.6$\pm$1.4 & 0.9$\pm$0.2 & 2.1$\pm$0.3 & 0.0$\pm$0.1 & 7.4$\pm$0.7 & 4.4$\pm$0.4 & 0.9$\pm$0.1 & 1.2$\pm$0.1 \\
RU Lup & 59810.03 & C & 147.0$\pm$0.4 & 37.0$\pm$3.7 & 19.6$\pm$2.0 & 11.1$\pm$1.2 & 0.5$\pm$0.2 & 1.4$\pm$0.2 & 0.1$\pm$0.1 & 5.9$\pm$0.6 & 2.8$\pm$0.3 & 0.7$\pm$0.1 & 0.8$\pm$0.1 \\
RU Lup & 59811.05 & C & 224.0$\pm$0.4 & 72.4$\pm$7.3 & 37.4$\pm$3.9 & 21.0$\pm$2.2 & 5.4$\pm$0.6 & 5.5$\pm$0.6 & 0.1$\pm$0.2 & 17.4$\pm$1.7 & 7.5$\pm$0.8 & 2.9$\pm$0.3 & 2.2$\pm$0.2 \\
RU Lup & 59812.02 & C & 222.0$\pm$0.4 & 69.2$\pm$6.9 & 32.0$\pm$3.3 & 19.3$\pm$2.0 & 4.9$\pm$0.6 & 4.7$\pm$0.5 & 0.2$\pm$0.2 & 16.7$\pm$1.7 & 7.5$\pm$0.8 & 3.0$\pm$0.3 & 2.3$\pm$0.2 \\
RU Lup & 59813.05 & C & 287.0$\pm$0.4 & 71.8$\pm$7.2 & 34.9$\pm$3.6 & 19.6$\pm$2.0 & 2.0$\pm$0.3 & 3.5$\pm$0.4 & 0.0$\pm$0.2 & 13.1$\pm$1.3 & 5.7$\pm$0.6 & 1.5$\pm$0.2 & 1.7$\pm$0.2 \\
RU Lup & 59814.00 & C & 194.0$\pm$0.4 & 51.6$\pm$5.2 & 26.4$\pm$2.8 & 15.5$\pm$1.6 & 1.1$\pm$0.3 & 2.4$\pm$0.3 & 0.0$\pm$0.2 & 7.9$\pm$0.8 & 4.5$\pm$0.5 & 1.1$\pm$0.1 & 1.2$\pm$0.1 \\
RU Lup & 59814.04 & E & 199.0$\pm$1.0 & 49.6$\pm$5.1 & 25.6$\pm$3.0 & 14.4$\pm$1.9 & 1.2$\pm$0.5 & 2.6$\pm$0.5 & 0.0$\pm$0.4 & 8.4$\pm$0.9 & 4.7$\pm$0.5 & 1.3$\pm$0.2 & 1.5$\pm$0.2 \\
RU Lup & 59815.05 & C & 268.0$\pm$0.4 & 84.3$\pm$8.4 & 42.3$\pm$4.3 & 24.8$\pm$2.5 & 4.8$\pm$0.6 & 5.4$\pm$0.6 & -0.1$\pm$0.2 & 19.9$\pm$2.0 & 9.2$\pm$0.9 & 2.9$\pm$0.3 & 2.4$\pm$0.3 \\
RU Lup & 59816.02 & C & 282.0$\pm$0.4 & 93.4$\pm$9.4 & 48.6$\pm$5.0 & 29.0$\pm$3.0 & 7.4$\pm$0.8 & 7.3$\pm$0.8 & 0.2$\pm$0.2 & 23.2$\pm$2.3 & 10.5$\pm$1.1 & 3.4$\pm$0.3 & 2.9$\pm$0.3 \\
RU Lup & 59817.04 & C & 300.0$\pm$0.4 & 93.1$\pm$9.3 & 48.6$\pm$4.9 & 28.8$\pm$2.9 & 5.1$\pm$0.6 & 5.9$\pm$0.7 & 0.0$\pm$0.2 & 20.8$\pm$2.1 & 10.0$\pm$1.0 & 2.9$\pm$0.3 & 2.7$\pm$0.3 \\
\enddata
\tablecomments{Fluxes are in units of 10$^{13}$ erg s$^{-1}$ cm$^{-2}$. Additional 10\% uncertainty due to continuum subtraction has been included in all uncertanties except \ha and \hb.Instrument names are abbreviated as: C: CHIRON, E: ESPRESSO, X: XSHOOTER, S: SOPHIE, U: UVES, T: TCES}
\end{deluxetable*}
\end{longrotatetable}

\begin{longrotatetable}
\movetabledown=10mm
\begin{deluxetable*}{c c c | c c c c c c c c c c c} \label{tab: Line Fluxes BP Tau}
\tabletypesize{\scriptsize}
\tablecaption{Line luminosities for BP Tau}
\tablehead{
\colhead{Object} & \colhead{MJD} & \colhead{Inst.} & \colhead{H${\alpha}$} & \colhead{H${\beta}$} & \colhead{H${\gamma}$} & \colhead{H${\delta}$} & \colhead{He I${4387}$} & \colhead{He I${4471}$} & \colhead{He I${4713}$} & \colhead{He I${5015}$} & \colhead{He I${5875}$} & \colhead{He I${6678}$} & \colhead{He I${7065}$}
}
\startdata
BP Tau & 59448.39 & X & 99.3$\pm$0.8 & 14.4$\pm$1.5 & 8.0$\pm$0.8 & 4.9$\pm$0.5 & -0.2$\pm$0.1 & 0.7$\pm$0.1 & 0.2$\pm$0.1 & 0.2$\pm$0.1 & 0.8$\pm$0.2 & 0.2$\pm$0.1 & 0.2$\pm$0.2 \\
BP Tau & 59452.35 & X & 71.3$\pm$0.6 & 9.7$\pm$1.0 & 5.5$\pm$0.6 & 3.6$\pm$0.4 & -0.1$\pm$0.1 & 0.6$\pm$0.1 & 0.2$\pm$0.1 & 0.2$\pm$0.1 & 0.9$\pm$0.1 & 0.2$\pm$0.1 & 0.2$\pm$0.1 \\
BP Tau & 59459.38 & E & 109.0$\pm$0.5 & 16.3$\pm$1.7 & 9.0$\pm$1.3 & 5.8$\pm$1.0 & -0.0$\pm$0.3 & 0.8$\pm$0.3 & 0.2$\pm$0.2 & 0.5$\pm$0.2 & 1.4$\pm$0.2 & 0.5$\pm$0.1 & 0.5$\pm$0.1 \\
BP Tau & 59460.37 & X & 82.4$\pm$0.5 & 12.7$\pm$1.3 & 7.0$\pm$0.7 & 4.5$\pm$0.5 & -0.1$\pm$0.1 & 0.7$\pm$0.1 & 0.1$\pm$0.1 & 0.3$\pm$0.1 & 0.9$\pm$0.1 & 0.3$\pm$0.1 & 0.2$\pm$0.1 \\
BP Tau & 59464.36 & E & 72.6$\pm$0.9 & 9.5$\pm$1.6 & 5.9$\pm$2.7 & 3.8$\pm$2.0 & 0.0$\pm$0.8 & 0.6$\pm$0.7 & 0.2$\pm$0.4 & 0.0$\pm$0.3 & 0.7$\pm$0.2 & 0.2$\pm$0.2 & 0.2$\pm$0.2 \\
BP Tau & 59467.38 & C & 58.1$\pm$0.3 & 15.0$\pm$1.7 & 8.5$\pm$1.4 & 2.7$\pm$0.7 & 0.9$\pm$0.3 & 0.4$\pm$0.3 & 0.0$\pm$0.2 & 0.3$\pm$0.2 & 0.6$\pm$0.1 & 0.2$\pm$0.1 & 0.1$\pm$0.1 \\
BP Tau & 59470.38 & C & 74.1$\pm$0.3 & 13.1$\pm$1.5 & 4.8$\pm$1.0 & 2.1$\pm$0.7 & -0.2$\pm$0.3 & 0.5$\pm$0.2 & -0.0$\pm$0.2 & 0.2$\pm$0.1 & 0.6$\pm$0.1 & 0.2$\pm$0.1 & -0.1$\pm$0.1 \\
BP Tau & 59845.35 & C & 83.2$\pm$0.3 & 12.9$\pm$1.3 & 5.5$\pm$0.6 & 1.9$\pm$0.3 & -0.1$\pm$0.1 & 0.5$\pm$0.1 & 0.2$\pm$0.1 & 0.2$\pm$0.1 & 1.0$\pm$0.1 & 0.3$\pm$0.1 & 0.1$\pm$0.1 \\
BP Tau & 59863.31 & C & 65.1$\pm$0.2 & 7.3$\pm$0.8 & 3.1$\pm$0.4 & 1.0$\pm$0.2 & -0.2$\pm$0.1 & 0.2$\pm$0.1 & 0.1$\pm$0.1 & 0.2$\pm$0.1 & 0.5$\pm$0.1 & 0.1$\pm$0.0 & 0.0$\pm$0.1 \\
BP Tau & 59868.29 & C & 62.1$\pm$0.2 & 8.1$\pm$0.9 & 3.1$\pm$0.5 & 0.8$\pm$0.2 & -0.2$\pm$0.1 & 0.2$\pm$0.1 & 0.1$\pm$0.1 & 0.1$\pm$0.1 & 0.6$\pm$0.1 & 0.1$\pm$0.0 & 0.0$\pm$0.1 \\
BP Tau & 59882.25 & C & 53.7$\pm$0.2 & 8.0$\pm$0.9 & 3.0$\pm$0.5 & 1.0$\pm$0.3 & -0.2$\pm$0.1 & 0.2$\pm$0.1 & 0.0$\pm$0.1 & 0.1$\pm$0.1 & 0.6$\pm$0.1 & 0.1$\pm$0.0 & 0.0$\pm$0.0 \\
BP Tau & 59893.21 & C & 28.6$\pm$0.2 & 3.2$\pm$0.5 & 1.4$\pm$0.5 & 0.3$\pm$0.3 & -0.6$\pm$0.2 & 0.1$\pm$0.1 & -0.1$\pm$0.1 & -0.2$\pm$0.1 & 0.5$\pm$0.1 & 0.1$\pm$0.0 & 0.0$\pm$0.0 \\
BP Tau & 59898.19 & C & 62.9$\pm$0.3 & 9.6$\pm$1.0 & 4.0$\pm$0.5 & 1.2$\pm$0.3 & -0.0$\pm$0.1 & 0.4$\pm$0.1 & 0.1$\pm$0.1 & 0.2$\pm$0.1 & 0.8$\pm$0.1 & 0.2$\pm$0.0 & 0.0$\pm$0.1 \\
BP Tau & 59907.19 & C & 66.6$\pm$0.2 & 11.8$\pm$1.2 & 3.6$\pm$0.5 & 1.2$\pm$0.3 & -0.3$\pm$0.1 & 0.7$\pm$0.1 & 0.2$\pm$0.1 & 0.3$\pm$0.1 & 0.9$\pm$0.1 & 0.2$\pm$0.0 & 0.2$\pm$0.1 \\
BP Tau & 59909.17 & C & 54.5$\pm$0.2 & 9.2$\pm$1.0 & 3.1$\pm$0.5 & 1.3$\pm$0.3 & -0.3$\pm$0.1 & 0.7$\pm$0.1 & 0.2$\pm$0.1 & 0.1$\pm$0.1 & 0.8$\pm$0.1 & 0.2$\pm$0.0 & 0.1$\pm$0.1 \\
BP Tau & 59910.17 & C & 67.5$\pm$0.3 & 12.4$\pm$1.3 & 4.9$\pm$0.6 & 2.1$\pm$0.3 & -0.3$\pm$0.1 & 0.8$\pm$0.1 & 0.3$\pm$0.1 & 0.4$\pm$0.1 & 1.3$\pm$0.1 & 0.4$\pm$0.1 & 0.2$\pm$0.1 \\
BP Tau & 59911.18 & C & 81.0$\pm$0.2 & 13.9$\pm$1.4 & 5.7$\pm$0.6 & 2.3$\pm$0.3 & -0.4$\pm$0.1 & 0.9$\pm$0.1 & 0.3$\pm$0.1 & 0.4$\pm$0.1 & 1.2$\pm$0.1 & 0.3$\pm$0.0 & 0.2$\pm$0.1 \\
BP Tau & 59912.15 & C & 65.1$\pm$0.2 & 10.4$\pm$1.1 & 4.4$\pm$0.5 & 1.6$\pm$0.2 & -0.4$\pm$0.1 & 0.8$\pm$0.1 & 0.2$\pm$0.1 & 0.2$\pm$0.1 & 0.8$\pm$0.1 & 0.1$\pm$0.0 & 0.1$\pm$0.1 \\
BP Tau & 59913.17 & C & 81.1$\pm$0.3 & 13.3$\pm$1.4 & 4.7$\pm$0.6 & 1.9$\pm$0.3 & -0.4$\pm$0.1 & 0.9$\pm$0.1 & 0.3$\pm$0.1 & 0.5$\pm$0.1 & 1.2$\pm$0.1 & 0.3$\pm$0.1 & 0.2$\pm$0.1 \\
BP Tau & 59914.18 & C & 72.7$\pm$0.3 & 11.5$\pm$1.2 & 3.9$\pm$0.6 & 1.7$\pm$0.3 & -0.5$\pm$0.1 & 1.0$\pm$0.1 & 0.4$\pm$0.1 & 0.2$\pm$0.1 & 1.0$\pm$0.1 & 0.1$\pm$0.0 & 0.1$\pm$0.1 \\
BP Tau & 59915.18 & C & 65.9$\pm$0.3 & 10.5$\pm$1.1 & 4.4$\pm$0.6 & 1.2$\pm$0.3 & -0.1$\pm$0.1 & 0.2$\pm$0.1 & 0.0$\pm$0.1 & 0.4$\pm$0.1 & 0.8$\pm$0.1 & 0.2$\pm$0.1 & 0.1$\pm$0.1 \\
BP Tau & 59916.19 & C & 55.2$\pm$0.3 & 8.7$\pm$0.9 & 3.2$\pm$0.5 & 1.0$\pm$0.3 & -0.2$\pm$0.1 & 0.1$\pm$0.1 & -0.0$\pm$0.1 & 0.2$\pm$0.1 & 0.6$\pm$0.1 & 0.1$\pm$0.0 & 0.0$\pm$0.1 \\
BP Tau & 59917.16 & C & 65.0$\pm$0.3 & 11.7$\pm$1.2 & 4.6$\pm$0.6 & 1.5$\pm$0.3 & -0.1$\pm$0.1 & 0.3$\pm$0.1 & 0.2$\pm$0.1 & 0.2$\pm$0.1 & 1.1$\pm$0.1 & 0.2$\pm$0.0 & 0.1$\pm$0.1 \\
BP Tau & 59918.16 & C & 67.4$\pm$0.3 & 11.5$\pm$1.2 & 4.8$\pm$0.6 & 1.8$\pm$0.3 & -0.1$\pm$0.1 & 0.5$\pm$0.1 & 0.2$\pm$0.1 & 0.4$\pm$0.1 & 1.0$\pm$0.1 & 0.3$\pm$0.1 & 0.1$\pm$0.1 \\
BP Tau & 59919.15 & C & 77.8$\pm$0.2 & 14.3$\pm$1.5 & 5.2$\pm$0.6 & 1.8$\pm$0.3 & -0.1$\pm$0.1 & 0.6$\pm$0.1 & 0.1$\pm$0.1 & 0.7$\pm$0.1 & 1.6$\pm$0.2 & 0.4$\pm$0.1 & 0.3$\pm$0.1 \\
BP Tau & 59922.14 & C & 49.0$\pm$0.3 & 7.8$\pm$0.9 & 2.6$\pm$0.8 & 1.8$\pm$0.5 & -0.9$\pm$0.2 & 1.1$\pm$0.2 & 0.5$\pm$0.1 & -0.1$\pm$0.1 & 0.8$\pm$0.1 & 0.0$\pm$0.0 & 0.2$\pm$0.1 \\
BP Tau & 59925.16 & C & 56.7$\pm$0.3 & 9.3$\pm$1.0 & 3.9$\pm$0.7 & 2.0$\pm$0.4 & -0.8$\pm$0.2 & 0.9$\pm$0.2 & 0.4$\pm$0.1 & 0.2$\pm$0.1 & 0.7$\pm$0.1 & 0.1$\pm$0.1 & 0.1$\pm$0.1 \\
BP Tau & 59926.14 & C & 71.8$\pm$0.3 & 11.8$\pm$1.2 & 4.8$\pm$0.6 & 2.5$\pm$0.4 & -0.6$\pm$0.1 & 0.8$\pm$0.1 & 0.3$\pm$0.1 & 0.2$\pm$0.1 & 1.0$\pm$0.1 & 0.3$\pm$0.1 & 0.1$\pm$0.1 \\
BP Tau & 59927.15 & C & 77.1$\pm$0.3 & 14.9$\pm$1.5 & 5.2$\pm$0.6 & 2.6$\pm$0.3 & -0.4$\pm$0.1 & 0.9$\pm$0.1 & 0.3$\pm$0.1 & 0.5$\pm$0.1 & 1.4$\pm$0.1 & 0.3$\pm$0.1 & 0.3$\pm$0.1 \\
BP Tau & 59928.11 & E & 60.4$\pm$0.9 & 8.8$\pm$1.8 & 5.4$\pm$2.6 & 3.4$\pm$2.0 & -0.1$\pm$0.8 & 0.5$\pm$0.7 & 0.1$\pm$0.5 & 0.3$\pm$0.4 & 0.7$\pm$0.2 & 0.2$\pm$0.2 & 0.2$\pm$0.2 \\
BP Tau & 59928.13 & C & 63.3$\pm$0.3 & 9.3$\pm$1.0 & 3.2$\pm$0.5 & 1.3$\pm$0.3 & -0.5$\pm$0.1 & 0.8$\pm$0.2 & 0.4$\pm$0.1 & 0.2$\pm$0.1 & 0.7$\pm$0.1 & 0.1$\pm$0.0 & 0.1$\pm$0.1 \\
BP Tau & 59929.14 & C & 50.0$\pm$0.2 & 7.9$\pm$0.9 & 3.7$\pm$0.5 & 1.6$\pm$0.3 & -0.6$\pm$0.1 & 0.7$\pm$0.1 & 0.3$\pm$0.1 & 0.0$\pm$0.1 & 0.5$\pm$0.1 & 0.0$\pm$0.0 & 0.0$\pm$0.1 \\
BP Tau & 59930.07 & E & 49.6$\pm$0.5 & 7.3$\pm$1.1 & 5.3$\pm$1.3 & 3.3$\pm$1.0 & -0.2$\pm$0.4 & 0.5$\pm$0.3 & 0.1$\pm$0.2 & 0.1$\pm$0.2 & 0.4$\pm$0.1 & 0.1$\pm$0.1 & 0.1$\pm$0.1 \\
BP Tau & 59930.13 & C & 53.2$\pm$0.3 & 7.7$\pm$0.9 & 3.2$\pm$0.6 & 2.0$\pm$0.4 & -0.5$\pm$0.2 & 1.0$\pm$0.2 & 0.4$\pm$0.1 & -0.0$\pm$0.1 & 0.6$\pm$0.1 & 0.0$\pm$0.0 & 0.1$\pm$0.1 \\
BP Tau & 59931.13 & C & 65.9$\pm$0.3 & 9.8$\pm$1.1 & 3.7$\pm$0.5 & 1.7$\pm$0.3 & -0.5$\pm$0.1 & 0.6$\pm$0.1 & 0.2$\pm$0.1 & 0.1$\pm$0.1 & 0.6$\pm$0.1 & 0.0$\pm$0.0 & 0.0$\pm$0.1 \\
BP Tau & 59932.13 & C & 54.8$\pm$0.3 & 8.3$\pm$0.9 & 2.7$\pm$0.5 & 1.0$\pm$0.3 & -0.3$\pm$0.1 & 0.6$\pm$0.1 & 0.2$\pm$0.1 & 0.1$\pm$0.1 & 0.5$\pm$0.1 & 0.1$\pm$0.0 & 0.1$\pm$0.1 \\
BP Tau & 59933.13 & C & 63.6$\pm$0.3 & 10.4$\pm$1.1 & 3.2$\pm$0.5 & 1.2$\pm$0.3 & 0.0$\pm$0.1 & 0.4$\pm$0.1 & 0.1$\pm$0.1 & 0.2$\pm$0.1 & 0.8$\pm$0.1 & 0.7$\pm$0.1 & 0.0$\pm$0.1 \\
BP Tau & 59933.17 & E & 59.7$\pm$0.7 & 9.0$\pm$1.4 & 5.8$\pm$1.8 & 3.6$\pm$1.4 & -0.2$\pm$0.5 & 0.4$\pm$0.4 & 0.0$\pm$0.3 & 0.2$\pm$0.2 & 0.5$\pm$0.2 & 0.1$\pm$0.1 & 0.2$\pm$0.1 \\
BP Tau & 59934.07 & E & 57.7$\pm$0.4 & 8.8$\pm$1.1 & 6.2$\pm$1.1 & 3.8$\pm$0.8 & -0.2$\pm$0.3 & 0.5$\pm$0.2 & 0.1$\pm$0.2 & 0.1$\pm$0.2 & 0.6$\pm$0.1 & 0.2$\pm$0.1 & 0.2$\pm$0.1 \\
BP Tau & 59934.12 & C & 64.0$\pm$0.3 & 10.9$\pm$1.2 & 3.6$\pm$0.5 & 1.4$\pm$0.3 & -0.4$\pm$0.1 & 0.7$\pm$0.1 & 0.1$\pm$0.1 & 0.1$\pm$0.1 & 0.8$\pm$0.1 & 0.2$\pm$0.0 & 0.1$\pm$0.1 \\
BP Tau & 59935.11 & C & 63.8$\pm$0.3 & 10.3$\pm$1.1 & 3.3$\pm$0.5 & 1.3$\pm$0.3 & -0.4$\pm$0.1 & 0.7$\pm$0.1 & 0.8$\pm$0.1 & 0.1$\pm$0.1 & 0.8$\pm$0.1 & 0.0$\pm$0.0 & 0.1$\pm$0.1 \\
BP Tau & 59936.08 & E & 62.1$\pm$0.6 & 9.4$\pm$1.3 & 6.2$\pm$1.5 & 3.9$\pm$1.2 & -0.1$\pm$0.4 & 0.6$\pm$0.4 & 0.1$\pm$0.2 & 0.2$\pm$0.2 & 0.7$\pm$0.1 & 0.2$\pm$0.1 & 0.2$\pm$0.1 \\
BP Tau & 59936.11 & C & 68.3$\pm$0.2 & 11.4$\pm$1.2 & 4.2$\pm$0.5 & 1.5$\pm$0.3 & -0.3$\pm$0.1 & 0.7$\pm$0.1 & 0.3$\pm$0.1 & 0.2$\pm$0.1 & 0.9$\pm$0.1 & 0.2$\pm$0.0 & 0.1$\pm$0.0 \\
BP Tau & 59937.12 & C & 57.9$\pm$0.2 & 10.8$\pm$1.1 & 4.8$\pm$0.6 & 1.9$\pm$0.3 & -0.2$\pm$0.1 & 0.8$\pm$0.1 & 0.1$\pm$0.1 & 0.1$\pm$0.1 & 0.9$\pm$0.1 & 0.2$\pm$0.0 & 0.1$\pm$0.1 \\
BP Tau & 59938.12 & C & 52.7$\pm$0.2 & 9.9$\pm$1.1 & 3.7$\pm$0.5 & 1.7$\pm$0.3 & -0.5$\pm$0.1 & 0.7$\pm$0.1 & 0.2$\pm$0.1 & 0.2$\pm$0.1 & 0.7$\pm$0.1 & 0.2$\pm$0.0 & 0.1$\pm$0.1 \\
BP Tau & 59939.11 & C & 44.1$\pm$0.3 & 7.3$\pm$0.9 & 2.8$\pm$0.6 & 1.6$\pm$0.4 & -0.4$\pm$0.2 & 0.8$\pm$0.2 & 0.3$\pm$0.1 & -0.0$\pm$0.1 & 0.7$\pm$0.1 & 0.3$\pm$0.1 & 0.0$\pm$0.1 \\
BP Tau & 59940.11 & C & 43.8$\pm$0.2 & 7.5$\pm$0.8 & 3.1$\pm$0.4 & 1.2$\pm$0.2 & 0.1$\pm$0.1 & 0.1$\pm$0.1 & 0.0$\pm$0.1 & 0.1$\pm$0.1 & 0.6$\pm$0.1 & 0.2$\pm$0.0 & -0.0$\pm$0.1 \\
BP Tau & 59941.13 & C & 41.1$\pm$0.2 & 6.6$\pm$0.7 & 3.2$\pm$0.4 & 0.8$\pm$0.2 & -0.0$\pm$0.1 & 0.3$\pm$0.1 & -0.0$\pm$0.1 & 0.1$\pm$0.1 & 0.5$\pm$0.1 & 0.1$\pm$0.0 & -0.0$\pm$0.1 \\
BP Tau & 59949.08 & C & 56.7$\pm$0.2 & 9.0$\pm$1.0 & 3.8$\pm$0.5 & 1.4$\pm$0.3 & -0.2$\pm$0.1 & 0.3$\pm$0.1 & 0.1$\pm$0.1 & 0.2$\pm$0.1 & 0.5$\pm$0.1 & 0.1$\pm$0.0 & 0.0$\pm$0.1 \\
BP Tau & 59950.08 & C & 70.5$\pm$0.3 & 11.7$\pm$1.2 & 4.7$\pm$0.6 & 1.8$\pm$0.3 & -0.2$\pm$0.1 & 0.3$\pm$0.1 & -0.0$\pm$0.1 & 0.6$\pm$0.1 & 1.1$\pm$0.1 & 0.3$\pm$0.1 & 0.1$\pm$0.1 \\
BP Tau & 59951.08 & C & 57.1$\pm$0.2 & 10.7$\pm$1.1 & 5.0$\pm$0.6 & 1.9$\pm$0.3 & -0.0$\pm$0.1 & 0.4$\pm$0.1 & 0.0$\pm$0.1 & 0.2$\pm$0.1 & 0.9$\pm$0.1 & 0.3$\pm$0.0 & 0.1$\pm$0.1 \\
BP Tau & 59952.08 & C & 55.3$\pm$0.2 & 11.1$\pm$1.2 & 5.2$\pm$0.7 & 2.1$\pm$0.3 & -0.1$\pm$0.1 & 0.4$\pm$0.1 & -0.0$\pm$0.1 & 0.3$\pm$0.1 & 1.0$\pm$0.1 & 0.2$\pm$0.0 & 0.1$\pm$0.1 \\
BP Tau & 59953.07 & C & 72.1$\pm$0.2 & 14.5$\pm$1.5 & 5.8$\pm$0.7 & 2.1$\pm$0.3 & -0.1$\pm$0.1 & 0.5$\pm$0.1 & 0.1$\pm$0.1 & 0.6$\pm$0.1 & 1.0$\pm$0.1 & 0.3$\pm$0.0 & 0.2$\pm$0.1 \\
\enddata
\tablecomments{Fluxes are in units of 10$^{13}$ erg s$^{-1}$ cm$^{-2}$. Additional 10\% uncertainty due to continuum subtraction has been included in all uncertanties except \ha and \hb.Instrument names are abbreviated as: C: CHIRON, E: ESPRESSO, X: XSHOOTER, S: SOPHIE, U: UVES, T: TCES}
\end{deluxetable*}
\end{longrotatetable}

\begin{longrotatetable}
\movetabledown=10mm
\begin{deluxetable*}{c c c | c c c c c c c c c c c} \label{tab: Line Fluxes GM Aur}
\tabletypesize{\scriptsize}
\tablecaption{Line luminosities for GM Aur}
\tablehead{
\colhead{Object} & \colhead{MJD} & \colhead{Inst.} & \colhead{H${\alpha}$} & \colhead{H${\beta}$} & \colhead{H${\gamma}$} & \colhead{H${\delta}$} & \colhead{He I${4387}$} & \colhead{He I${4471}$} & \colhead{He I${4713}$} & \colhead{He I${5015}$} & \colhead{He I${5875}$} & \colhead{He I${6678}$} & \colhead{He I${7065}$}
}
\startdata
GM Aur & 59498.99 & S & 72.4$\pm$0.3 & 4.8$\pm$0.6 & 3.2$\pm$0.7 & 1.0$\pm$0.6 & -0.1$\pm$0.2 & 0.1$\pm$0.2 & -0.1$\pm$0.1 & 0.1$\pm$0.1 & 0.9$\pm$0.1 & 0.1$\pm$0.0 & --\\
GM Aur & 59499.98 & S & 70.5$\pm$0.4 & 3.8$\pm$0.6 & 2.8$\pm$0.8 & 1.3$\pm$0.7 & -0.4$\pm$0.2 & 0.3$\pm$0.2 & 0.0$\pm$0.2 & -0.1$\pm$0.1 & 0.8$\pm$0.1 & 0.1$\pm$0.1 & --\\
GM Aur & 59501.13 & S & 71.7$\pm$0.4 & 5.4$\pm$0.8 & 4.3$\pm$0.8 & 2.4$\pm$0.7 & -0.3$\pm$0.2 & 0.4$\pm$0.2 & -0.0$\pm$0.2 & 0.0$\pm$0.1 & 1.0$\pm$0.1 & 0.1$\pm$0.1 & --\\
GM Aur & 59502.03 & S & 61.4$\pm$0.2 & 4.0$\pm$0.5 & 3.5$\pm$0.5 & 2.1$\pm$0.4 & -0.4$\pm$0.1 & 0.3$\pm$0.1 & -0.0$\pm$0.1 & -0.1$\pm$0.1 & 0.7$\pm$0.1 & 0.0$\pm$0.0 & --\\
GM Aur & 59502.96 & T & 52.2$\pm$0.6 & 3.5$\pm$1.0 & --& --& --& --& -0.1$\pm$0.3 & -0.1$\pm$0.2 & 0.2$\pm$0.1 & -0.0$\pm$0.1 & -0.0$\pm$0.1 \\
GM Aur & 59503.11 & S & 64.3$\pm$0.3 & 4.8$\pm$0.6 & 3.9$\pm$0.6 & 2.0$\pm$0.5 & -0.4$\pm$0.2 & 0.2$\pm$0.1 & -0.0$\pm$0.1 & -0.1$\pm$0.1 & 0.5$\pm$0.1 & -0.0$\pm$0.0 & --\\
GM Aur & 59504.08 & S & 69.6$\pm$0.2 & 5.1$\pm$0.6 & 3.9$\pm$0.5 & 2.2$\pm$0.4 & -0.4$\pm$0.1 & 0.2$\pm$0.1 & -0.0$\pm$0.1 & -0.1$\pm$0.1 & 0.6$\pm$0.1 & -0.0$\pm$0.0 & --\\
GM Aur & 59504.34 & X & 77.4$\pm$0.7 & 5.9$\pm$0.7 & 3.4$\pm$0.5 & 1.6$\pm$0.2 & -0.3$\pm$0.1 & 0.3$\pm$0.1 & 0.0$\pm$0.1 & -0.1$\pm$0.1 & 0.5$\pm$0.1 & -0.0$\pm$0.1 & 0.1$\pm$0.1 \\
GM Aur & 59505.02 & S & 67.5$\pm$0.2 & 5.1$\pm$0.6 & 3.8$\pm$0.6 & 2.0$\pm$0.4 & -0.4$\pm$0.1 & 0.3$\pm$0.1 & -0.0$\pm$0.1 & -0.1$\pm$0.1 & 0.8$\pm$0.1 & 0.0$\pm$0.0 & --\\
GM Aur & 59505.11 & T & 59.4$\pm$0.5 & 5.2$\pm$1.1 & --& --& --& --& -0.1$\pm$0.3 & -0.2$\pm$0.2 & 0.1$\pm$0.1 & 0.0$\pm$0.1 & 0.0$\pm$0.1 \\
GM Aur & 59505.96 & S & 74.6$\pm$0.3 & 4.6$\pm$0.6 & 2.6$\pm$0.5 & 0.6$\pm$0.4 & -0.2$\pm$0.1 & 0.1$\pm$0.1 & -0.1$\pm$0.1 & 0.0$\pm$0.1 & 0.8$\pm$0.1 & 0.1$\pm$0.0 & --\\
GM Aur & 59507.09 & T & 38.3$\pm$0.8 & 1.8$\pm$1.4 & --& --& --& --& -0.1$\pm$0.4 & -0.1$\pm$0.3 & 0.3$\pm$0.2 & 0.1$\pm$0.1 & 0.0$\pm$0.2 \\
GM Aur & 59509.23 & E & 111.0$\pm$0.6 & 13.0$\pm$1.5 & 7.1$\pm$1.4 & 4.0$\pm$1.1 & -0.3$\pm$0.4 & 0.8$\pm$0.3 & 0.0$\pm$0.2 & 0.1$\pm$0.2 & 0.7$\pm$0.1 & 0.1$\pm$0.1 & 0.4$\pm$0.1 \\
GM Aur & 59510.04 & S & 94.6$\pm$0.3 & 9.5$\pm$1.0 & 6.7$\pm$0.8 & 4.2$\pm$0.6 & -0.4$\pm$0.1 & 0.5$\pm$0.1 & -0.0$\pm$0.1 & 0.0$\pm$0.1 & 1.0$\pm$0.1 & 0.1$\pm$0.0 & --\\
GM Aur & 59510.09 & T & 78.0$\pm$0.6 & 7.8$\pm$1.2 & --& --& --& --& -0.0$\pm$0.3 & 0.0$\pm$0.2 & 0.4$\pm$0.1 & 0.1$\pm$0.1 & 0.1$\pm$0.1 \\
GM Aur & 59510.14 & T & 80.4$\pm$0.6 & 8.2$\pm$1.3 & --& --& --& --& -0.0$\pm$0.3 & 0.0$\pm$0.2 & 0.5$\pm$0.1 & 0.1$\pm$0.1 & 0.0$\pm$0.1 \\
GM Aur & 59511.97 & T & 47.5$\pm$0.6 & 1.7$\pm$1.1 & --& --& --& --& -0.1$\pm$0.3 & -0.1$\pm$0.2 & 0.1$\pm$0.1 & -0.1$\pm$0.1 & 0.0$\pm$0.1 \\
GM Aur & 59511.99 & S & 47.0$\pm$0.3 & 1.5$\pm$0.4 & 1.4$\pm$0.4 & 0.3$\pm$0.4 & -0.5$\pm$0.1 & 0.0$\pm$0.1 & -0.1$\pm$0.1 & -0.2$\pm$0.1 & 0.5$\pm$0.1 & -0.0$\pm$0.0 & --\\
GM Aur & 59512.01 & T & 46.5$\pm$0.5 & 1.9$\pm$0.9 & --& --& --& --& -0.1$\pm$0.2 & -0.1$\pm$0.2 & 0.1$\pm$0.1 & -0.0$\pm$0.1 & -0.0$\pm$0.1 \\
GM Aur & 59513.03 & T & 54.5$\pm$0.6 & 4.4$\pm$1.1 & --& --& --& --& -0.1$\pm$0.3 & 0.0$\pm$0.2 & 0.4$\pm$0.1 & 0.0$\pm$0.1 & 0.1$\pm$0.1 \\
GM Aur & 59513.08 & T & 46.7$\pm$0.7 & 3.1$\pm$1.3 & --& --& --& --& -0.2$\pm$0.3 & -0.1$\pm$0.3 & 0.2$\pm$0.1 & 0.0$\pm$0.1 & 0.1$\pm$0.1 \\
GM Aur & 59513.10 & S & 64.9$\pm$0.3 & 5.5$\pm$0.7 & 4.2$\pm$0.7 & 2.4$\pm$0.6 & -0.4$\pm$0.2 & 0.3$\pm$0.2 & -0.0$\pm$0.1 & -0.0$\pm$0.1 & 1.0$\pm$0.1 & 0.1$\pm$0.0 & --\\
GM Aur & 59514.07 & S & 64.8$\pm$0.3 & 5.5$\pm$0.6 & 4.2$\pm$0.6 & 2.6$\pm$0.4 & -0.3$\pm$0.1 & 0.4$\pm$0.1 & 0.1$\pm$0.1 & -0.0$\pm$0.1 & 0.9$\pm$0.1 & 0.1$\pm$0.0 & --\\
GM Aur & 59515.04 & T & 56.8$\pm$0.4 & 4.9$\pm$0.9 & --& --& --& --& -0.0$\pm$0.2 & -0.1$\pm$0.2 & 0.3$\pm$0.1 & 0.0$\pm$0.1 & 0.0$\pm$0.1 \\
GM Aur & 59515.08 & T & 55.5$\pm$0.4 & 5.0$\pm$0.9 & --& --& --& --& -0.0$\pm$0.2 & -0.1$\pm$0.2 & 0.3$\pm$0.1 & 0.0$\pm$0.1 & 0.1$\pm$0.1 \\
GM Aur & 59515.09 & S & 58.4$\pm$0.4 & 4.5$\pm$0.6 & 4.0$\pm$0.7 & 2.1$\pm$0.6 & -0.4$\pm$0.2 & 0.3$\pm$0.2 & -0.0$\pm$0.1 & -0.1$\pm$0.1 & 0.7$\pm$0.1 & 0.0$\pm$0.1 & --\\
GM Aur & 59516.01 & S & 60.1$\pm$0.3 & 5.1$\pm$0.7 & 4.2$\pm$0.7 & 2.3$\pm$0.6 & -0.4$\pm$0.2 & 0.3$\pm$0.2 & -0.0$\pm$0.1 & -0.1$\pm$0.1 & 0.7$\pm$0.1 & 0.0$\pm$0.1 & --\\
GM Aur & 59553.13 & E & 57.6$\pm$0.5 & 3.9$\pm$0.8 & 2.6$\pm$1.6 & 1.3$\pm$1.1 & -0.2$\pm$0.5 & 0.3$\pm$0.4 & 0.1$\pm$0.2 & -0.2$\pm$0.2 & 0.1$\pm$0.1 & -0.0$\pm$0.1 & 0.1$\pm$0.1 \\
GM Aur & 59554.17 & E & 56.2$\pm$0.8 & 4.2$\pm$1.4 & 3.3$\pm$2.5 & 1.8$\pm$1.7 & 0.6$\pm$0.8 & 0.3$\pm$0.7 & 0.1$\pm$0.4 & -0.1$\pm$0.3 & 0.2$\pm$0.2 & 0.0$\pm$0.1 & 0.2$\pm$0.1 \\
GM Aur & 59555.19 & E & 59.8$\pm$0.4 & 6.4$\pm$0.8 & 4.2$\pm$1.1 & 2.2$\pm$0.9 & -0.2$\pm$0.3 & 0.6$\pm$0.3 & 0.1$\pm$0.2 & 0.0$\pm$0.1 & 0.9$\pm$0.1 & 0.3$\pm$0.1 & 0.4$\pm$0.1 \\
GM Aur & 59556.15 & X & 65.6$\pm$0.4 & 7.6$\pm$0.8 & 4.8$\pm$0.5 & 2.9$\pm$0.3 & -0.2$\pm$0.1 & 0.5$\pm$0.1 & 0.1$\pm$0.1 & 0.0$\pm$0.1 & 1.0$\pm$0.1 & 0.2$\pm$0.1 & 0.2$\pm$0.1 \\
GM Aur & 59556.19 & E & 66.3$\pm$0.5 & 7.1$\pm$1.1 & 4.9$\pm$1.8 & 2.9$\pm$1.4 & -0.2$\pm$0.5 & 0.5$\pm$0.4 & 0.1$\pm$0.3 & 0.0$\pm$0.2 & 0.7$\pm$0.1 & 0.2$\pm$0.1 & 0.3$\pm$0.1 \\
GM Aur & 59558.86 & T & 50.7$\pm$0.5 & 4.8$\pm$1.1 & --& --& --& --& -0.0$\pm$0.3 & -0.1$\pm$0.2 & 0.4$\pm$0.1 & 0.1$\pm$0.1 & 0.1$\pm$0.1 \\
GM Aur & 59558.90 & T & 48.3$\pm$0.6 & 4.5$\pm$1.1 & --& --& --& --& -0.1$\pm$0.3 & -0.0$\pm$0.2 & 0.3$\pm$0.1 & 0.0$\pm$0.1 & 0.0$\pm$0.1 \\
GM Aur & 59559.91 & T & 36.4$\pm$0.5 & 2.4$\pm$0.9 & --& --& --& --& -0.1$\pm$0.2 & -0.1$\pm$0.2 & 0.2$\pm$0.1 & 0.0$\pm$0.1 & 0.0$\pm$0.1 \\
GM Aur & 59559.95 & T & 35.9$\pm$0.5 & 2.5$\pm$1.0 & --& --& --& --& -0.1$\pm$0.2 & -0.1$\pm$0.2 & 0.2$\pm$0.1 & 0.0$\pm$0.1 & 0.0$\pm$0.1 \\
GM Aur & 59910.17 & E & 71.6$\pm$0.5 & 7.6$\pm$1.0 & 4.7$\pm$1.1 & 2.3$\pm$0.8 & -0.4$\pm$0.3 & 0.4$\pm$0.3 & 0.0$\pm$0.2 & -0.1$\pm$0.2 & 0.5$\pm$0.1 & 0.1$\pm$0.1 & 0.2$\pm$0.1 \\
GM Aur & 59910.21 & C & 75.5$\pm$0.2 & 8.3$\pm$0.9 & 2.9$\pm$0.4 & 1.0$\pm$0.2 & -0.5$\pm$0.1 & 0.6$\pm$0.1 & 0.1$\pm$0.1 & -0.2$\pm$0.1 & 0.6$\pm$0.1 & 0.1$\pm$0.0 & 0.1$\pm$0.0 \\
GM Aur & 59911.19 & C & 67.1$\pm$0.2 & 5.5$\pm$0.7 & 2.0$\pm$0.4 & 0.6$\pm$0.2 & -0.5$\pm$0.1 & 0.5$\pm$0.1 & 0.2$\pm$0.1 & -0.2$\pm$0.1 & 0.4$\pm$0.1 & -0.0$\pm$0.0 & 0.1$\pm$0.0 \\
GM Aur & 59912.20 & C & 71.1$\pm$0.2 & 7.1$\pm$0.8 & 2.6$\pm$0.4 & 0.8$\pm$0.2 & -0.5$\pm$0.1 & 0.4$\pm$0.1 & 0.1$\pm$0.1 & -0.2$\pm$0.1 & 0.5$\pm$0.1 & -0.1$\pm$0.0 & 0.1$\pm$0.0 \\
GM Aur & 59913.18 & C & 93.9$\pm$0.3 & 9.2$\pm$1.0 & 2.4$\pm$0.5 & 0.5$\pm$0.3 & -0.5$\pm$0.1 & 0.7$\pm$0.1 & 0.3$\pm$0.1 & -0.2$\pm$0.1 & 0.7$\pm$0.1 & 0.0$\pm$0.0 & 0.1$\pm$0.1 \\
GM Aur & 59914.19 & C & 77.2$\pm$0.2 & 7.1$\pm$0.8 & 0.8$\pm$0.4 & 0.3$\pm$0.3 & -0.4$\pm$0.1 & 0.5$\pm$0.1 & 0.3$\pm$0.1 & -0.1$\pm$0.1 & 0.7$\pm$0.1 & -0.1$\pm$0.0 & 0.1$\pm$0.0 \\
GM Aur & 59915.13 & E & 68.3$\pm$1.2 & 6.9$\pm$2.3 & 4.4$\pm$3.0 & 2.5$\pm$2.1 & -0.3$\pm$0.9 & 0.4$\pm$0.9 & 0.1$\pm$0.6 & -0.1$\pm$0.5 & 0.5$\pm$0.3 & 0.0$\pm$0.2 & 0.2$\pm$0.2 \\
GM Aur & 59915.20 & C & 69.9$\pm$0.3 & 7.5$\pm$0.9 & 2.0$\pm$0.5 & 0.6$\pm$0.3 & 1.0$\pm$0.2 & 0.0$\pm$0.1 & -0.1$\pm$0.1 & -0.2$\pm$0.1 & 0.7$\pm$0.1 & 0.2$\pm$0.0 & 0.1$\pm$0.0 \\
GM Aur & 59916.16 & E & 65.2$\pm$0.9 & 7.3$\pm$1.8 & 4.7$\pm$2.4 & 2.4$\pm$1.6 & -0.3$\pm$0.7 & 0.4$\pm$0.6 & 0.1$\pm$0.5 & -0.1$\pm$0.4 & 0.5$\pm$0.2 & -0.0$\pm$0.2 & 0.2$\pm$0.2 \\
GM Aur & 59916.21 & C & 70.3$\pm$0.3 & 8.0$\pm$0.9 & 3.0$\pm$0.5 & 0.3$\pm$0.2 & -0.2$\pm$0.1 & 0.1$\pm$0.1 & -0.0$\pm$0.1 & -0.1$\pm$0.1 & 0.7$\pm$0.1 & 0.0$\pm$0.0 & 0.1$\pm$0.1 \\
GM Aur & 59917.19 & C & 76.1$\pm$0.2 & 8.8$\pm$1.0 & 3.3$\pm$0.5 & 0.8$\pm$0.3 & -0.3$\pm$0.1 & 0.1$\pm$0.1 & -0.0$\pm$0.1 & -0.1$\pm$0.1 & 0.7$\pm$0.1 & -0.0$\pm$0.0 & 0.1$\pm$0.1 \\
GM Aur & 59918.19 & C & 70.8$\pm$0.2 & 7.4$\pm$0.8 & 2.9$\pm$0.5 & 0.7$\pm$0.3 & -0.2$\pm$0.1 & 0.0$\pm$0.1 & -0.1$\pm$0.1 & -0.1$\pm$0.1 & 0.7$\pm$0.1 & 0.0$\pm$0.0 & 0.1$\pm$0.0 \\
GM Aur & 59919.18 & C & 69.8$\pm$0.2 & 6.9$\pm$0.8 & 2.2$\pm$0.4 & 0.4$\pm$0.2 & -0.2$\pm$0.1 & 0.2$\pm$0.1 & -0.0$\pm$0.1 & -0.1$\pm$0.1 & 0.6$\pm$0.1 & 0.1$\pm$0.0 & 0.1$\pm$0.0 \\
GM Aur & 59920.18 & C & 52.9$\pm$0.2 & 5.2$\pm$0.6 & 1.5$\pm$0.4 & 0.4$\pm$0.3 & -0.4$\pm$0.1 & 0.2$\pm$0.1 & -0.1$\pm$0.1 & -0.2$\pm$0.1 & 0.5$\pm$0.1 & 0.0$\pm$0.0 & 0.1$\pm$0.0 \\
GM Aur & 59922.18 & C & 41.3$\pm$0.3 & 3.8$\pm$0.6 & 0.0$\pm$0.7 & 0.5$\pm$0.5 & -1.4$\pm$0.3 & 0.7$\pm$0.2 & 0.2$\pm$0.1 & -0.7$\pm$0.1 & 0.4$\pm$0.1 & -0.2$\pm$0.0 & 0.1$\pm$0.0 \\
GM Aur & 59925.18 & C & 65.9$\pm$0.3 & 6.4$\pm$0.8 & 2.5$\pm$0.5 & 0.6$\pm$0.3 & -0.7$\pm$0.1 & 1.0$\pm$0.2 & 0.3$\pm$0.1 & -0.2$\pm$0.1 & 0.7$\pm$0.1 & 0.1$\pm$0.0 & 0.1$\pm$0.1 \\
GM Aur & 59926.17 & C & 75.6$\pm$0.3 & 7.0$\pm$0.8 & 2.2$\pm$0.6 & 0.7$\pm$0.4 & -0.5$\pm$0.2 & 0.8$\pm$0.2 & 0.2$\pm$0.1 & -0.2$\pm$0.1 & 0.8$\pm$0.1 & 0.1$\pm$0.0 & 0.2$\pm$0.1 \\
GM Aur & 59927.18 & C & 65.4$\pm$0.2 & 5.5$\pm$0.7 & 1.8$\pm$0.4 & 0.4$\pm$0.2 & -0.5$\pm$0.1 & 0.7$\pm$0.1 & 0.1$\pm$0.1 & -0.2$\pm$0.1 & 0.9$\pm$0.1 & 0.0$\pm$0.0 & 0.2$\pm$0.1 \\
GM Aur & 59928.15 & C & 73.2$\pm$0.3 & 6.3$\pm$0.7 & 1.6$\pm$0.5 & 0.2$\pm$0.3 & -0.5$\pm$0.1 & 0.6$\pm$0.2 & 0.4$\pm$0.1 & -0.3$\pm$0.1 & 0.7$\pm$0.1 & 0.1$\pm$0.0 & 0.2$\pm$0.1 \\
GM Aur & 59929.16 & C & 93.4$\pm$0.3 & 10.1$\pm$1.1 & 3.8$\pm$0.5 & 1.2$\pm$0.3 & -0.6$\pm$0.1 & 0.8$\pm$0.1 & 0.3$\pm$0.1 & -0.1$\pm$0.1 & 1.0$\pm$0.1 & 0.1$\pm$0.0 & 0.2$\pm$0.1 \\
GM Aur & 59931.14 & C & 91.2$\pm$0.3 & 9.6$\pm$1.0 & 3.1$\pm$0.5 & 1.5$\pm$0.3 & -0.6$\pm$0.1 & 0.8$\pm$0.2 & 0.3$\pm$0.1 & -0.1$\pm$0.1 & 1.0$\pm$0.1 & 0.1$\pm$0.0 & 0.2$\pm$0.1 \\
\enddata
\tablecomments{Fluxes are in units of 10$^{13}$ erg s$^{-1}$ cm$^{-2}$. Additional 10\% uncertainty due to continuum subtraction has been included in all uncertainties except \ha and \hb.Instrument names are abbreviated as: C: CHIRON, E: ESPRESSO, X: XSHOOTER, S: SOPHIE, U: UVES, T: TCES}
\end{deluxetable*}
\end{longrotatetable}

%% file: main.bbl
\begin{thebibliography}{}
\expandafter\ifx\csname natexlab\endcsname\relax\def\natexlab#1{#1}\fi
\providecommand{\url}[1]{\href{#1}{#1}}
\providecommand{\dodoi}[1]{doi:~\href{http://doi.org/#1}{\nolinkurl{#1}}}
\providecommand{\doeprint}[1]{\href{http://ascl.net/#1}{\nolinkurl{http://ascl.net/#1}}}
\providecommand{\doarXiv}[1]{\href{https://arxiv.org/abs/#1}{\nolinkurl{https://arxiv.org/abs/#1}}}

\bibitem[{{Alcal{\'a}} {et~al.}(2014){Alcal{\'a}}, {Natta}, {Manara}, {Spezzi},
  {Stelzer}, {Frasca}, {Biazzo}, {Covino}, {Randich}, {Rigliaco}, {Testi},
  {Comer{\'o}n}, {Cupani}, \& {D'Elia}}]{Alcala2014}
{Alcal{\'a}}, J.~M., {Natta}, A., {Manara}, C.~F., {et~al.} 2014, \aap, 561,
  A2, \dodoi{10.1051/0004-6361/201322254}

\bibitem[{{Alcal{\'a}} {et~al.}(2017){Alcal{\'a}}, {Manara}, {Natta}, {Frasca},
  {Testi}, {Nisini}, {Stelzer}, {Williams}, {Antoniucci}, {Biazzo}, {Covino},
  {Esposito}, {Getman}, \& {Rigliaco}}]{Alcala2017}
{Alcal{\'a}}, J.~M., {Manara}, C.~F., {Natta}, A., {et~al.} 2017, \aap, 600,
  A20, \dodoi{10.1051/0004-6361/201629929}

\bibitem[{{Alencar} {et~al.}(2012){Alencar}, {Bouvier}, {Walter}, {Dougados},
  {Donati}, {Kurosawa}, {Romanova}, {Bonfils}, {Lima}, {Massaro}, {Ibrahimov},
  \& {Poretti}}]{Alencar2012}
{Alencar}, S.~H.~P., {Bouvier}, J., {Walter}, F.~M., {et~al.} 2012, \aap, 541,
  A116, \dodoi{10.1051/0004-6361/201118395}

\bibitem[{{Allard}(2014)}]{btsettl}
{Allard}, F. 2014, in Exploring the Formation and Evolution of Planetary
  Systems, ed. M.~{Booth}, B.~C. {Matthews}, \& J.~R. {Graham}, Vol. 299,
  271--272, \dodoi{10.1017/S1743921313008545}

\bibitem[{{Armeni} {et~al.}(in prep.){Armeni}, {Stelzer}, {Frasca}, {Manara},
  {Walter}, {Campbell-White}, {Alcal\'a}, {Gameiro}, \& {Gangi}}]{Armeniinprep}
{Armeni}, A., {Stelzer}, B., {Frasca}, A., {et~al.} in prep., \aap

\bibitem[{{Blinova} {et~al.}(2016){Blinova}, {Romanova}, \&
  {Lovelace}}]{Blinova2016}
{Blinova}, A.~A., {Romanova}, M.~M., \& {Lovelace}, R.~V.~E. 2016, \mnras, 459,
  2354, \dodoi{10.1093/mnras/stw786}

\bibitem[{{Bouvier} {et~al.}(2023){Bouvier}, {Sousa}, {Pouilly}, {Almenara},
  {Donati}, {Alencar}, {Frasca}, {Grankin}, {Carmona}, {Pantolmos}, {Zaire},
  {Bonfils}, {Bayo}, {Rebull}, {Alonso-Santiago}, {Gameiro}, {Cook}, \&
  {Artigau}}]{Bouvier2023}
{Bouvier}, J., {Sousa}, A., {Pouilly}, K., {et~al.} 2023, \aap, 672, A5,
  \dodoi{10.1051/0004-6361/202245342}

\bibitem[{{Calvet} \& {Gullbring}(1998)}]{Calvet1998}
{Calvet}, N., \& {Gullbring}, E. 1998, \apj, 509, 802, \dodoi{10.1086/306527}

\bibitem[{{Dekker} {et~al.}(2000){Dekker}, {D'Odorico}, {Kaufer}, {Delabre}, \&
  {Kotzlowski}}]{UVES}
{Dekker}, H., {D'Odorico}, S., {Kaufer}, A., {Delabre}, B., \& {Kotzlowski}, H.
  2000, in Society of Photo-Optical Instrumentation Engineers (SPIE) Conference
  Series, Vol. 4008, Optical and IR Telescope Instrumentation and Detectors,
  ed. M.~{Iye} \& A.~F. {Moorwood}, 534--545, \dodoi{10.1117/12.395512}

\bibitem[{{Donati} {et~al.}(2007){Donati}, {Jardine}, {Gregory}, {Petit},
  {Bouvier}, {Dougados}, {M{\'e}nard}, {Collier Cameron}, {Harries}, {Jeffers},
  \& {Paletou}}]{Donati2007}
{Donati}, J.~F., {Jardine}, M.~M., {Gregory}, S.~G., {et~al.} 2007, \mnras,
  380, 1297, \dodoi{10.1111/j.1365-2966.2007.12194.x}

\bibitem[{{Donati} {et~al.}(2008){Donati}, {Jardine}, {Gregory}, {Petit},
  {Paletou}, {Bouvier}, {Dougados}, {M{\'e}nard}, {Collier Cameron}, {Harries},
  {Hussain}, {Unruh}, {Morin}, {Marsden}, {Manset}, {Auri{\`e}re}, {Catala}, \&
  {Alecian}}]{Donati2008}
---. 2008, \mnras, 386, 1234, \dodoi{10.1111/j.1365-2966.2008.13111.x}

\bibitem[{{Donati} {et~al.}(2010){Donati}, {Skelly}, {Bouvier}, {Gregory},
  {Grankin}, {Jardine}, {Hussain}, {M{\'e}nard}, {Dougados}, {Unruh},
  {Mohanty}, {Auri{\`e}re}, {Morin}, {Far{\`e}s}, \& {MAPP
  Collaboration}}]{Donati2010}
{Donati}, J.~F., {Skelly}, M.~B., {Bouvier}, J., {et~al.} 2010, \mnras, 409,
  1347, \dodoi{10.1111/j.1365-2966.2010.17409.x}

\bibitem[{{Donati} {et~al.}(2011){Donati}, {Gregory}, {Alencar}, {Bouvier},
  {Hussain}, {Skelly}, {Dougados}, {Jardine}, {M{\'e}nard}, {Romanova}, \&
  {Unruh}}]{Donati2011}
{Donati}, J.~F., {Gregory}, S.~G., {Alencar}, S.~H.~P., {et~al.} 2011, \mnras,
  417, 472, \dodoi{10.1111/j.1365-2966.2011.19288.x}

\bibitem[{{Dupree} {et~al.}(2014){Dupree}, {Brickhouse}, {Cranmer}, {Berlind},
  {Strader}, \& {Smith}}]{Dupree2014}
{Dupree}, A.~K., {Brickhouse}, N.~S., {Cranmer}, S.~R., {et~al.} 2014, \apj,
  789, 27, \dodoi{10.1088/0004-637X/789/1/27}

\bibitem[{{Espaillat} {et~al.}(2008){Espaillat}, {Muzerolle}, {Hern{\'a}ndez},
  {Brice{\~n}o}, {Calvet}, {D'Alessio}, {McClure}, {Watson}, {Hartmann}, \&
  {Sargent}}]{Espaillat2008}
{Espaillat}, C., {Muzerolle}, J., {Hern{\'a}ndez}, J., {et~al.} 2008, \apjl,
  689, L145, \dodoi{10.1086/595869}

\bibitem[{{Espaillat} {et~al.}(2021){Espaillat}, {Robinson}, {Romanova},
  {Thanathibodee}, {Wendeborn}, {Calvet}, {Reynolds}, \& {Muzerolle}}]{Nature}
{Espaillat}, C.~C., {Robinson}, C.~E., {Romanova}, M.~M., {et~al.} 2021, \nat,
  597, 41, \dodoi{10.1038/s41586-021-03751-5}

\bibitem[{{Espaillat} {et~al.}(2022){Espaillat}, {Herczeg}, {Thanathibodee},
  {Pittman}, {Calvet}, {Arulanantham}, {France}, {Serna}, {Hern{\'a}ndez},
  {K{\'o}sp{\'a}l}, {Walter}, {Frasca}, {Fischer}, {Johns-Krull}, {Schneider},
  {Robinson}, {Edwards}, {{\'A}brah{\'a}m}, {Fang}, {Erkal}, {Manara},
  {Alcal{\'a}}, {Alecian}, {Alexander}, {Alonso-Santiago}, {Antoniucci},
  {Ardila}, {Banzatti}, {Benisty}, {Bergin}, {Biazzo}, {Brice{\~n}o},
  {Campbell-White}, {Cleeves}, {Coffey}, {Eisl{\"o}ffel}, {Facchini}, {Fedele},
  {Fiorellino}, {Froebrich}, {Gangi}, {Giannini}, {Grankin}, {G{\"u}nther},
  {Guo}, {Hartmann}, {Hillenbrand}, {Hinton}, {Kastner}, {Koen}, {Mauc{\'o}},
  {Mendigut{\'\i}a}, {Nisini}, {Panwar}, {Principe}, {Robberto},
  {Sicilia-Aguilar}, {Valenti}, {Wendeborn}, {Williams}, {Xu}, \&
  {Yadav}}]{Espaillat2022}
{Espaillat}, C.~C., {Herczeg}, G.~J., {Thanathibodee}, T., {et~al.} 2022, \aj,
  163, 114, \dodoi{10.3847/1538-3881/ac479d}

\bibitem[{{Fiorellino} {et~al.}(2021){Fiorellino}, {Manara}, {Nisini},
  {Ramsay}, {Antoniucci}, {Giannini}, {Biazzo}, {Alcal{\`a}}, \&
  {Fedele}}]{Fiorellino2021}
{Fiorellino}, E., {Manara}, C.~F., {Nisini}, B., {et~al.} 2021, \aap, 650, A43,
  \dodoi{10.1051/0004-6361/202039264}

\bibitem[{{Fischer} {et~al.}(2023){Fischer}, {Hillenbrand}, {Herczeg},
  {Johnstone}, {Kospal}, \& {Dunham}}]{PPVII10}
{Fischer}, W.~J., {Hillenbrand}, L.~A., {Herczeg}, G.~J., {et~al.} 2023, in
  Astronomical Society of the Pacific Conference Series, Vol. 534, Protostars
  and Planets VII, ed. S.~{Inutsuka}, Y.~{Aikawa}, T.~{Muto}, K.~{Tomida}, \&
  M.~{Tamura}, 355, \dodoi{10.48550/arXiv.2203.11257}

\bibitem[{{Frasca} {et~al.}(2017){Frasca}, {Biazzo}, {Alcal{\'a}}, {Manara},
  {Stelzer}, {Covino}, \& {Antoniucci}}]{Frasca2017}
{Frasca}, A., {Biazzo}, K., {Alcal{\'a}}, J.~M., {et~al.} 2017, \aap, 602, A33,
  \dodoi{10.1051/0004-6361/201630108}

\bibitem[{{Frasca} {et~al.}(2016){Frasca}, {Molenda-{\.Z}akowicz}, {De Cat},
  {Catanzaro}, {Fu}, {Ren}, {Luo}, {Shi}, {Wu}, \& {Zhang}}]{Frasca2016}
{Frasca}, A., {Molenda-{\.Z}akowicz}, J., {De Cat}, P., {et~al.} 2016, \aap,
  594, A39, \dodoi{10.1051/0004-6361/201628337}

\bibitem[{{Gangi} {et~al.}(2022){Gangi}, {Antoniucci}, {Biazzo}, {Frasca},
  {Nisini}, {Alcal{\'a}}, {Giannini}, {Manara}, {Giunta}, {Harutyunyan},
  {Munari}, \& {Vitali}}]{Gangi2022}
{Gangi}, M., {Antoniucci}, S., {Biazzo}, K., {et~al.} 2022, \aap, 667, A124,
  \dodoi{10.1051/0004-6361/202244042}

\bibitem[{{Gravity Collaboration} {et~al.}(2020){Gravity Collaboration},
  {Garcia Lopez}, {Natta}, {Caratti o Garatti}, {Ray}, {Fedriani},
  {Koutoulaki}, {Klarmann}, {Perraut}, {Sanchez-Bermudez}, {Benisty},
  {Dougados}, {Labadie}, {Brandner}, {Garcia}, {Henning}, {Caselli}, {Duvert},
  {de Zeeuw}, {Grellmann}, {Abuter}, {Amorim}, {Baub{\"o}ck}, {Berger},
  {Bonnet}, {Buron}, {Cl{\'e}net}, {Coud{\'e} Du Foresto}, {de Wit}, {Eckart},
  {Eisenhauer}, {Filho}, {Gao}, {Garcia Dabo}, {Gendron}, {Genzel},
  {Gillessen}, {Habibi}, {Haubois}, {Haussmann}, {Hippler}, {Hubert},
  {Horrobin}, {Jimenez Rosales}, {Jocou}, {Kervella}, {Kolb}, {Lacour}, {Le
  Bouquin}, {L{\'e}na}, {Ott}, {Paumard}, {Perrin}, {Pfuhl}, {Ramirez}, {Rau},
  {Rousset}, {Scheithauer}, {Shangguan}, {Stadler}, {Straub}, {Straubmeier},
  {Sturm}, {van Dishoeck}, {Vincent}, {von Fellenberg}, {Widmann}, {Wieprecht},
  {Wiest}, {Wiezorrek}, {Woillez}, {Yazici}, \& {Zins}}]{Gravity2020}
{Gravity Collaboration}, {Garcia Lopez}, R., {Natta}, A., {et~al.} 2020, \nat,
  584, 547, \dodoi{10.1038/s41586-020-2613-1}

\bibitem[{{Gravity Collaboration} {et~al.}(2023){Gravity Collaboration},
  {Wojtczak}, {Labadie}, {Perraut}, {Tessore}, {Soulain}, {Ganci}, {Bouvier},
  {Dougados}, {Al{\'e}cian}, {Nowacki}, {Cozzo}, {Brandner}, {Caratti O
  Garatti}, {Garcia}, {Garcia Lopez}, {Sanchez-Bermudez}, {Amorim}, {Benisty},
  {Berger}, {Bourdarot}, {Caselli}, {Cl{\'e}net}, {de Zeeuw}, {Davies},
  {Drescher}, {Duvert}, {Eckart}, {Eisenhauer}, {Eupen},
  {F{\"o}rster-Schreiber}, {Gendron}, {Gillessen}, {Grant}, {Grellmann},
  {Hei{\ss}el}, {Henning}, {Hippler}, {Horrobin}, {Hubert}, {Jocou},
  {Kervella}, {Lacour}, {Lapeyr{\`e}re}, {Le Bouquin}, {L{\'e}na}, {Lutz},
  {Mang}, {Ott}, {Paumard}, {Perrin}, {Scheithauer}, {Shangguan}, {Shimizu},
  {Spezzano}, {Straub}, {Straubmeier}, {Sturm}, {van Dishoeck}, {Vincent}, \&
  {Widmann}}]{Gravity2023}
{Gravity Collaboration}, {Wojtczak}, J.~A., {Labadie}, L., {et~al.} 2023, \aap,
  669, A59, \dodoi{10.1051/0004-6361/202244675}

\bibitem[{{Gullbring} {et~al.}(1998){Gullbring}, {Hartmann}, {Brice{\~n}o}, \&
  {Calvet}}]{Gullbring1998}
{Gullbring}, E., {Hartmann}, L., {Brice{\~n}o}, C., \& {Calvet}, N. 1998, \apj,
  492, 323, \dodoi{10.1086/305032}

\bibitem[{{Hartigan} {et~al.}(1989){Hartigan}, {Hartmann}, {Kenyon}, {Hewett},
  \& {Stauffer}}]{Hartigan1989}
{Hartigan}, P., {Hartmann}, L., {Kenyon}, S., {Hewett}, R., \& {Stauffer}, J.
  1989, \apjs, 70, 899, \dodoi{10.1086/191361}

\bibitem[{{Hartmann} {et~al.}(2016){Hartmann}, {Herczeg}, \&
  {Calvet}}]{Hartmann2016}
{Hartmann}, L., {Herczeg}, G., \& {Calvet}, N. 2016, \araa, 54, 135,
  \dodoi{10.1146/annurev-astro-081915-023347}

\bibitem[{{Hartmann} {et~al.}(1994){Hartmann}, {Hewett}, \&
  {Calvet}}]{Hartmann1994}
{Hartmann}, L., {Hewett}, R., \& {Calvet}, N. 1994, \apj, 426, 669,
  \dodoi{10.1086/174104}

\bibitem[{{Herbert} {et~al.}(2023){Herbert}, {Froebrich}, \&
  {Scholz}}]{Herbert2023}
{Herbert}, C., {Froebrich}, D., \& {Scholz}, A. 2023, \mnras, 520, 5433,
  \dodoi{10.1093/mnras/stac3051}

\bibitem[{{Herczeg} \& {Hillenbrand}(2008)}]{Herczeg2008}
{Herczeg}, G.~J., \& {Hillenbrand}, L.~A. 2008, \apj, 681, 594,
  \dodoi{10.1086/586728}

\bibitem[{{Herczeg} {et~al.}(2023){Herczeg}, {Chen}, {Donati}, {Dupree},
  {Walter}, {Hillenbrand}, {Johns-Krull}, {Manara}, {G{\"u}nther}, {Fang},
  {Schneider}, {Valenti}, {Alencar}, {Venuti}, {Alcal{\'a}}, {Frasca},
  {Arulanantham}, {Linsky}, {Bouvier}, {Brickhouse}, {Calvet}, {Espaillat},
  {Campbell-White}, {Carpenter}, {Chang}, {Cruz}, {Dahm}, {Eisl{\"o}ffel},
  {Edwards}, {Fischer}, {Guo}, {Henning}, {Ji}, {Jose}, {Kastner}, {Launhardt},
  {Principe}, {Robinson}, {Serna}, {Siwak}, {Sterzik}, \&
  {Takasao}}]{Herczeg2023}
{Herczeg}, G.~J., {Chen}, Y., {Donati}, J.-F., {et~al.} 2023, \apj, 956, 102,
  \dodoi{10.3847/1538-4357/acf468}

\bibitem[{{Ingleby} {et~al.}(2013){Ingleby}, {Calvet}, {Herczeg}, {Blaty},
  {Walter}, {Ardila}, {Alexander}, {Edwards}, {Espaillat}, {Gregory},
  {Hillenbrand}, \& {Brown}}]{Ingleby2013}
{Ingleby}, L., {Calvet}, N., {Herczeg}, G., {et~al.} 2013, \apj, 767, 112,
  \dodoi{10.1088/0004-637X/767/2/112}

\bibitem[{{J{\"o}nsson} {et~al.}(2020){J{\"o}nsson}, {Holtzman}, {Allende
  Prieto}, {Cunha}, {Garc{\'\i}a-Hern{\'a}ndez}, {Hasselquist}, {Masseron},
  {Osorio}, {Shetrone}, {Smith}, {Stringfellow}, {Bizyaev}, {Edvardsson},
  {Majewski}, {M{\'e}sz{\'a}ros}, {Souto}, {Zamora}, {Beaton}, {Bovy}, {Donor},
  {Pinsonneault}, {Poovelil}, \& {Sobeck}}]{Jonsson2020}
{J{\"o}nsson}, H., {Holtzman}, J.~A., {Allende Prieto}, C., {et~al.} 2020, \aj,
  160, 120, \dodoi{10.3847/1538-3881/aba592}

\bibitem[{{Long} {et~al.}(2011){Long}, {Romanova}, {Kulkarni}, \&
  {Donati}}]{Long2011}
{Long}, M., {Romanova}, M.~M., {Kulkarni}, A.~K., \& {Donati}, J.~F. 2011,
  \mnras, 413, 1061, \dodoi{10.1111/j.1365-2966.2010.18193.x}

\bibitem[{{Manara} {et~al.}(2023){Manara}, {Ansdell}, {Rosotti}, {Hughes},
  {Armitage}, {Lodato}, \& {Williams}}]{Manara2023}
{Manara}, C.~F., {Ansdell}, M., {Rosotti}, G.~P., {et~al.} 2023, in
  Astronomical Society of the Pacific Conference Series, Vol. 534, Protostars
  and Planets VII, ed. S.~{Inutsuka}, Y.~{Aikawa}, T.~{Muto}, K.~{Tomida}, \&
  M.~{Tamura}, 539, \dodoi{10.48550/arXiv.2203.09930}

\bibitem[{{Manara} {et~al.}(2013){Manara}, {Beccari}, {Da Rio}, {De Marchi},
  {Natta}, {Ricci}, {Robberto}, \& {Testi}}]{Manara2013}
{Manara}, C.~F., {Beccari}, G., {Da Rio}, N., {et~al.} 2013, \aap, 558, A114,
  \dodoi{10.1051/0004-6361/201321866}

\bibitem[{{Manara} {et~al.}(2020){Manara}, {Natta}, {Rosotti}, {Alcal{\'a}},
  {Nisini}, {Lodato}, {Testi}, {Pascucci}, {Hillenbrand}, {Carpenter},
  {Scholz}, {Fedele}, {Frasca}, {Mulders}, {Rigliaco}, {Scardoni}, \&
  {Zari}}]{Manara2020}
{Manara}, C.~F., {Natta}, A., {Rosotti}, G.~P., {et~al.} 2020, \aap, 639, A58,
  \dodoi{10.1051/0004-6361/202037949}

\bibitem[{{Manara} {et~al.}(2021){Manara}, {Frasca}, {Venuti}, {Siwak},
  {Herczeg}, {Calvet}, {Hernandez}, {Tychoniec}, {Gangi}, {Alcal{\'a}},
  {Boffin}, {Nisini}, {Robberto}, {Briceno}, {Campbell-White},
  {Sicilia-Aguilar}, {McGinnis}, {Fedele}, {K{\'o}sp{\'a}l}, {{\'A}brah{\'a}m},
  {Alonso-Santiago}, {Antoniucci}, {Arulanantham}, {Bacciotti}, {Banzatti},
  {Beccari}, {Benisty}, {Biazzo}, {Bouvier}, {Cabrit}, {Caratti o Garatti},
  {Coffey}, {Covino}, {Dougados}, {Eisl{\"o}ffel}, {Ercolano}, {Espaillat},
  {Erkal}, {Facchini}, {Fang}, {Fiorellino}, {Fischer}, {France}, {Gameiro},
  {Garcia Lopez}, {Giannini}, {Ginski}, {Grankin}, {G{\"u}nther}, {Hartmann},
  {Hillenbrand}, {Hussain}, {James}, {Koutoulaki}, {Lodato}, {Mauc{\'o}},
  {Mendigut{\'\i}a}, {Mentel}, {Miotello}, {Oudmaijer}, {Rigliaco}, {Rosotti},
  {Sanchis}, {Schneider}, {Spina}, {Stelzer}, {Testi}, {Thanathibodee}, {Vink},
  {Walter}, {Williams}, \& {Zsidi}}]{Manara2021}
{Manara}, C.~F., {Frasca}, A., {Venuti}, L., {et~al.} 2021, \aap, 650, A196,
  \dodoi{10.1051/0004-6361/202140639}

\bibitem[{{Miotello} {et~al.}(2023){Miotello}, {Kamp}, {Birnstiel}, {Cleeves},
  \& {Kataoka}}]{PPVII14}
{Miotello}, A., {Kamp}, I., {Birnstiel}, T., {Cleeves}, L.~C., \& {Kataoka}, A.
  2023, in Astronomical Society of the Pacific Conference Series, Vol. 534,
  Protostars and Planets VII, ed. S.~{Inutsuka}, Y.~{Aikawa}, T.~{Muto},
  K.~{Tomida}, \& M.~{Tamura}, 501, \dodoi{10.48550/arXiv.2203.09818}

\bibitem[{{Muzerolle} {et~al.}(1998{\natexlab{a}}){Muzerolle}, {Calvet}, \&
  {Hartmann}}]{Muzerolle1998}
{Muzerolle}, J., {Calvet}, N., \& {Hartmann}, L. 1998{\natexlab{a}}, \apj, 492,
  743, \dodoi{10.1086/305069}

\bibitem[{{Muzerolle} {et~al.}(2001){Muzerolle}, {Calvet}, \&
  {Hartmann}}]{Muzerolle2001}
---. 2001, \apj, 550, 944, \dodoi{10.1086/319779}

\bibitem[{{Muzerolle} {et~al.}(1998{\natexlab{b}}){Muzerolle}, {Hartmann}, \&
  {Calvet}}]{Muzerolle1998b}
{Muzerolle}, J., {Hartmann}, L., \& {Calvet}, N. 1998{\natexlab{b}}, \aj, 116,
  2965, \dodoi{10.1086/300636}

\bibitem[{{Natta} {et~al.}(2004){Natta}, {Testi}, {Muzerolle}, {Randich},
  {Comer{\'o}n}, \& {Persi}}]{Natta2004}
{Natta}, A., {Testi}, L., {Muzerolle}, J., {et~al.} 2004, \aap, 424, 603,
  \dodoi{10.1051/0004-6361:20040356}

\bibitem[{{Nelissen} {et~al.}(2023){Nelissen}, {Natta}, {McGinnis}, {Pittman},
  {Delvaux}, \& {Ray}}]{Nelissen2023}
{Nelissen}, M., {Natta}, A., {McGinnis}, P., {et~al.} 2023, \aap, 677, A64,
  \dodoi{10.1051/0004-6361/202347231}

\bibitem[{{Nguyen} {et~al.}(2012){Nguyen}, {Brandeker}, {van Kerkwijk}, \&
  {Jayawardhana}}]{Nguyen2012}
{Nguyen}, D.~C., {Brandeker}, A., {van Kerkwijk}, M.~H., \& {Jayawardhana}, R.
  2012, \apj, 745, 119, \dodoi{10.1088/0004-637X/745/2/119}

\bibitem[{{Pepe} {et~al.}(2021){Pepe}, {Cristiani}, {Rebolo}, {Santos},
  {Dekker}, {Cabral}, {Di Marcantonio}, {Figueira}, {Lo Curto}, {Lovis},
  {Mayor}, {M{\'e}gevand}, {Molaro}, {Riva}, {Zapatero Osorio}, {Amate},
  {Manescau}, {Pasquini}, {Zerbi}, {Adibekyan}, {Abreu}, {Affolter}, {Alibert},
  {Aliverti}, {Allart}, {Allende Prieto}, {{\'A}lvarez}, {Alves}, {Avila},
  {Baldini}, {Bandy}, {Barros}, {Benz}, {Bianco}, {Borsa}, {Bourrier},
  {Bouchy}, {Broeg}, {Calderone}, {Cirami}, {Coelho}, {Conconi}, {Coretti},
  {Cumani}, {Cupani}, {D'Odorico}, {Damasso}, {Deiries}, {Delabre},
  {Demangeon}, {Dumusque}, {Ehrenreich}, {Faria}, {Fragoso}, {Genolet},
  {Genoni}, {G{\'e}nova Santos}, {Gonz{\'a}lez Hern{\'a}ndez}, {Hughes},
  {Iwert}, {Kerber}, {Knudstrup}, {Landoni}, {Lavie}, {Lillo-Box}, {Lizon},
  {Maire}, {Martins}, {Mehner}, {Micela}, {Modigliani}, {Monteiro}, {Monteiro},
  {Moschetti}, {Murphy}, {Nunes}, {Oggioni}, {Oliveira}, {Oshagh}, {Pall{\'e}},
  {Pariani}, {Poretti}, {Rasilla}, {Rebord{\~a}o}, {Redaelli}, {Santana
  Tschudi}, {Santin}, {Santos}, {S{\'e}gransan}, {Schmidt}, {Segovia},
  {Sosnowska}, {Sozzetti}, {Sousa}, {Span{\`o}}, {Su{\'a}rez Mascare{\~n}o},
  {Tabernero}, {Tenegi}, {Udry}, \& {Zanutta}}]{ESPRESSO}
{Pepe}, F., {Cristiani}, S., {Rebolo}, R., {et~al.} 2021, \aap, 645, A96,
  \dodoi{10.1051/0004-6361/202038306}

\bibitem[{{Pittman} {et~al.}(2022){Pittman}, {Espaillat}, {Robinson},
  {Thanathibodee}, {Calvet}, {Wendeborn}, {Hern{\'a}ndez}, {Manara}, {Walter},
  {{\'A}brah{\'a}m}, {Alcal{\'a}}, {Alencar}, {Arulanantham}, {Cabrit},
  {Eisl{\"o}ffel}, {Fiorellino}, {France}, {Gangi}, {Grankin}, {Herczeg},
  {K{\'o}sp{\'a}l}, {Mendigut{\'\i}a}, {Serna}, \& {Venuti}}]{Pittman2022}
{Pittman}, C.~V., {Espaillat}, C.~C., {Robinson}, C.~E., {et~al.} 2022, \aj,
  164, 201, \dodoi{10.3847/1538-3881/ac898d}

\bibitem[{{Robinson} \& {Espaillat}(2019)}]{RE19}
{Robinson}, C.~E., \& {Espaillat}, C.~C. 2019, \apj, 874, 129,
  \dodoi{10.3847/1538-4357/ab0d8d}

\bibitem[{{Roman-Duval} {et~al.}(2020){Roman-Duval}, {Proffitt}, {Taylor},
  {Monroe}, {Fischer}, {Fischer}, {Fullerton}, {Aloisi}, {Britt}, {Busko},
  {Carlberg}, {De Rosa}, {Jedrzejewski}, {Lockwood}, {Frazer}, {Hernandez},
  {James}, {Oliveira}, {Plesha}, {Riedel}, {Riley}, {Sahnow}, {Sankrit},
  {Shaw}, {Smith}, {Sohn}, {Som}, {Ubeda}, \& {Welty}}]{ULLYSES}
{Roman-Duval}, J., {Proffitt}, C.~R., {Taylor}, J.~M., {et~al.} 2020, Research
  Notes of the American Astronomical Society, 4, 205,
  \dodoi{10.3847/2515-5172/abca2f}

\bibitem[{{Sabotta} {et~al.}(2019){Sabotta}, {Kabath}, {Korth}, {Guenther},
  {Dupkala}, {Grziwa}, {Klocova}, \& {Skarka}}]{Sabotta2019}
{Sabotta}, S., {Kabath}, P., {Korth}, J., {et~al.} 2019, \mnras, 489, 2069,
  \dodoi{10.1093/mnras/stz2232}

\bibitem[{{Sicilia-Aguilar} {et~al.}(2020){Sicilia-Aguilar}, {Manara}, {de
  Boer}, {Benisty}, {Pinilla}, \& {Bouvier}}]{SiciliaAguilar2020}
{Sicilia-Aguilar}, A., {Manara}, C.~F., {de Boer}, J., {et~al.} 2020, \aap,
  633, A37, \dodoi{10.1051/0004-6361/201936565}

\bibitem[{{Siwak} {et~al.}(2014){Siwak}, {Rucinski}, {Matthews}, {Guenther},
  {Kuschnig}, {Moffat}, {Rowe}, {Sasselov}, \& {Weiss}}]{Siwak2014}
{Siwak}, M., {Rucinski}, S.~M., {Matthews}, J.~M., {et~al.} 2014, \mnras, 444,
  327, \dodoi{10.1093/mnras/stu1304}

\bibitem[{{Siwak} {et~al.}(2018){Siwak}, {Ogloza}, {Moffat}, {Matthews},
  {Rucinski}, {Kallinger}, {Kuschnig}, {Cameron}, {Weiss}, {Rowe}, {Guenther},
  \& {Sasselov}}]{Siwak2018}
{Siwak}, M., {Ogloza}, W., {Moffat}, A. F.~J., {et~al.} 2018, \mnras, 478, 758,
  \dodoi{10.1093/mnras/sty1220}

\bibitem[{{Soubiran} {et~al.}(2018){Soubiran}, {Jasniewicz}, {Chemin},
  {Zurbach}, {Brouillet}, {Panuzzo}, {Sartoretti}, {Katz}, {Le Campion},
  {Marchal}, {Hestroffer}, {Th{\'e}venin}, {Crifo}, {Udry}, {Cropper},
  {Seabroke}, {Viala}, {Benson}, {Blomme}, {Jean-Antoine}, {Huckle}, {Smith},
  {Baker}, {Damerdji}, {Dolding}, {Fr{\'e}mat}, {Gosset}, {Guerrier}, {Guy},
  {Haigron}, {Jan{\ss}en}, {Plum}, {Fabre}, {Lasne}, {Pailler}, {Panem},
  {Riclet}, {Royer}, {Tauran}, {Zwitter}, {Gueguen}, \& {Turon}}]{Soubrian2018}
{Soubiran}, C., {Jasniewicz}, G., {Chemin}, L., {et~al.} 2018, \aap, 616, A7,
  \dodoi{10.1051/0004-6361/201832795}

\bibitem[{{Sousa} {et~al.}(2023){Sousa}, {Bouvier}, {Alencar}, {Donati},
  {Dougados}, {Alecian}, {Carmona}, {Rebull}, {Cook}, {Artigau}, {Fouqu{\'e}},
  \& {Doyon}}]{Sousa2023}
{Sousa}, A.~P., {Bouvier}, J., {Alencar}, S.~H.~P., {et~al.} 2023, \aap, 670,
  A142, \dodoi{10.1051/0004-6361/202244720}

\bibitem[{{Thanathibodee} {et~al.}(2019){Thanathibodee}, {Calvet}, {Muzerolle},
  {Brice{\~n}o}, {Hern{\'a}ndez}, \& {Mauc{\'o}}}]{Thanathibodee2019}
{Thanathibodee}, T., {Calvet}, N., {Muzerolle}, J., {et~al.} 2019, \apj, 884,
  86, \dodoi{10.3847/1538-4357/ab4127}

\bibitem[{{Thanathibodee} {et~al.}(2023){Thanathibodee}, {Molina}, {Serna},
  {Calvet}, {Hern{\'a}ndez}, {Muzerolle}, \&
  {Franco-Hern{\'a}ndez}}]{Thanathibodee2023}
{Thanathibodee}, T., {Molina}, B., {Serna}, J., {et~al.} 2023, \apj, 944, 90,
  \dodoi{10.3847/1538-4357/acac84}

\bibitem[{{Thanathibodee} {et~al.}(2018){Thanathibodee}, {Calvet}, {Herczeg},
  {Brice{\~n}o}, {Clark}, {Reiter}, {Ingleby}, {McClure}, {Mauc{\'o}}, \&
  {Hern{\'a}ndez}}]{Thanathibodee2018}
{Thanathibodee}, T., {Calvet}, N., {Herczeg}, G., {et~al.} 2018, \apj, 861, 73,
  \dodoi{10.3847/1538-4357/aac5e9}

\bibitem[{{Tokovinin} {et~al.}(2013){Tokovinin}, {Fischer}, {Bonati},
  {Giguere}, {Moore}, {Schwab}, {Spronck}, \& {Szymkowiak}}]{CHIRON}
{Tokovinin}, A., {Fischer}, D.~A., {Bonati}, M., {et~al.} 2013, \pasp, 125,
  1336, \dodoi{10.1086/674012}

\bibitem[{{van Dokkum}(2001)}]{VanDokkum2001}
{van Dokkum}, P.~G. 2001, \pasp, 113, 1420, \dodoi{10.1086/323894}

\bibitem[{{Venuti} {et~al.}(2017){Venuti}, {Bouvier}, {Cody}, {Stauffer},
  {Micela}, {Rebull}, {Alencar}, {Sousa}, {Hillenbrand}, \&
  {Flaccomio}}]{Venuti2017}
{Venuti}, L., {Bouvier}, J., {Cody}, A.~M., {et~al.} 2017, \aap, 599, A23,
  \dodoi{10.1051/0004-6361/201629537}

\bibitem[{{Vernet} {et~al.}(2011){Vernet}, {Dekker}, {D'Odorico}, {Kaper},
  {Kjaergaard}, {Hammer}, {Randich}, {Zerbi}, {Groot}, {Hjorth}, {Guinouard},
  {Navarro}, {Adolfse}, {Albers}, {Amans}, {Andersen}, {Andersen}, {Binetruy},
  {Bristow}, {Castillo}, {Chemla}, {Christensen}, {Conconi}, {Conzelmann},
  {Dam}, {de Caprio}, {de Ugarte Postigo}, {Delabre}, {di Marcantonio},
  {Downing}, {Elswijk}, {Finger}, {Fischer}, {Flores}, {Fran{\c{c}}ois},
  {Goldoni}, {Guglielmi}, {Haigron}, {Hanenburg}, {Hendriks}, {Horrobin},
  {Horville}, {Jessen}, {Kerber}, {Kern}, {Kiekebusch}, {Kleszcz}, {Klougart},
  {Kragt}, {Larsen}, {Lizon}, {Lucuix}, {Mainieri}, {Manuputy}, {Martayan},
  {Mason}, {Mazzoleni}, {Michaelsen}, {Modigliani}, {Moehler}, {M{\o}ller},
  {Norup S{\o}rensen}, {N{\o}rregaard}, {P{\'e}roux}, {Patat}, {Pena}, {Pragt},
  {Reinero}, {Rigal}, {Riva}, {Roelfsema}, {Royer}, {Sacco}, {Santin},
  {Schoenmaker}, {Spano}, {Sweers}, {Ter Horst}, {Tintori}, {Tromp}, {van
  Dael}, {van der Vliet}, {Venema}, {Vidali}, {Vinther}, {Vola}, {Winters},
  {Wistisen}, {Wulterkens}, \& {Zacchei}}]{XSHOOTER}
{Vernet}, J., {Dekker}, H., {D'Odorico}, S., {et~al.} 2011, \aap, 536, A105,
  \dodoi{10.1051/0004-6361/201117752}

\bibitem[{{Wendeborn} {et~al.}(2024{\natexlab{a}}){Wendeborn}, {Espaillat},
  {Lopez}, {Thanathibodee}, {Robinson}, {Pittman}, {Calvet}, {Flors}, {Walter},
  {K{\'o}sp{\'a}l}, {Grankin}, {Mendigut{\'\i}a}, {G{\"u}nther},
  {Eisl{\"o}ffel}, {Guo}, {France}, {Fiorellino}, {Fischer}, {{\'A}brah{\'a}m},
  \& {Herczeg}}]{PaperI}
{Wendeborn}, J., {Espaillat}, C.~C., {Lopez}, S., {et~al.} 2024{\natexlab{a}},
  arXiv e-prints, arXiv:2405.21038, \dodoi{10.48550/arXiv.2405.21038}

\bibitem[{{Wendeborn} {et~al.}(2024{\natexlab{b}}){Wendeborn}, {Espaillat},
  {Thanathibodee}, {Robinson}, {Pittman}, {Calvet}, {K{\'o}sp{\'a}l},
  {Grankin}, {Walter}, {Guo}, \& {Eisl{\"o}ffel}}]{PaperII}
{Wendeborn}, J., {Espaillat}, C.~C., {Thanathibodee}, T., {et~al.}
  2024{\natexlab{b}}, arXiv e-prints, arXiv:2405.21071,
  \dodoi{10.48550/arXiv.2405.21071}

\bibitem[{{Whittet} {et~al.}(2004){Whittet}, {Shenoy}, {Clayton}, \&
  {Gordon}}]{Whittet2004}
{Whittet}, D.~C.~B., {Shenoy}, S.~S., {Clayton}, G.~C., \& {Gordon}, K.~D.
  2004, \apj, 602, 291, \dodoi{10.1086/380837}

\bibitem[{{Williams} \& {Cieza}(2011)}]{Williams2011}
{Williams}, J.~P., \& {Cieza}, L.~A. 2011, \araa, 49, 67,
  \dodoi{10.1146/annurev-astro-081710-102548}

\bibitem[{{Zhu} {et~al.}(2023){Zhu}, {Stone}, \& {Calvet}}]{Zhu2023}
{Zhu}, Z., {Stone}, J.~M., \& {Calvet}, N. 2023, arXiv e-prints,
  arXiv:2309.15318, \dodoi{10.48550/arXiv.2309.15318}

\bibitem[{{Zhu} {et~al.}(2024){Zhu}, {Stone}, \& {Calvet}}]{Zhu2024}
---. 2024, \mnras, 528, 2883, \dodoi{10.1093/mnras/stad3712}

\bibitem[{{Zsidi} {et~al.}(2022){Zsidi}, {Manara}, {K{\'o}sp{\'a}l}, {Hussain},
  {{\'A}brah{\'a}m}, {Alecian}, {B{\'o}di}, {P{\'a}l}, \& {Sarkis}}]{Zsidi2022}
{Zsidi}, G., {Manara}, C.~F., {K{\'o}sp{\'a}l}, {\'A}., {et~al.} 2022, \aap,
  660, A108, \dodoi{10.1051/0004-6361/202142203}

\end{thebibliography}
